# Elucidating Guest-Host Mechanisms in ZIF-L for Tuneable Highly Luminescent 2D Materials

*Dylan A. Sherman,[a] Lorenzo Donà,[b] Cyril Besnard,[a] Lars Mester,[c] Ben Slater,[a] Judit Farrando-Pérez,[d] Christopher Allen,[e,f] Joaquín Silvestre-Albero,[d] and Jin-Chong Tan[a*]*

[a] *Multifunctional Materials & Composites (MMC) Laboratory, Department of Engineering Science, University of Oxford, Parks Road, Oxford OX1 3PJ, United Kingdom.*

[b] *Department of Chemistry, NIS and INSTM Reference Centre, University of Turin, via Pietro Giuria 7, Torino 10125, Italy.*

[c] *Attocube Systems AG, Eglfinger Weg 2, DE-85540 Haar, Germany.*

[d] *Laboratorio de Materiales Avanzados, Departamento de Química Inorgánica-Instituto Universitario de Materiales, Universidad de Alicante, Ap. 99, E-03080 Alicante, Spain.*

[e] *Electron Physical Science Imaging Centre (ePSIC), Diamond Light Source Ltd., OX11 0DE, United Kingdom.*

[f] *Department of Materials, University of Oxford, Parks Road, Oxford, OX1 3PH, United Kingdom.*

** Corresponding author's e-mail: jin-chong.tan@eng.ox.ac.uk*

**Abstract**

3D metal-organic frameworks (MOFs) are well known effective hosts for luminescent guests that improve tuneability, photostability and material fabricability. 2D guest@MOF systems, while less explored, offer competitive advantages over 3D guest@MOF systems due to their optical transparency, inter-layer spacing, and vertical thinness. This work examines luminescent organic dyes@ZIF-L to establish the underlying mechanisms of guest incorporation in ZIF-L and demonstrate the advantages of the 2D ZIF-L architecture for the material's functional luminescence. Analysing a case study system, fluorescein@ZIF-L (F@ZIF-L), using nanoscale FTIR, diffraction, and topology mapping, confirmed that fluorescein (F) resided in the ZIF-L framework cavities. Supported by surface energy simulations, the extent of guest incorporation was found to be indicated by a morphological continuum, from the characteristic leaf-shaped ZIF-L to rectangular F@ZIF-L. By modifying synthesis temperature and solvent ratios, the luminescent properties of F@ZIF-L could be rationally tuned in terms of guest loading (% mol) and arrangement (i.e. guest monomer to aggregate ratio). When optimised, F@ZIF-L exhibited tuneable emission chromaticity, 99.7% photoluminescent quantum yield, minimal guest leaching in solution over 12 months, and high photostability. Perylene@ZIF-L exhibited unique white light emitting properties, with CIE coordinates (0.33, 0.34) arising from a combination of yellow $\alpha$-phase excimer and blue monomeric perylene emission. Finally, oriented luminescent thin films of guest@ZIF-L materials were grown on malleable Zn foils, demonstrating an *in situ* fabrication technique. Together, the work highlights the potential for luminescent dye@ZIF-L systems in developing tuneable and resilient luminescent components of next-generation optoelectronics, sensors, and lighting systems.

## Introduction

Functional luminescent materials that can be carefully tuned are critical for new generation nanoscale sensors and indicators, solid-state lighting systems, and optoelectronics.[1] White light emission (WLE), for example, requires precise attainment of Commission Internationale d l'Eclairage (CIE) coordinates at (0.33, 0.33).[2,3] While many organic dyes, such as fluorescein (F), perylene, and coumarins, can exhibit strong functional fluorescence for these purposes, as free molecules these properties are challenging to control. Fluorescein, for example, is non-emissive in the solid state due to aggregation caused quenching (ACQ) effect.[4,5]

Metal-organic frameworks (MOFs) have been employed as effective 3D structures to host organic luminescent dyes, providing a tuneable composite system of control through host-guest interactions.[6] Due to the high porosity and customisability afforded by MOF materials, a myriad of dye@MOF systems have been developed that are tailored to optimise the properties of the luminescent dye, leading to stronger emission and improved photoluminescent quantum yield (PLQY).[7] Zeolitic imidazolate frameworks (ZIFs) in particular offer large cages to encapsulate organic dyes, while also exhibiting high structural, thermal and chemical stability, attributable in part to their zeolite-type pore architecture.[8]

3D frameworks do have limitations, however, including limited optical transparency, challenging fabricability, dye leaching, self-quenching through reabsorption, and delayed sensing due to limited target molecular diffusion through the MOF superstructure.[9,10] 2D guest@MOF materials have been proven to overcome some of these issues by offering additional properties including increased external surface area, number of active sites, interlayer spacing for diffusion and molecular transport, orientability, and in some cases, thinness.[11] One particularly promising candidate is the ZIF-L ($Zn(mim)_2 \cdot (Hmim)_{1/2} \cdot (H_2O)_{3/2}$, $C_{10}H_{16}N_5O_{3/2}Zn$; Hmim = 2-methylimidazole), which is constructed by 2 modified layers that originate from the sodalite (SOD) topology of ZIF-8.[12] In each asymmetric unit, there are two crystallographically unique Zn(II) ions, each coordinated by four N atoms, four Hmim ligands and one free Hmim molecule.[13] The free Hmim, and additional terminal ligands, interact in the interlayer spacing *via* hydrogen bonds and *van der Waals* interactions, critical to preventing the adjacent 2D layers from linking to form ZIF-8 during synthesis, but also creating strong interlayer bonding.[12] The framework contains two large pore cavities: one cushion type cavity of 9.4 x 7.0 x 5.3 Å and a smaller cavity of 3.6 x 2.8 x 2.3 Å.[12] Notably, the terminal and free Hmim ligands create flexibility of the framework, with longer molecules such as $CO_2$ causing significant channel widening compared to $N_2$ (6.48 Å vs 6.13 Å).[14]

ZIF-L combines the stability and large cavities typical of ZIF materials with a 2D morphology, creating a denser framework than ZIF-8 (1.4 g/cm$^3$ vs 0.94 g/cm$^3$)[15] that is reported to have high thermal and chemical stability, negligible cytotoxicity and high surface area with exposed active sites.[12] The material was initially explored for its high $CO_2$ adsorption capacity (higher than ZIF-8 at room temperature).[12,16,17] It has also been proven to be effective as a filter for wastewater treatment,[18,19,20,21,22,23] catalyst for polymerisation,[24] and as a membrane to separate gas molecules,[13] with the interlayer space facilitating mass diffusion. Usefully, ZIF-L can be grown along the *b*-axis to create oriented heteroepitaxial growth for thin film coatings, attributed to the *van der Drift* growth model where to obtain sufficient growth space the crystals preferentially grow upwards, preferring orientation along the longest axis.[19,25,26,27,28] Oriented films are critical for high performance thin-film optoelectronics and sensors.

Despite ZIF-L being an appealing structure for guest@host engineering, due to the large cavities, interlayer channels, and framework flexibility, very few reports of guest@ZIF-L solid-state systems exist, and luminescent ZIF-L materials are limited. While ZIF-L exhibits emission under 365 nm in the violet-deep blue region due to the $\pi$–$\pi$* transition of the Hmim ligand, it is typically not intense or sensitive enough to provide useful functionality alone.[29,30] Phosphate has been doped in ZIF-L to enhance the natural fluorescence of the framework to enable $Fe^{3+}$ detection.[31] There are also four reports of carbon dot@ZIF-L composites to create fluorescent probes for detecting cyclines,[32] metal ions,[30,33] or be used as components of antimicrobial hydrogels to identify and adsorb $Cu^{2+}$ ions.[34] While these developments show the effectiveness of ZIF-L composites for luminescent sensors, works have not attempted to understand the extent of guest encapsulation into ZIF-L or the underlying mechanisms of guest incorporation. Indeed, a number of works identify ZIF-L particles widening on guest encapsulation, but note further investigation is required to interpret these observations.[32]

Herein, we report a collection of highly luminescent dye@ZIF-L materials. Fluorescein (F) was selected as the candidate guest given its character as a strong green emitter and the need to be ‘turned on’ in the solid state by leveraging host-guest interactions to prevent ACQ. The dimensions of F’s molecular core (8.5 x 6.9 x 5.0 Å) are theoretically compatible with the flexible ZIF-L pore, and F itself has a degree of flexibility *via* rotation of the phenyl substituent. Additionally, we synthesised perylene@ZIF-L that exhibits WLE upon encapsulation. The luminescence of F@ZIF-L and perylene@ZIF-L was found to be highly tuneable based on guest loading and synthesis conditions that varied the arrangement of guests in the 2D host framework. Employing *ab initio* density functional theory (DFT) calculations and nanoscale analysis techniques, the mechanisms behind guest encapsulation in ZIF-L were elucidated, revealing that the degree of guest confinement into ZIF-L pores is indicated by a 2D morphological progression from the pointed leaf shaped particles to wider ovals, then rectangles. Finally, exploiting the orientable growth of ZIF-L, we demonstrate the facile growth of luminescent thin films on Zn foils to produce functional fluorescent lighting surfaces.

**Materials synthesis**

*In situ* guest encapsulation approach was used to synthesise F@ZIF-L. The typical ZIF-L synthesis,[12,35] was modified by dissolving guest molecules into the reagent mixture before stirring (see Methods in Supporting Information). Typically, ZIF-L synthesis is $H_2O$ based, as water can hydrolyze and postpone growth into a third dimension to form ZIF-8.[36] After the N amine of the Hmim ligand forms hydrogen bonds with water, it induces the formation of further H-bonded Hmim ligands, linking sodalite layers to form ZIF-L. However, the solubility of F in water is very low: at the 80 mL scale used for synthesis, only 0.10 mg of F fully dissolved. F is more soluble in organic solvents such as MeOH and EtOH. These solvents, however, cause the amine H to dissociate and coordinate with $Zn^{2+}$ ions, to then form 4-coordinated $Zn^{2+}$ *via* the dehydrogenated Hmim and assemble ZIF-8.[11] We therefore tested the addition of very dilute quantities of common organic solvents to the $H_2O$ (80 mL). Powder X-ray diffraction (PXRD) confirmed that toluene, isopropyl alcohol, acetone, acetonitrile and tetrahydrofuran all formed ZIF-8 at any concentration tested, while dimethylformamide (DMF) and ethanol (EtOH) formed ZIF-L but only up to 2 mL and 5 mL, respectively, after which ZIF-8 formed (Figure S1). Further testing was undertaken with methanol (in MeOH, F solubility of ≈ 10 mg/mL), confirming up to 10 mL can be added and still form pure phase ZIF-L, while pure phase ZIF-8 forms from 20 mL of MeOH (Figure S2).

A range of F@ZIF-L samples were synthesised: 5 using $H_2O$ only (up to 0.5 mg F content during synthesis), 16 using a constant MeOH content of 4 mL but varying F content during synthesis (from 0.01 mg to 80 mg), 3 sets of samples (0.1, 0.3, 0.5 mg) each with varying MeOH contents (1, 2, 4, 8 mL), and various samples with 0.3 mg F synthesis content with varied temperatures (from the standard 30 °C to 35, 40, 45, and 50 °C). The guest synthesis content was based on the range of guest quantities reported for ZIF-L carbon dot composites (from 1.5 mg to 25 mg).[30,32,34] All samples, after drying, formed an average of 500 mg F@ZIF-L (85% yield based on Zn content), which aligns with expected material yields from previous reports.[11]

**F@ZIF-L Bulk Structure and Composition**

PXRD patterns of F@ZIF-L up to and including 10 mg of F during synthesis (4 mL MeOH) exhibit sharp Bragg peaks and align with the experimental and simulated patterns of ZIF-L, confirming the retention of the ZIF-L crystalline framework (Figures 1a, S3). From 20 mg of F, a mixed ZIF-L/ZIF-8 phase results, with only ZIF-8 forming at 80 mg of F (Figure S4). Attenuated total reflectance-Fourier transform infrared (ATR-FTIR) and Raman spectra of F@ZIF-L with up to 5 mg of F further confirmed the retention of the ZIF-L framework, with spectra aligning with the vibrational bands of ZIF-L without any additional bands present (Figures 1b, S5-S6).

Beyond 10mg of F, a series of additional vibrational bands were observable (Figure S7). These increased in intensity with increased guest loading and align well with the FTIR spectra of solid state F, and reported simulation FTIR spectra of the dianion of F.[37] For example, at 918 $cm^{-1}$ a peak forms attributable to the out-of-plane bending of the CH group on the benzenecarboxylate unit of F. This indicates the presence of aggregated F species either on the surface of F particles, or as a mixture with F particles. Field emission scanning electron (FE-SEM) micrographs of 10 mg, 20 mg and 40 mg F@ZIF-L confirmed the latter, with large growths of F forming between ZIF-L particles (Figure S8). Nearfield infrared nanospectroscopy (nanoFTIR) was used to probe one of these aggregates, revealing the nanoFTIR spectra aligned well with the nanoFTIR spectra of the pristine F as a solid powder (Figures S9-S10). Meanwhile, the FE-SEM of the 80 mg F sample exhibits an equiaxial crystal morphology comparable to ZIF-8 (Figure S11). For these reasons, we excluded all samples of F@ZIF-L synthesised from 10 mg or more F from further studies, preferencing instead materials that exhibit behaviours typical of well incorporated guest@host systems.

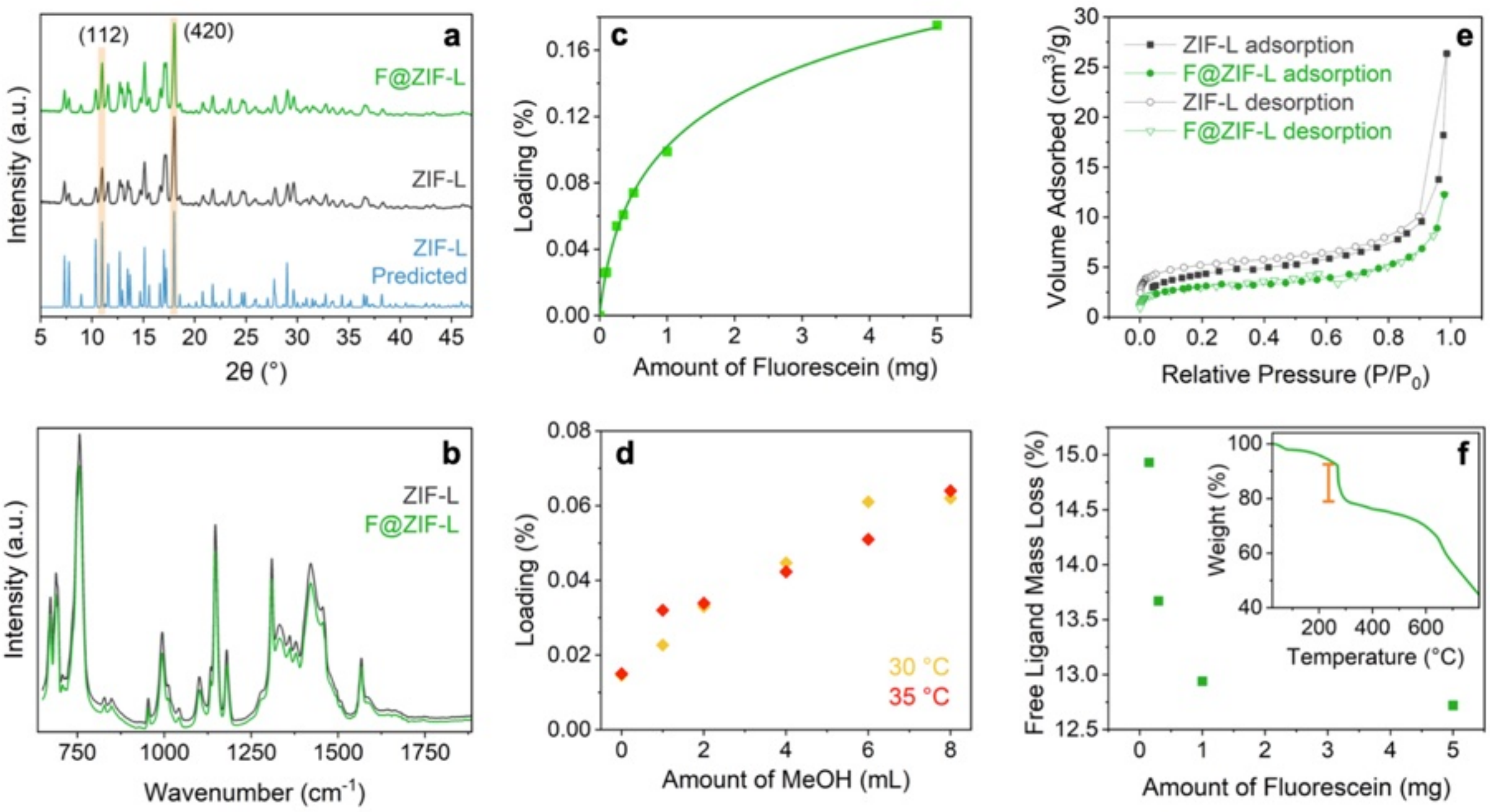

**Figure 1**. a) PXRD patterns of F@ZIF-L (0.3 mg, F) and ZIF-L, compared to the simulated XRD pattern generated from a crystallographic information file (CIF code, CCDC 1509273).[12] b) ATR-FTIR of F@ZIF-L (0.3 mg, F) and ZIF-L. c) Guest loading (% mol) of F in ZIF-L compared to synthesis quantities of F determined by $^1$H NMR digestion. d) Guest loading of F in ZIF-L from 0.3 mg synthesis quantity when MeOH synthesis content was varied. e) $N_2$ sorption isotherms at 77 K of F@ZIF-L (0.3 mg, F) compared to ZIF-L. f) TGA of F@ZIF-L (0.3 mg, F) (inset) with TGA determined free ligand mass loss (orange bar) from F@ZIF-L samples employing various F content during synthesis.

The guest loading of F being confined in F@ZIF-L was estimated using $^1$H NMR digest and the signal ratio from the 2-methylimidazolate ligand (singlet at 7.3 ppm corresponding to the two methine protons of the imidazole ring) and F (doublet at 8.3 ppm corresponding to a single proton in the ortho position relative to the carboxy group) (Figure S12). A logarithmic relationship ($R^2$ = 0.998) between guest content during synthesis and loading % mol was determined, with a maximum guest loading of 0.16 ± 0.01% mol (Figure 1c). A F@ZIF-8 material also reported a logarithmic guest loading.[37] Introducing MeOH content (up to 8 mL) into the synthesis further increased guest loading more linearly compared to $H_2O$ only synthesis (Figure 1d). When synthesised at 35 °C, loading remained similar in quantity to that observed at 30 °C and exhibited a strong correlation with increased MeOH during synthesis (Figure 1d).

$N_2$ adsorption and desorption at 77 K was investigated to examine porosity. Isotherms revealed a reduction in maximum volume of $N_2$ adsorbed from 26 cm$^3$/g for ZIF-L to 12 cm$^3$/g for F@ZIF-L (0.3 mg, F) (Figure 1e). Similar reduction in adsorption capacity of $N_2$ is well reported for guest@ZIF systems, including phosphate@ZIF-L, and is indicative of effective guest incorporation into the host framework.[31] Furthermore, the Brunauer-Emmett-Teller (BET) surface area was measured to be 15 m$^2$/g for ZIF-L, aligning with reported values,[22] while a 30% reduction to 10 m$^2$/g was observed for F@ZIF-L (Table S1). The materials exhibited some microporosity, as expected for ZIF-L, with a sharp increase in adsorption at low pressure ($P/P_0$ < 0.1).[30] A 65% reduction of microporous volume was observed for F@ZIF-L. These differences in surface area and micropores are comparable to those reported for the phosphate and carbon dot@ZIF-L composite materials (Table S1).[30,31,32,33]

Thermogravimetric analysis (TGA) appears consistent with ZIF-L reported data:[12] an initial loss at 100 °C from solvent ($H_2O$), then at 250 - 300 °C a second mass loss due to removal of the weakly linked Hmim molecules between the layers of ZIF-L, followed by framework decomposition at 550 °C to form ZnO (Figure 1f, S13). F is reported to exhibit thermal instability from 330 °C, after the second thermal loss step for ZIF-L.[37] While typical reported loss due to trapped solvent is around 4.3%,[15] F@ZIF-L materials only exhibit a 2% weight loss. Moreover, the loss due to weakly bound Hmim molecules reduces from an expected 15% loss for ZIF-L as guest loading increases to 12.72% (Figure 1f).[30,37] Close observation of ATR-FTIR spectra revealed that the broad band around 3250 $cm^{-1}$, assignable to the free Hmim,[38] gradually decreased with increased guest loading (Figures S5, S7). These data indicate that with increased F used during ZIF-L synthesis, there is (1) less solvent within the F@ZIF-L framework and (2) a progressive reduction of free Hmim (that protrude into open framework space), both indicate that instead of F guest molecules being incorporated into the porous space of the ZIF-L framework. In the absence of Hmim, F could act as a similar directing unit to control framework growth, forming similar intermolecular bonds, including H-bonds, that supress growth between the sodalite layers to yield ZIF-8.

**Morphology of Particles**

Atomic force microscopy (AFM) and FE-SEM data were correlated to observe particle morphological changes for F@ZIF-L (Figure 2a-q). ZIF-L appears in the well-known

'leaf' morphology with average dimensions of 7.33 x 2.76 μm (Figure 2a-b). Synthesis of F@ZIF-L with low guest loadings ($H_2O$ only) produced the expected ZIF-L morphology but with smaller particle sizes (Figure S14, 2o). Increasing MeOH content during synthesis incrementally widened the particle at all guest loading concentrations, so that an oval then curved rectangular particle shape formed, along with decreasing overall particle size (Figure 2c-h, 2o). Similar particle widening was reported for carbon dot@ZIF-L materials.[32,34] Synthesising ZIF-L with the same increased MeOH content, but no guest content, resulted in the same particle widening, clarifying the change is primarily a result of the MeOH content (Figure S15). Surfaces of all particles are consistent with the arc topography of ZIF-L (Figure 2p), with no indication of guest surface aggregation.

Adjusting synthesis temperature, however, resulted in unique morphological changes attributable to guest content. Between 35-45 °C, smaller rectangular particles were synthesised when both F (0.1 mg to 0.5 mg) and MeOH (6 mL) were used in synthesis (Figure 2m-o, S16). As guest loading increased, the rectangle became longer but narrower (Figures S17-S18). From ≥ 50 °C and/or F content > 0.5 mg, an elongated hexagon formed (Figures S16-S17). From 35 °C, when MeOH was included during synthesis without F, only the elongated hexagonal particle was observed (Figure S17), while increasing temperature with F but without MeOH resulted in the typical leaf shaped morphology (Figure S19). At 35 °C when F content remained constant (0.3 mg), but MeOH content was adjusted from 0-8 mL, the particle morphed from a leaf shape to pill shape, then towards a rectangle by 6 mL (Figure S19). These data highlight that while MeOH is required to attain the ideal rectangular morphology, it is not sufficient alone, rather a combination of MeOH and F is necessary. The rectangular particles also exhibited uniformly flat surfaces, in contrast to the gradually curved topology of ZIF-L and leaf-shaped F@ZIF-L (Figures 2p, S20 for AFM mechanical phase), indicating more complete crystal growth and a structure with more effective guest encapsulation.

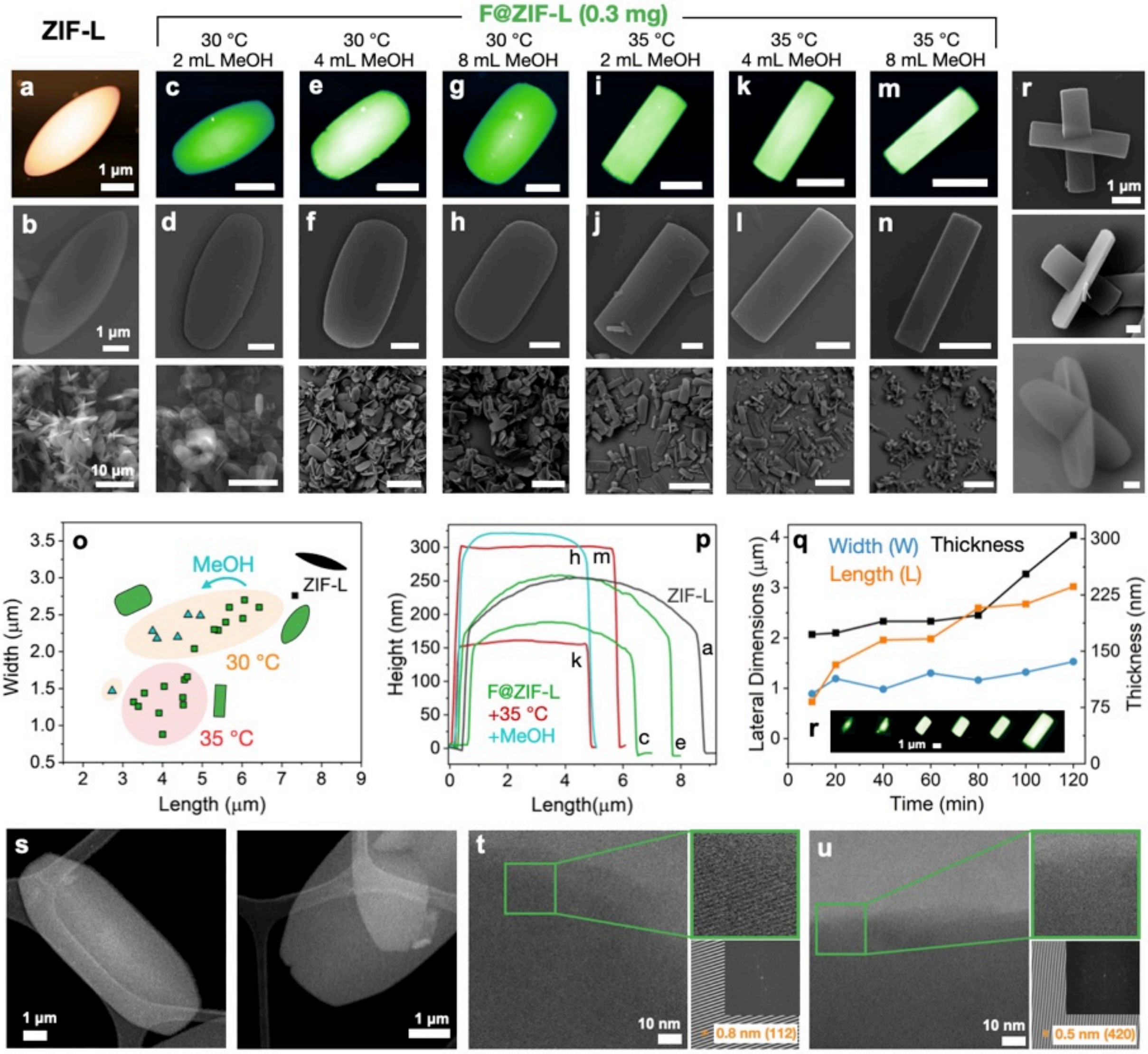

**Figure 2.** AFM and FE-SEM images, both of a single particle (above) and wider field (below) of ZIF-L (a-b), and F@ZIF-L (c-m). o) Lateral dimensions of F@ZIF-L samples (green) and ZIF-L (black) compared, highlighting clusters of data from reactions at 35 °C (red), 30 °C (orange) and 30 °C with added MeOH (blue, triangles). p) Height profiles of F@ZIF-L and ZIF-L from representative AFM images. q) Lateral dimensions and thickness of F@ZIF-L during 120-minute synthesis (inset: images of particle growth during synthesis at each 20-minute interval). s) HAADF-STEM images of F@ZIF-L (0.3 mg, F). t-u) BF-STEM images of F@ZIF-L particle edges, showing uniform domains of crystallinity. Right, above: sample zoomed in section of a nanosized domain. Right, below: Fourier transform showing *d*-spacing and the corresponding (*hkl*) lattice plane of ZIF-L.

Morphology of the rectangular F@ZIF-L was tracked during the 120-minute synthesis (35 °C, 0.3 mg F, 6 mL MeOH) (Figure 2q). Data reveals initial growth along the *b*-axis, consistent with previously reported ZIF-L growth data.[28,38] The particle formed in the first 60 minutes, after which an approximately linear increase in all dimensions was progressively observed.

Of note, all particle morphologies appear to exist in a windmill-type particle form, involving the amalgamation of 2 particles (Figure 2r). FE-SEM revealed the central intersection of the two particles was often perpendicular, and that ‘slits’ formed in some single F@ZIF-L particles that appear to be suitable growth sites for the perpendicular windmill particles (Figure S21). These windmill morphologies were observed to be a minimal percentage of F@ZIF-L samples, for which we estimated to be at most 1 in every 100 particles, observed.

**Local Structure and Energetics Influencing 2D Crystal Morphology**

To rationalise the single-crystal morphology based on structural energetics, we further establish that individual particles of F@ZIF-L do exhibit good crystallinity at the nanoscale. High-angle annular dark-field imaging scanning transmission electron microscopy (HAADF-STEM) was employed to probe the local nanostructure of F@ZIF-L particles (Figure 2s). Studies report ZIF-L is highly sensitive to electron dose, and exhibits a loss of crystallinity at >25 e/A$^2$, followed by linker decomposition at >100 e/A$^2$.[15,39] Henceforth, by maintaining a low beam intensity of 10-20 e/A$^2$, HAADF micrographs were successfully collected of F@ZIF-L (Figure 2s) and ZIF-L (Figure S22). HAADF images of F@ZIF-L confirmed the more rectangular morphology. Bright field (BF) STEM micrographs of F@ZIF-L particle edges displayed uniform crystallinity across the entire particle view (Figure 2t-u). Fourier transform was used to determine the *d*-spacing of the crystalline lattice, what was found to be ~0.8 nm in Figure 2t, and ~0.5 nm in Figure 2u (see also Figure S23). These spacings correspond to the (112) and (420) crystal lattice planes, respectively, using the (*hkl*) Miller indices; they represent the two most intense Bragg reflections observed in PXRD patterns (Figure 1a). These high-resolution STEM data confirmed the local crystalline structure of F@ZIF-L, and that this is consistent with the bulk polycrystalline data from PXRD.

Previous studies have rationalised the unique leaf-shape morphology of ZIF-L based on surface energies.[13] In brief, energies are influenced by the dangling bonds, arising from terminal and free Hmim ligands in the structure. The surface energies of the (100) and (110) facets are similar, each requiring one single bond per Zn to be broken, while the energy doubled for the (010) facet due to both a single and double coordination bond required to be broken per Zn centre (Figure 3a). The more energetically favourable assembly, therefore, minimises the presence of (010), leading to the pointed ZIF-L ends. The ZIF-L curve was previously determined to be comprised of the (100) edges along the *b*-axis and (110) comprising most of the particle curvature, towards the pointed end, connected by small surfaces of the (320), (310), and (210) facets.[13]

The most extreme F@ZIF-L morphology change, a rectangular particle (top view), contrastingly exhibits only the (100) and (010) edges along the *b*- and *a*-axis respectively, with (001) on the top and bottom surfaces perpendicular to the *c*-axis (Figure 3a). The rounded rectangular and oval-shaped F@ZIF-L particles show the incremental increase in (010) along the *a*-axis, reducing the presence of the (110) curvature contribution linking (010) to (100). This suggests the inclusion of guest molecules has varied surface energetics extensively, along with evidence that guests are well incorporated into the crystalline framework itself, rather than mere surface adhesion. Indeed, Figure 3b shows that the (110) surface intersects diagonally through the ZIF-L pore, breaking the pore, so the sequential removal of (110) surfaces observed in F@ZIF-L particles indicates an energetic preference to maintain enclosed full pore structures upon guest incorporation, instead favouring the typically higher energy (010) surface.

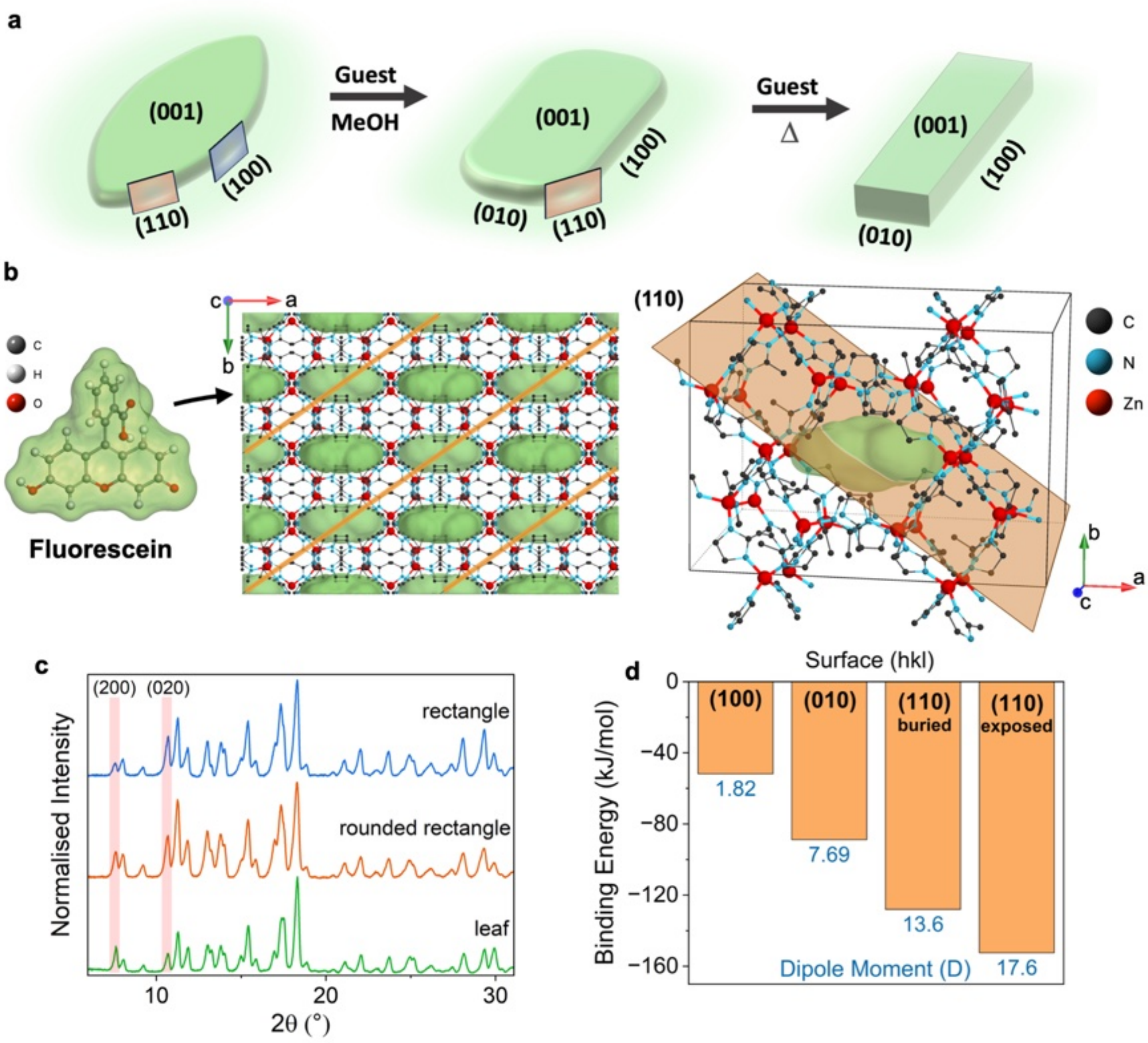

**Figure 3**. a) Sketch of the primary F@ZIF-L morphologies, indicating exposed surfaces with (*hkl*) lattice planes (orange plane shown for (110) and blue plane shown for (100) where particles have no straight

edge). b) Left: Molecular structure of fluorescein. Centre: (001) cross section of ZIF-L indicating pore cavity in green, showing staggered pores in/out of the *c*-axis along the *a*-axis. Orange lines indicate the (110) planes. Right: component of ZIF-L unit cell, representing pore cavity in green and the (110) plane in orange. c) PXRD patterns of F@ZIF-L with varying morphologies. d) Binding energy and dipole moments (blue) of F (dianion) on various surfaces of ZIF-L, determined from periodic DFT calculations (PBEsol0-3c) using slab models of 9-12 Å in thickness.

PXRD of F@ZIF-L samples synthesised at 35 °C with varying morphology reflects these changes at a bulk scale (Figure 3c). Specifically, the (020) reflection increases in intensity, while the (200) peak is reduced, indicating increased growth of the {010} family of planes in the rectangular morphologies (top view). Generally, after normalisation to the most intense reflection (420), all reflections also appeared more intense and sharper in samples with rectangular particles, indicating increased crystallinity. These changes were also observed in the diffraction patterns of the 30 °C F@ZIF-L (0.3 mg, F) series when 8 mL of MeOH was used (creating the widest more rectangular particle) (Figure S24), while synthesis with only $H_2O$ that exhibited no variation from the leaf morphology did not exhibit any changes in diffraction (Figure S24).

To test this hypothesis, employing *ab initio* density functional theory (DFT) calculations, we modelled the adsorption of fluorescein on three different facets of ZIF-L, namely the (110), (010), and (110) surfaces using slab models of 9-12 Å thickness. The calculations were carried out with the CRYSTAL23 periodic DFT code,[40] at PBEsol0-3c level of theory (see Methods, SI). Based on reported pH effects on fluorescein and a measured pH of 8.5-9 during F@ZIF-L synthesis,[37] along with observed NMR peaks from F@ZIF-L digest, it was assumed that fluorescein is incorporated into F@ZIF-L in its dianionic form, so $Zn^{2+}$ was used as a counterion (feasible to arise from minor defects in the framework). Importantly, two binding sites were possible for F on the (110) surface: a 'buried' site within the pore and an 'exposed' site at the centre of the pore if fully enclosed (i.e. along the surface diagonal) (Figure S25). As expected, the binding energy of F was most favourable on the (110) surface (-152.4 kJ/mol), being in contrast significantly higher for (010) and (100) (-88.8 kJ/mol and -51.8 kJ/mol, respectively) (Figure 3d). These theoretical data suggest that F has a strong energetic preference to bind to the pore opening site of ZIF-L at the (110) surface over other exposed surfaces during ZIF-L self-assembly. The dipole moment resulting from adsorption at the (110) surface, however, is significantly higher than for (010) and (100) (Figure 3d). This creates instability, which would engage reconstruction or further self-assembly during F@ZIF-L growth to remove exposure of the surface. Hence, a particle morphology results that contains minimal (110) in favour instead of (010), contrasted to the guest-free ZIF-L.

This leads to a critical finding: exposed ZIF-L pores during synthesis are energetically favourable sites of guest adsorption, and a more rectangular ZIF-L crystal after synthesis indicates more complete incorporation of guest particles into the framework pores. This allows for further deduction as to why temperature and solvent varied the morphology of F@ZIF-L. Temperature is known to be an influential factor in MOF morphology.[36,41] MeOH molecules can be trapped in the ZIF-L framework, and likely result in the particle broadening seen for MeOH/$H_2O$ synthesised ZIF-L. F is therefore in competition with MeOH for encapsulation. At a higher temperature, MeOH is more volatile, and confinement is less energetically favourable. Additionally, F solubility increases with temperature and heat increases the amount of energy available to the reaction system.[42] Collectively, we reason that these factors likely favour the energetics of adsorbing F onto surfaces constructing pore space, leading to more effective incorporation and the observed rectangular morphology.

**Near Field Vibrational Nanospectroscopy**

Near field infrared nanospectroscopy (nanoFTIR) conducted on a scattering-type scanning near-field optical microscopy (s-SNOM) has been established as an effective technique to detect local vibrational response if guest molecules are located on the surface of MOF materials.[43] NanoFTIR spectra of F@ZIF-L were therefore measured for various guest loadings then compared to the nanoFTIR spectra of ZIF-L and fluorescein (Figure 4a-f). Observing the F@ZIF-L (0.3 mg, F) spectra (Figure 4a), all vibrational bands can be seen to align well with those of ZIF-L including the out-of-plane Hmim ring bending mode at 750 $cm^{-1}$, the in-plane bending mode of C-H at 1146 $cm^{-1}$, and the entire imidazole ring stretching bands from 1300-1500 $cm^{-1}$.[24,30,31] No additional peaks were observable that could be attributed to the expected vibrations from fluorescein (F), indicating effective encapsulation of F in the ZIF-L framework. The spectra is also consistent with bulk material ATR-FTIR spectra.

Comparing spectra across all guest loadings up to 5 mg (10 mg and above were excluded due to the presence of mixed F aggregate particles and ZIF-L), also revealed no observable bands arising from F (Figure 4b). While the spectra exhibited consistent bands, the 997 and 1141 $cm^{-1}$ bands exhibited a considerable increase in intensity compared to the 750 $cm^{-1}$ band as guest loading increased (Figure 4d), indicating the reduction in signal from out-of-plane ring bending relative to in-plane mode. This is evidence of guest presence in pores and framework cavities, given guests would sterically hinder the extent of out-of-plane bending modes throughout the framework.

Across all morphologies, uniform optical amplitude signal was observed (Figure 4c). For the three distinct morphologies (leaf, rounded rectangle, rectangle) line scans comprised of 9 spectra were collected across the [010] axis of the particles (Figure 4e, S26). All characteristic peaks, particularly 750 $cm^{-1}$, of the ZIF-L framework remained consistent along the axis. To further probe the ends of the rectangular and rounded rectangular particles, 2D maps of 10 x 10 scans (spatial resolution ~ 20 nm) were also measured to show uniformity of the 750 $cm^{-1}$ signal across the scanned region (Figure 4f, S27).

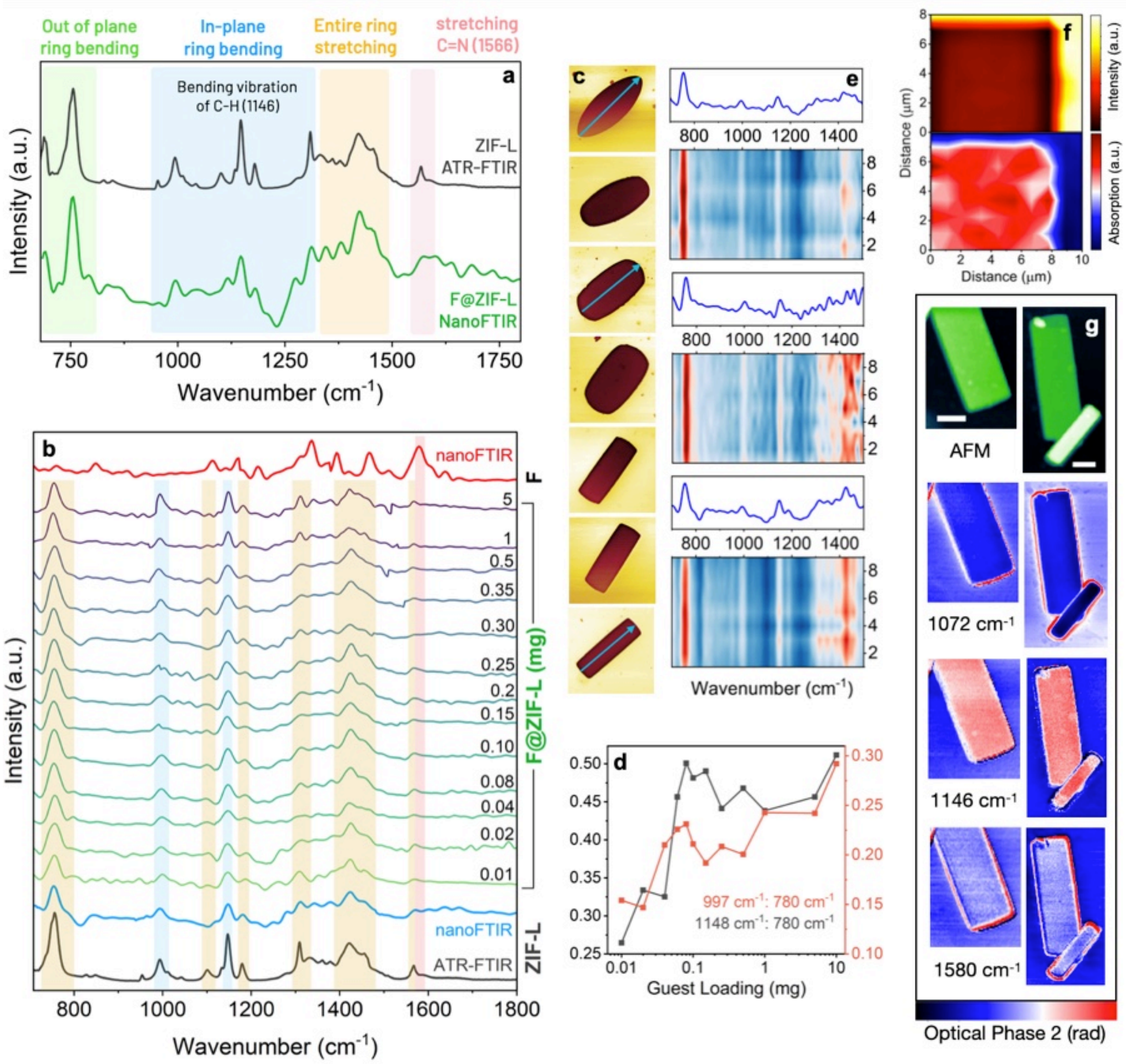

**Figure 4**. a) NanoFTIR measured on a single crystal of F@ZIF-L (0.3 mg, F), compared with the ATR-FTIR of a bulk sample of ZIF-L. b) NanoFTIR of F@ZIF-L samples with varying guest loading (indicating in mg). c) Optical amplitude images of F@ZIF-L particles measured by s-SNOM, confirming consistent intensity across particles of 2nd harmonic optical phase signal (O2P) *via* nanoFTIR. d) Ratio of vibrational bands in-plane:out-of-plane relative to guest loading, derived from spectra in b. e) 9-point line scans of three F@ZIF-L particles (blue lines in c indicate position of scans), with colour indicating intensity of O2P signal (equivalent to infrared absorption) across each spectrum. f) 10x10 nanoFTIR hyperspectral map of F@ZIF-L rectangular particle (0.3 mg, F) showing intensity of 780 $cm^{-1}$ peak from each spectrum. g) AFM (top, scale bar indicating 1 μm) and PsHet imaging of three F@ZIF-L (0.5 mg, F) rectangular particles across three different wavelengths, colour indicating the O2P signal contrasts (red means higher absorption for specific $cm^{-1}$).

Finally, s-SNOM pseudoheterodyne (PsHet) infrared nanoimaging was used to assess the intensity of near field signal observable at specific wavelengths (Figure 4g). By using a quantum cascade laser (QCL) as laser source compared with the broadband laser employed for nanoFTIR, illumination wavelength of the sample with PsHet QCL can be tuned to an absorption band of interest.[43] A widefield fluorescence microscope was attached to the s-SNOM instrument also so that particles examined could first be confirmed as exhibiting the expected fluorescence of F@ZIF-L (see photophysical properties below). PsHet nanoimaging by tuning the irradiation source to 1146 $cm^{-1}$, the characteristic C-H bending vibration of ZIF-L, showed homogeneity of strong absorption and reflectance across the entire F@ZIF-L particle observed (0.5 mg, 35 °C, 6 mL MeOH). Minimal contrast was obtained when the particle was illuminated at 1580 $cm^{-1}$, where the strongest F band occurs (the closest ZIF-L band being C=N at 1566 $cm^{-1}$). At most, a comparatively very weak but homogenous signal was observable across the entire particle (phase shift: 0.13-0.2 rad). Absorbance at the edge of the crystals is disregarded as shadowing effect and noise, typical edge artefacts seen in SNOM imaging, and a phenomenon also observed at a reference wavelength of 1072 $cm^{-1}$. This supports homogenous incorporation of F into the ZIF-L framework, and no apparent regions of concentrated guest aggregation examined at the local nanoscale.

**Luminescent Behaviour of F@ZIF-L**

The photophysical properties of 15 F@ZIF-L samples with varying guest loading but consistent MeOH reaction quantity (6 mL) were first examined (Figure 5a-h). All samples exhibited strong solid-state fluorescence with emission properties typical of F: a single emission band from 500-650 nm with a corresponding 450-550 nm excitation band (Figure 5a-e).[4,44,45] Importantly, this excitation and emission range does not overlap with emission from ZIF-L itself, avoiding issues of band mixing (Figure S28). The emission behaviour of F@ZIF-L is distinct from simply mixing ZIF-L with F, meaning as a physical mixture with the ratios from the F@ZIF-L synthesis, which produced negligible emission due to aggregation-caused quenching (ACQ) (Figure S29).[5] It required a mass ratio of 1:1 of ZIF-L:F to observe low intensity F emission. This confirms that in F@ZIF-L the F molecules have been incorporated into the framework to an extent that overcomes ACQ to exhibit intense fluorescence.

MeOH was found to be an important variable in syntheses to ensure proper dissolution of F to then enable framework encapsulation. Indeed, synthesis using only $H_2O$ yielded five samples with comparably low emission intensity (Figure S30). Emission increased from 0.01 to 0.1 mg F synthesis content, after which intensity stabilised despite further increasing F content, indicating maximum solubility was reached in $H_2O$ (Figure S30). Emission intensity at 0.01 mg was twice as strong when MeOH was used compared with $H_2O$ only.

Examining the 15 F@ZIF-L samples (6 mL MeOH) when excited at 470 nm, the emission $\lambda_{max}$ red shifted with 0.01 to 20 mg F content from 534 to 555 nm (Figures 5a-b, S31), resulting in emission chromaticity adjusting from pale green to bright green, then yellow (Figures 5g-h, S32). Emission intensity increased 336% from the lowest loading (0.01 mg) to 0.2 mg (Figure 5a). Further guest loading caused a decrease in intensity, until comparatively negligible emission was observed (Figure 5b) for ≥ 10mg. Photoluminescent quantum yield (PLQY) followed emission trends, initially increasing from 0.01 to 0.08 mg F content to reach a maximum 99.95% from 84.76% (Figure 5f). PLQY then decreased exponentially with guest loading to less than 2% for ≥10mg. Excitation spectra observed at 600 nm for up to 5 mg F content, showed a primary excitation band at 510 nm, which followed the intensity variations seen emission spectra (Figures 5d, S33). Notably, these values are red shifted compared to reported F@ZIF-8 emission with $\lambda_{max}$ = 515-535 nm and excitation with $\lambda_{max}$ = 499 nm,[37] suggesting 2D ZIF-L more favourably stabilises F emission pathways compared to the analogous 3D ZIF-8. From ≥10mg a distinct excitation spectral profile was observed, with a sharp excitation band at 533 nm and less intense shoulder bands from 450-525 nm (Figure 5e). These distinct changes in optical properties, along with minimal emission, at higher guest content arise from the oversaturation of guest relative to host and aggregated F particles forming, as observed in SEM and ATR-FTIR. In contrast, the ideal guest loading for the best photophysical properties appears to occur between 0.1-0.2 mg of F during synthesis (0.02-0.04 % mol loading).

F is known to aggregate,[46] including in ZIF-8,[37] so such behaviour was more precisely probed by examining fluorescence lifetime using time-correlated single-photon counting (TCSPC). Data was collected for each sample at emission $\lambda_{max}$ (540 nm) along with 520 and 560 nm (Figures 5i-j, S34). Data indicated multiexponential decay behaviour, modelled with three-time components: $\tau_1 \approx$ 0.4-0.8 ns, $\tau_2 \approx$ 1.1-3.4 ns, and $\tau_3 \approx$ 3.1 – 6.0 ns (Table S2). It is well agreed in literature that $\tau_3$ is attributable to the F monomer, $\tau_2$ to the J-aggregated F (head-to-tail), and $\tau_1$ to the H-aggregated F (head-to-head) (Figure 5i).[37,47] At lower guest loadings (0.01-0.2 mg), monomer and J-aggregate components dominate (Figure 2i), leading to strong emission and a red shift attributable to the longer wavelength emission from J-aggregates (Figure 5a-b).[48] This corresponds to the broadening of excitation bands at longer wavelengths (Figure 5d), which is also indicative of J-aggregate formation.[47] These data indicate effective encapsulation of F in ZIF-L, creating a combination of isolated species likely residing in ZIF-L pores, along with interactions between species in adjacent pores (end-to-end) or due to a staggered overlap. As guest loading increases from the 0.2 mg sample, monomer content decreases while H-aggregates increase, which quench emission as seen in spectra (Figure 5b). A higher energy shoulder is also seen forming in excitation spectra, a further indicator of H-aggregate formation (Figure 5d-e).[47] This suggests 0.2 mg is near the host interior saturation limit for F, after which excess guest more easily aggregates as head-to-head stacked particles, potentially on particle surfaces or defective regions of the ZIF-L crystalline framework. H-aggregation is so significant for ≥10mg (32% contribution) that emission is effectively quenched.

Lifetime values also provide insight into guest incorporation. F@ZIF-L exhibits a longer monomeric lifetime decay component (5-6 ns for lower guest loadings) compared to the typical F lifetime of 4.08 ns (Figure 5j). This is caused by the confinement effect, often seen for dye@MOF materials, where non-radiative processes are supressed due to encapsulation, resulting in increased lifetime and quantum yield.[6] The lifetime of the monomer and J-aggregate components decrease as guest loading increases. This behaviour has been reported in dye@MOF systems previously and was attributed to increased probability of interactions between F species across adjacent pores and through framework channels or structural defects.[47,49]

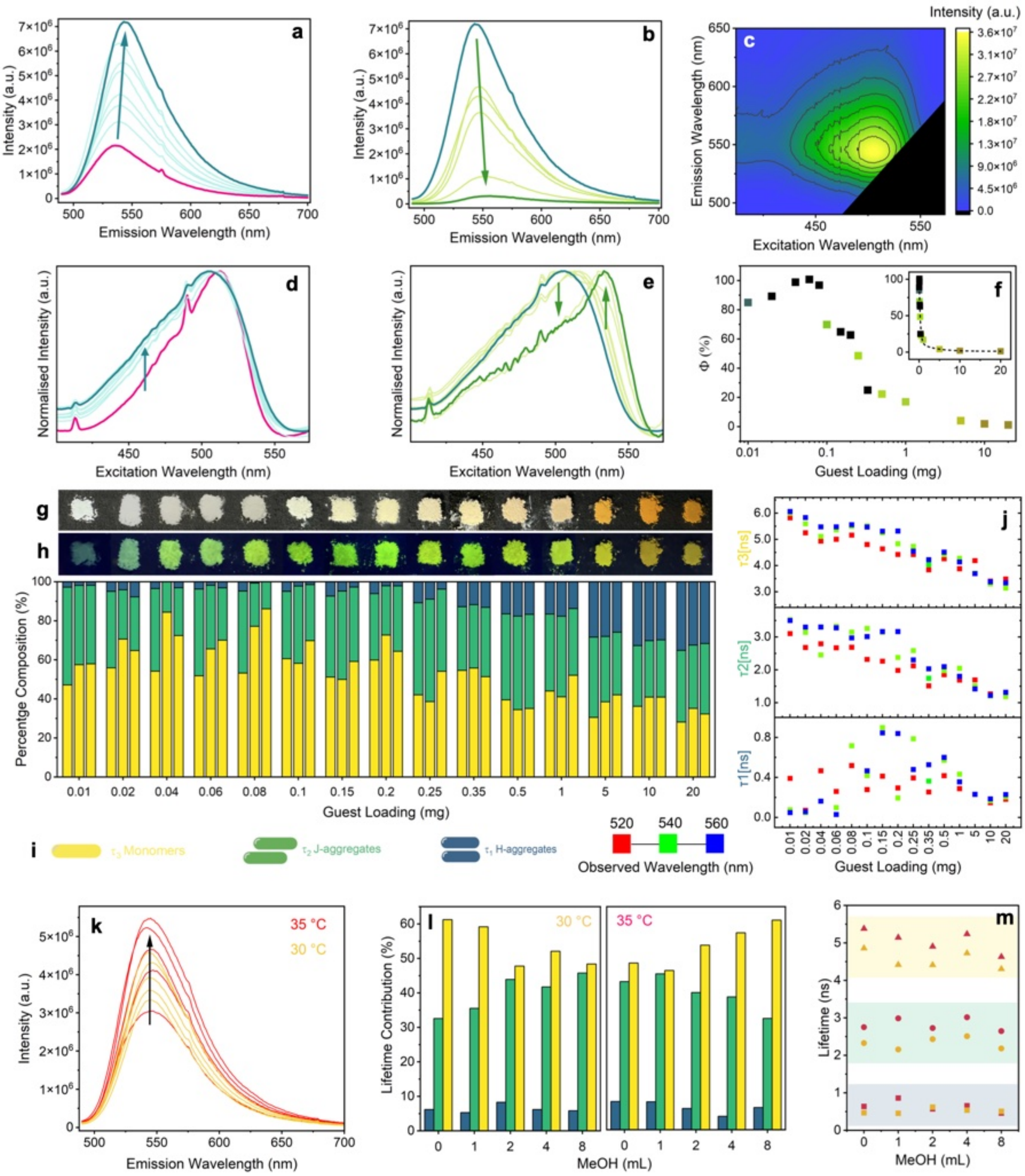

**Figure 5**. a-b) Emission spectra of F@ZIF-L (30 °C, 4 mL MeOH) excited at 470 nm with guest loadings from 0.01 (pink), to 0.2 mg (blue), then to 20 mg (green). Arrows indicate spectral progression with increasing guest loading. c) Emission map of F@ZIF-L (0.3 mg, F). d-e) Excitation spectra observed at 600 nm of F@ZIF-L samples as in (a-b). Arrows indicate spectral progressions with increasing guest loading. f) Photoluminescent quantum yield of F@ZIF-L samples. g-h) F@ZIF-L samples shown under ambient (g) and 365 nm UV light (h). i) Lifetime decay parameters observed at 520, 540 and 560 nm (left to right for each sample) with a sketch of corresponding F arrangements below (yellow = monomers, green = J-aggregates, blue = H-aggregates). j) Component lifetime values for all F@ZIF-L samples, with observed wavelength indicated by data point colour (red = 520 nm, green = 540 nm, blue = 560 nm). k) Emission spectra of F@ZIF-L (0.3 mg) synthesised at 30 °C (yellow) and 35 °C (red) with increasing MeOH content (0, 1, 2, 4, 8 mL). Arrow indicates increase in intensity resulting from increasing MeOH content. l) Lifetime decay parameters observed at 540 nm for F@ZIF-L (0.3 mg) with varying temperature and MeOH content during synthesis, and m) the corresponding lifetime parameter values.

**Optimising F@ZIF-L Luminescence by Tuning Reaction Conditions**

The effect of guest loading and emission behaviour observed above remained consistent at different MeOH content (1, 2, 4, and 8 mL) (Figure S35). Increasing MeOH content at a constant guest loading, interestingly, increased emission intensity (and corresponding excitation spectra intensity) (Figure S35). At 0.1 mg, from 1 mL to 8 mL of MeOH, a 2.3x increase resulted, while for 0.3 and 0.5 mg a 1.5x intensity increase occurred. Emission profiles remained consistent when the data was normalised, indicating no additional emission bands resulting from MeOH content (Figure S35). TCSPC lifetime data showed that at 0.1 mg (Figures S36, Table S3), increasing methanol resulted in a reduction in monomers and increase in J-aggregates, while H-aggregates remained consistent. As NMR indicated guest loading increased with MeOH content, this may simply be the result of increased F, imitating the data observed at higher F synthesis content examined earlier. Introducing more bulky MeOH solvent molecules may also disrupt the self-assembly of the framework, introduce more structural defects and open pore sites, increasing the overlap of adjacent guests to interact as J-aggregates. The same effect is observable, but with a lesser reduction in monomers at 0.3 mg. For 0.5 mg, the composition remains consistent across MeOH loading. This supports the premise that loading is already oversaturated at this guest concentration, so any benefits MeOH provides to guest incorporation are less relevant. Hence, while guest loading is a dominant force influencing emission guest properties, at lower guest content the variability of MeOH can have significant effects also.

Noting the morphology variance achieved with temperature adjustment, the resultant fluorescent variations were also examined at a guest loading set to 0.3 mg. With increasing MeOH content, the samples synthesised at 35 °C exhibited more intense emission and stronger excitation spectra relative to the same MeOH used at 30 °C (Figure 5k, Figure S37). Given guest loadings were similar at each temperature as MeOH content varies, this improved emission efficiency must instead be attributable to the pathways of emission and guest arrangement. Indeed, lifetime data indicates clear opposing trends (Figure 5l, Table S4): while at 30 °C monomer content decreases and J-aggregates increase with increasing MeOH content, at 35 °C monomer content increases but J-aggregates decrease with MeOH content. H-aggregates remain consistent; expected given they arise from surface-species guest loading oversaturation. The lifetime of monomer and J-aggregate components are observed to increase by nearly 1 ns at 35 °C, compared to 30 °C (Figure 5m). This suggests improved confinement, increasing emission intensity due to increased suppression of non-radiative decay pathways. This could arise from the improved framework structural completeness, providing more complete pore space for the confinement of guests, hence increased monomeric species.

**Fluorescence Microscopy**

Single particles of F@ZIF-L were individually examined under fluorescence confocal microscopy. Using lambda scanning (3 nm resolution), all particles excited with a 488 nm laser showed consistent emission spectra to the ‘bulk’ sample of F@ZIF-L, confirming the luminescent behaviour observed in bulk resulted from F@ZIF-L crystals (Figure 6a). The increasing band broadness across the sample spectra also aligns with the increasing proportion of aggregates present in each sample based on bulk analysis. Particles are reported both with colouring simulated from lambda scanning to reflect emission seen by eye, along with artificial colouring from averaging *z*-stack scanning (14 slices across 7 μm) (Figure 6b). While the simulated emission colour appears homogenous across the particles, the *z*-stack scan revealed more nuanced behaviour: both as single particles and windmill-type particles exhibited stronger emission at particle ends ((010) surfaces with curves of predominantly (110)), with weaker emission (30% of ends) throughout the particle (Figure 6c).

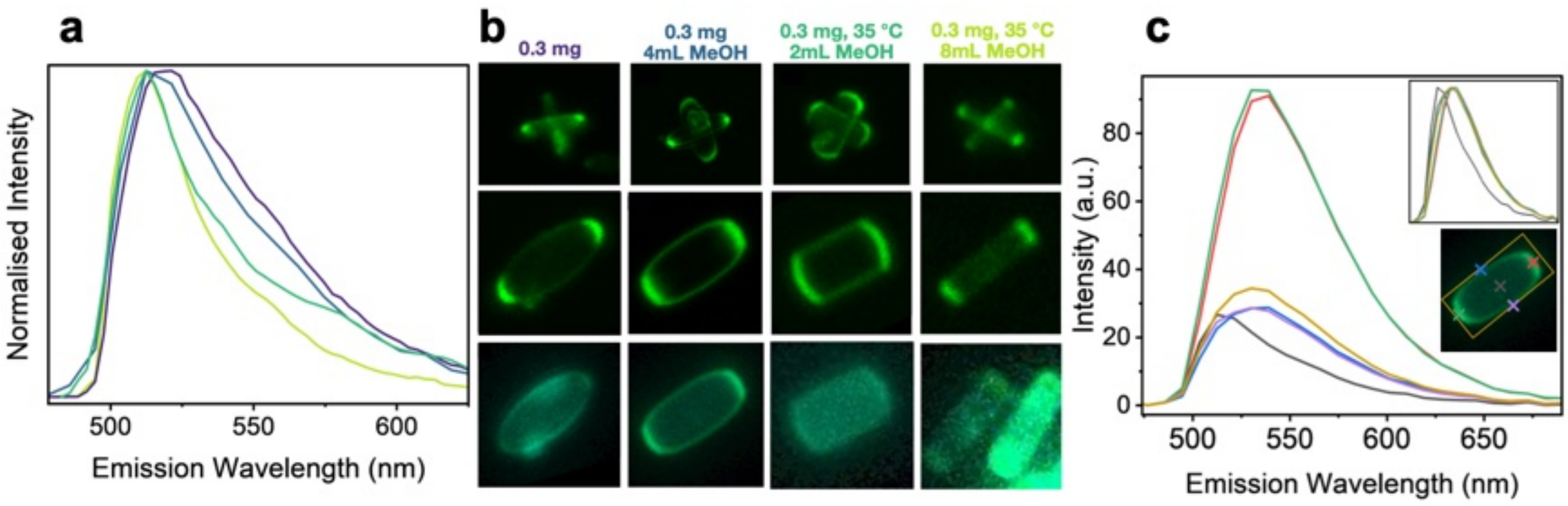

**Figure 6**. a) Emission spectra from individual F@ZIF-L particles excited at 488 nm, the colour of each spectra matching the sample description in (b). b) Fluorescence microscopy of F@ZIF-L particles with artificial colouring to represent signal intensity (above, middle), and predicted emission colour based on emission profile (bottom). c) Emission intensity variation across a single 0.3 mg F@ZIF-L (4 mL MeOH) particle, with normalised spectra inset. Colours of spectra match the positions on the particle.

A similar emission distribution across MOF crystals has been attributed to incomplete diffusion of guests during post-synthesis modification.[50] Here, however, we employed *in situ* encapsulation of F as the framework self assembles. s-SNOM and nanoFTIR data also clearly demonstrate a consistent and strong signal from the ZIF-L framework across the entire particle, eliminating the possibility of fragmented edges that may expose more guest material. One justification for the variability in emission intensity across particles is guest arrangement: at the crystal edges, it is more feasible for the more J-aggregates to form where framework features are incomplete or contain defects, intensifying emission. Through the particle, monomeric guest positioning is more expected. We do not discount the possibility of edge effects and light scattering also, given the 100-150 nm thinness of particles, leading to the appearance of intense emission at the particle edges. Similar effects were observed in luminescent microrods of lanthanide MOFs, where active optical waveguides resulted in brighter emission spots at the tip of each particle by propagating emission waves through the particle.[51]

**Photostability**

Photostability was examined using three guest loadings, namely 0.04, 0.30, and 1 mg of F per 80 mL of synthesis solvent (Figure S38). Data shows similar degradation, (35.5 - 39.3%) over 24-hour exposure to 150 W concentrated UV irradiation from a Xenon lamp at the absorption maximum of each F@ZIF-L sample. Compared with a typical 4 W LED cell, these results reflect an accelerated simulation of long-term use. Importantly, the decay did not exhibit significant exponential loss in the initial minutes of exposure, a typical feature of surface attached dye species on MOF materials, and chromaticity did not alter after exposure (Figure S38). F@ZIF-L was dispersed in MeOH and after 13 months remained luminescent (Figure S39). The MeOH did not exhibit fluorescence, and only negligible trace quantities of F were detected in the MeOH (0.071% of the intensity of F@ZIF-L, calculated to be 0.014 ppm), indicating robust entrapment of F in the ZIF-L framework (Figure S39).

**Other Luminescent Guest@ZIF-L Systems**

With an understanding of optimal guest loading and synthesis strategies, the wide applicability of the guest@ZIF-L design to create luminescent functional materials was demonstrated by attempting to incorporate 4 different organic dye molecules into ZIF-L that theoretically have suitable dimensions for the largest ZIF-L pore. Of these, perylene@ZIF-L and 7-hydroxycoumarin@ZIF-L (HC@ZIF-L) were successfully synthesised. Pyrene and coronene were not successfully incorporated, primarily due to incompatible solvents for dissolution. PXRD confirmed the retention of the ZIF-L crystalline framework (Figure S40), while FTIR spectra indicated no vibrational bands from guest molecules (Figure S41). FE-SEM showed all particles exhibited widening, as for F@ZIF-L (Figure 7a-e), and a shift from pointed ends to rounded rectangular corners, thereby indicating effective incorporation of guest@ZIF-L.

Perylene and HC are luminescent, emitting green/yellow and blue respectively in the solid state.[52] When encapsulated in ZIF-L, HC@ZIF-L retained the emission behaviour of HC but with an enhancement in emission intensity, partially resulting from the additional contribution of ZIF-L Hmim emission overlapping in the same region (Figures 7c, S42). Remarkably, perylene@ZIF-L emits a near ideal white light with CIE coordinates (0.33, 0.34) when synthesised in $H_2O$/DMF (2 mL) solvent mix and with higher guest loadings (0.1 % mol measured by NMR digest) (Figures 7f-j, S43)). In contrast, lower guest loadings (0.03 % mol) using the same solvent mix resulted in a light blue emission and using $H_2O$ (very low solubility of perylene) a darker blue emitter resulted with 0.009 % mol guest loading. Notably the $H_2O$ synthesised perylene@ZIF-L retained the leaf ZIF-L morphology, indicating minimal guest loading compared with DMF syntheses (Figure S43).

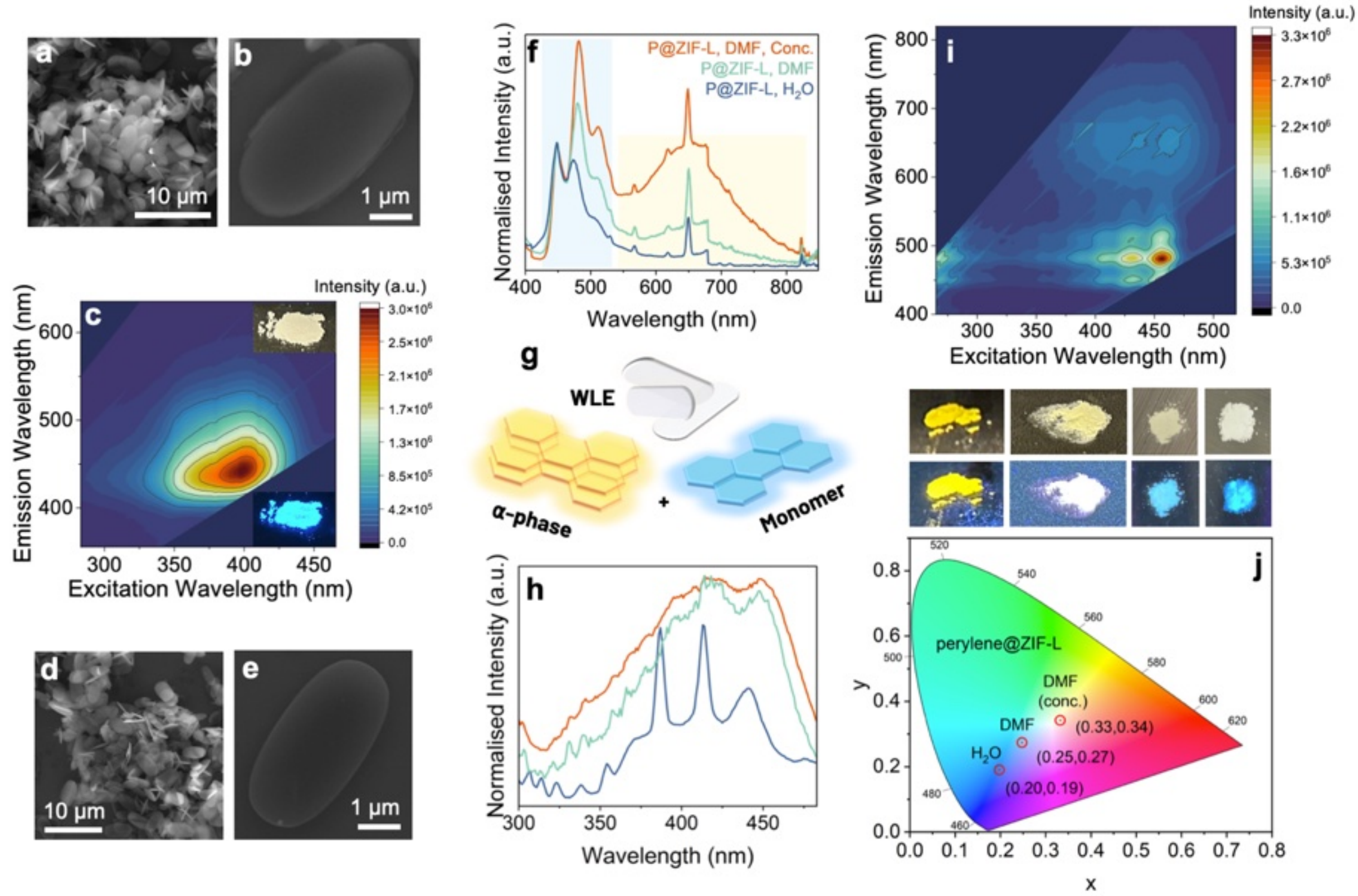

**Figure 7**. a-b) FE-SEM of HC@ZIF-L showing rounded rectangular particles. c) Emission map of HC@ZIF-L. d-e) FE-SEM of perylene@ZIF-L showing rounded rectangular particles. f) Emission spectra of perylene@ZIF-L (excited at 380 nm) highlighting the monomer (blue) and excimer (yellow) regions. g) Schematic illustrating WLE perylene@ZIF-L particles, formed from arrangements of single perylene monomers emitting in the blue region and $\alpha$-phase dimers emitting in the yellow-orange region. h) Excitation spectra of perylene@ZIF-L samples, according to colour key of (f), observed at 500 nm. i) Emission map of perylene@ZIF-L (DMF, with concentrated guest), j) emission chromaticity of perylene@ZIF-L, showing near ideal WLE coordinates, with corresponding photographs (above) of the samples under ambient and a 365-nm UV light (below).

While white light emitting (WLE) perylene has not been reported, other perylene@MOF systems such as ZIF-8 provide theoretical explanations for the observed WLE phenomenon in ZIF-L.[49,53,54] The emission spectra of the perylene@ZIF-L samples (excited across 350-475 nm) contain two components: sharp peaks around 450-500 nm, followed by a broad intense band from 600-700 nm (Figure 7i). The first band is assignable to emission of single monomeric perylene molecules, and is rare to observe in the solid state due to needing to isolate molecules. The second arises due to self-trapped exciton (STE) emission, with *α*-phase perylene units (dimeric stacked molecules) emitting as an E-state excimer (Figure 7g).[52,55] This state was observed when entrapping perylene in MIL-68 due to confinement inducing the dimeric structure. Uniquely, perylene@ZIF-L is formed of both dimeric *α*-phase units and monomers, a similar arrangement seen for F. This combination creates a broadband emission distributed appropriately to form WLE. For lower guest loaded perylene@ZIF-L samples, the *α*-phase contribution reduces until non-existent, indicating the preference for perylene to remain as monomers (Figure 7f). Excitation spectra observed at 700 nm support this, with the lowest guest loading perylene@ZIF-L exhibiting discrete sharp bands corresponding to monomer emission, in contrast to a broad structureless band observed at higher guest loading, indicative of the *α*-phase dimeric emission.

This finding aligns with a crystallographic and steric understanding: dimensions of perylene align closely with the pore size of ZIF-L (9.3 x 6.7 vs 9.4 v 7 Å) and with 5.3 Å of height per pore. It is therefore possible for dimeric perylene to form, especially considering the flexibility of ZIF-L, and the pore space would encourage the precise stacked placement required for the *α*-phase. However, less strain and a more energetically favourable, and stable, material would be produced by isolating a single perylene species in each pore. Interestingly, perylene adsorption was modelled on the (100), (010), and (110) surfaces of ZIF-L using the same method as for F (Figure S45). The data showed consistent trends, indicating an even stronger preference than F for binding at (110) over (010) or (100), along with a 27% lower binding energy at (110) over F@ZIF-L. Given the calculated energy of perylene molecules to aggregate is

−129.7 kJ/mol, the adsorption at (110) in ZIF-L is a highly favourable energetic process, either as a monomer or dimeric unit.

### Heteroepitaxial Growth of F@ZIF-L Films

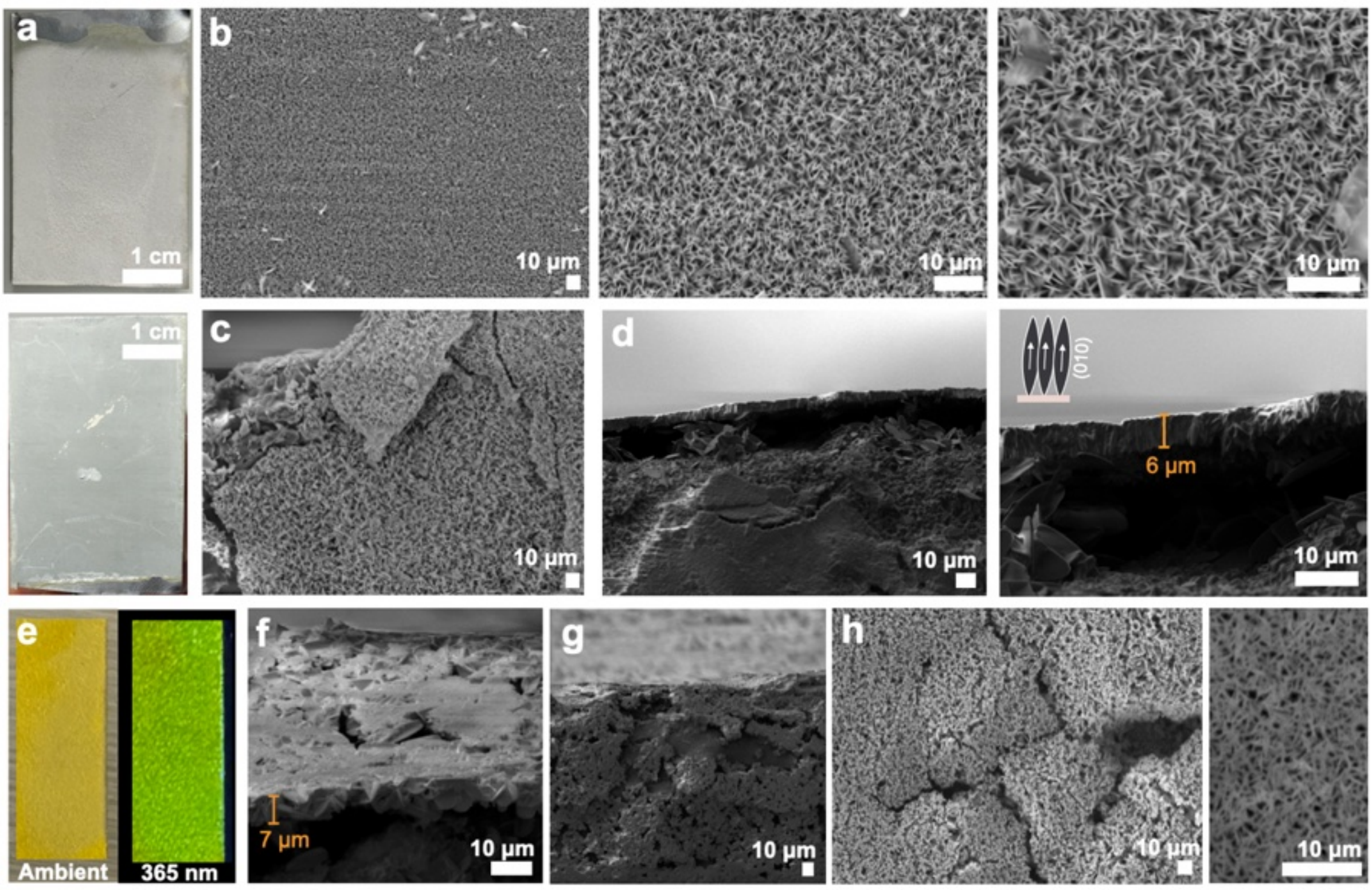

**Figure 8**. a) Images of ZIF-L coated Zn foil. b) FE-SEM of ZIF-L film at various levels of magnification. c) FE-SEM at the edge of the ZIF-L film. d) FE-SEM of ZIF-L film highlighting crystal orientation and typical thickness. e) F@ZIF-L layer on Zn foil under ambient and UV (365 nm) light. (f-h) FE-SEM of F@ZIF-L film showing uniformity of orientation and thickness near a film edge.

We finally briefly demonstrate a secondary benefit to the ZIF-L morphology: creating oriented films. Immersing Zn foil in the reagent mixture for ZIF-L produced well oriented thin-films of ZIF-L around 5 µm thick,[26] comparable to the length of a single ZIF-L particle along its *b*-axis (Figure 8a-d). By including F in the synthesis, an oriented thin-film of F@ZIF-L was formed of 6-8 µm thickness (Figure 8e-h). The incorporation of F into the ZIF-L framework, rather than merely coating a ZIF-L film was confirmed by FE-SEM images, which showed the rounder F@ZIF-L particles forming the film itself and no regions of F particle aggregation (Figure 8f-h). The film exhibited strong fluorescence of F@ZIF-L character, further confirming guest-host interactions (Figure 8e).

## Conclusions

This study has established fundamental principles directing the formation of solid-state guest@ZIF-L materials. Employing fluorescein as a case study molecule, systematic studies revealed a morphological continuum that was indicative of guest incorporation: as guest loading increased, particles adjusted from the characteristic leaf shape of ZIF-L to rounded rectangles then rectangular particles. From crystallographic data and DFT theory it was concluded that these variations arise due to limiting the presence of the (110) plane, which intersects the pore cavities of ZIF-L and is the preferential binding site of guests in ZIF-L out of the possible exposed surfaces, preferring instead to expose (010). This meant greater retention of pore space in the particle, and higher crystallinity. Indeed, nanoFTIR confirmed the lack of surface guest species on ZIF-L particles.

Encapsulating F in ZIF-L prevents ACQ, turning on the molecule to form highly luminescent F@ZIF-L particles. Particles with a rectangular morphology were found to be more luminescent and proportionally composed of F arranged as monomers, indicative of effective guest incorporation. HC@ZIF-L exhibited blue emission, while encapsulating perylene@ZIF-L produced either blue emitting materials or an ideal white-light emitter with CIE (0.33, 0.34) depending on the guest loading. This result derived (as for F) from guest arrangement, being a combination of blue emitting monomer perylene units and yellow emitting dimer $\alpha$-phase excimers. The materials showed high photostability and guest retention over long-term testing, attributable to the dense 2D layers of ZIF-L that are tightly interconnected by hydrogen bonding. A final advantage of the 2D morphology was unlocked by growing uniformly oriented luminescent thin films of F@ZIF-L directly on a Zn foil substrate.

Together, the findings demonstrate the advantageous functionality accessible by incorporating luminescent guests into ZIF-L, as an exemplar of a 2D MOF host, to produce stable, tuneable and highly effective lighting materials. The now elucidated nanosheet growth mechanisms offer applicability to organic guests generally, however, and a myriad of opportunities exist to encapsulate organic molecules with sensing, electroluminescent, electrochromic potential, or more, to create and customise a new generation of chromatic-based functional 2D ZIF-L materials.

## Declarations

This submitted manuscript is the final completed version of the original study described in the DPhil Thesis of the first author, D.A.S., for details see references [56] and [57].

## Acknowledgments

D.A.S. acknowledges the scholarships from the General Sir John Monash Foundation and the Clarendon Fund. J.C.T. thank the ERC Consolidator Grant (PROMOFS 771575) and EPSRC (EP/R511742/1) for funding the research. L.D. gratefully acknowledges the Gauss Centre for Supercomputing e.V. (https://www.gauss-centre.eu/) for providing computing time on the GCS Supercomputer SuperMUC-NG at Leibniz Super Computing Centre (https://www.lrz.de/) and the support from the Project CH4.0 under the MUR program “Dipartimenti di Eccellenza 2023-2027” (CUP: D13C22003520001). The authors would like to acknowledge Dr Jana Koth, Facility Manager at the Wolfson Imaging Centre (MRC Weatherall Institute of Molecular Medicine) for the provision and operation of fluorescence microscopy equipment. We thank Diamond Light Source for access and support in use of the electron Physical Science Imaging Centre (Instrument E02 and proposal number MG31944) that contributed to the results presented here. We are grateful to Prof. Bartolomeo Civalleri from the University of Turin for his support on DFT calculations using the CRYSTAL code.

**Author Contributions**

Conceptualisation and methodology were developed by D.A.S. and J.-C.T. Synthesis was performed by D.A.S. PXRD, AFM, nanoFTIR, ATR-FTIR, TGA, and FS-5 fluorescence spectroscopic measurements and data analyses were performed by D.A.S. *Ab initio* DFT calculations employing the CRYSTAL23 code were performed by L.D. FE-SEM characterisation was performed by C.B. and D.S. TEM characterisation was performed by C.A. with analysis of the data by D.A.S. PsHet s-SNOM imaging was conducted by L.M. Synthesis of perylene@ZIF-L and corresponding characterisation by SEM, FTIR, $^{1}$H NMR and PXRD was carried out by B.S. and D.A.S. Nitrogen adsorption studies were performed by J.F.-P. under the guidance of J.S.-A. Manuscript was drafted by D.A.S., edited by J.-C.T., and subsequently reviewed by all authors. Project supervision and funding acquisition by J.-C.T.

**Data Statement**

The data that support the findings of this study are available in the supplementary material of this article. Any other data will be made available on request.

# *Supporting Information*

*for*

# Elucidating Guest-Host Mechanisms in ZIF-L for Tuneable Highly Luminescent 2D Materials

*Dylan A. Sherman,[a] Lorenzo Donà,[b] Cyril Besnard,[a] Lars Mester,[c] Ben Slater,[a] Judit Farrando-Pérez,[d] Christopher S. Allen,[e,f] Joaquín Silvestre-Albero,[d] and Jin-Chong Tan[a*]*

*[a] Multifunctional Materials & Composites (MMC) Laboratory, Department of Engineering Science, University of Oxford, Parks Road, Oxford OX1 3PJ, United Kingdom.*

*[b] Department of Chemistry, NIS and INSTM Reference Centre, University of Turin, via Pietro Giuria 7, Torino 10125, Italy.*

*[c] Attocube Systems AG, Eglfinger Weg 2, DE-85540 Haar, Germany.*

*[d] Laboratorio de Materiales Avanzados, Departamento de Química Inorgánica-Instituto Universitario de Materiales, Universidad de Alicante, Ap. 99, E-03080 Alicante, Spain*

*[e] Electron Physical Science Imaging Centre (EPSIC), Diamond Light Source Ltd., OX11 0DE, United Kingdom.*

*[f] Department of Materials, University of Oxford, Parks Road, Oxford, OX1 3PH, United Kingdom.*

** Corresponding author's e-mail: jin-chong.tan@eng.ox.ac.uk*

**Table of Contents**

## Methods

All reagents, solvents, and chromophore guests were commercially acquired from Sigma-Aldrich, Fisher Scientific and Alfa Aesar, and used as received.

Synthesis of ZIF-L followed reported procedures.[1,2] 0.586 g of $Zn(NO_3)_2.6H_2O$ and 1.298 g of Hmim (2-methylimidazole) were dissolved in 40 mL of deionised water (DI) respectively. The aqueous solution of $Zn(NO_3)_2$ was then stirred into the solution of Hmim, and the resulting mixture stirred at 30 °C for 2 h. The precipitated ZIF-L was collected by centrifugation at 8000 rpm for 10 minutes, washed by DI water three times and dried for 12 hours at 50 °C.

Synthesis of guest@ZIF-L followed a modified ZIF-L synthesis procedure. The guest (fluorescein, perylene, or 7-hydroxycoumarin) was first added to the aqueous solution of Hmim along with any additional solvent content (MeOH or DMF). This mixture was then added to the aqueous solution of $Zn(NO_3)_2$, and the resulting mixture stirred at 30 °C for 2 h. The precipitated guest@ZIF-L was collected by centrifugation at 8000 rpm for 10 minutes, washed by MeOH twice then DI water repeatedly until the DI did not appear luminescent (typically 5-6 x 45 mL). The final product was dried for 12 hours at 50 °C.

Powder X-Ray Diffraction (PXRD) patterns were collected using a Rigaku MiniFlex diffractometer equipped with a Cu K$\alpha$ source and step size of 0.0025° at a scan rate of 0.04° $min^{-1}$. Samples were prepared using a 0.1 mm shallow-well glass sample holder.

Attenuated Total Reflectance Fourier Transform Infrared Spectroscopy (ATR-FTIR) measurements were performed using a Nicolet iS10 FTIR spectrometer. High-resolution synchrotron radiation infrared measurements of the mid-IR (MIR) and far-IR (FIR) spectra were performed at the B22 MIRIAM beamline in Diamond Light Source. FIR and MIR spectra were collected under vacuum, using the ATR module installed on the Bruker Vertex 80 V FTIR bench equipped with the DLaTGS detector. For improved signal-to-noise ratio in FIR, a liquid helium-cooled bolometer detector was used.

Atomic Force Microscopy (AFM) imaging was performed on a neaSNOM instrument (neaspec GmbH) operating in tapping mode. Height topography images were collected using the Scout350 probe (NuNano), with a nominal tip radius of 5 nm, a spring constant of 42 N $m^{-1}$, and resonant frequency of 350 kHz.

Near-field optical FTIR (nano-FTIR) spectra were collected using the neaSNOM instrument (neaspec GmbH) with AFM tapping-mode. The platinum-coated tip (NanoAndMore GmbH, cantilever resonance frequency 250 kHz and nominal tip radius ≈20 nm) was illuminated by a Toptica broadband mid-infrared (MIR) femtosecond differential frequency generation (DFG) laser. Individual sample crystals were analysed at 10 unique regions, with each spectrum acquired from an average of 20 Fourier-processed interferograms at 10 $cm^{-1}$ spectral resolution, 2048 points per interferogram, and an 18 ms integration time. The sample spectrum was normalized to a reference spectrum measured on a Si surface to reconstruct the final nano-FTIR amplitude and phase. The continuous broadband MIR spectra were attained by combining two illumination sources. All measurements were carried out under ambient conditions (≈40% RH).

Fluorescence Microscopy images were collected using a Zeiss LSM780 with a 10x confocal lens. Lambda scanning (at a resolution of either 9 or 3 nm) was used to image fluorescence with laser lines of either 405 nm (diode source), 488 nm (argon multiline 25 mW), or 543 nm (HeNe 1 mW). Data was analysed using Zeiss ZEN 3.9.

s-SNOM pseudoheterodyne (PsHet) imaging were performed at Neaspec GmbH. A tuneable quantum cascade laser (QCL Daylight solutions) coupled to the neaSNOM microscope was employed as the source of monochromatic irradiation, with output powers tuned to approximately 2 mW. For each scan, the pixel integration time was set as 16 ms. The tip was operating at a frequency of 253 kHz. The recorded signal was demodulated at the third harmonic through a PsHet detection mode. A fluorescence microscope was attached to the lens of the unit so particles could be examined for fluorescence first before proceeding to PsHet imaging.

Raman Spectroscopy was performed using a Bruker MultiRAM Raman spectrometer with sample compartment D418, equipped with a Nd-YAG laser (1064 nm) and a LN-Ge diode as a detector. The laser power used for sample excitation was 50 mW, and 64 scans were accumulated at a resolution of 1 $cm^{-1}$.

Thermogravimetric analysis (TGA) was acquired using a TA Instruments Q50 TGA machine equipped with a platinum sample holder under an $N_2$ inert atmosphere at a heating rate of 10°C $min^{-1}$ from 30 to 750 °C.

Scanning Electron Microscopy (SEM) was obtained at 10 keV under high vacuum using a SEM Tescan Lyra 3 (Tescan, Czech Republic) with secondary and backscattered electron imaging (SEI and BSE respectively) using a voltage from 10 to 15 keV. Samples were drop-cast onto Si wafer or a polished Al specimen stub that was coated with gold (Au) with a thickness of 12 nm using the SC7620 sputter coater (Quorum Technologies) at 20 mA plasma current for 1.5 min.

Spectrofluorimetric Measurements: Steady–state fluorescence and reflectance spectra, photoluminescence quantum yield (PLQY), and time-correlated single photon counting (TCSPC) emission decay data were recorded using the FS-5 spectrofluorometer (Edinburgh Instruments) equipped with the appropriate modules for each specific experiment. For TCSPC measurements, a 365 nm EPLED picosecond pulsed laser source was used. Lifetime fitting of the time constants from decay data was performed using the Fluoracle software. Excitation spectra exhibit artefacts (sharp peaks) from equipment operation that were not removed in presented data.

Diffuse Reflected Spectroscopy: A 2600 UV-Vis spectrophotometer (Shimadzu) was used to measure the absorption spectra and calculate the Kubelka-Munk (KM) function. Time-dependent UV-irradiation was conducted using a 365-nm handheld UV lamp (UVA, 6W).

HAADF-STEM was conducted on a JEOL JEM-2100Plus microscope operated at an accelerating voltage of 200 keV.

Nuclear Magnetic Resonance (NMR) Guest Loading Analysis: 40 mg of each sample were dissolved in a solution composed of 500 µL methanol-d4 and 50 µL $DCl/D_2O$ (35 wt.%). All NMR spectroscopy was collected at 298 K using a Bruker Avance spectrometer operating at 600 MHz, equipped with a BBO cryoprobe. Data were collected using a relaxation delay of 20 s, with 128 k points and a sweep width of 19.8 ppm, giving a digital resolution of 0.18 Hz. Data was processed using Bruker Topspin with a line broadening of 1 Hz and 2 rounds of zero-filling. The loading amount was calculated from the molar ratio of Hmim to guest. For fluorescein (F), the doublet at ≈8.30 ppm was used, which corresponds to the single proton in the ortho position relative to the carboxyl group. Global spectral deconvolution (in the MestReNova software package) was used to pick and integrate the peaks.

Computational Details: *Ab initio* density functional theory (DFT) calculations were carried out with the CRYSTAL23 periodic code,[3] at PBEsol0-3c level of theory.[4] The PBEsol0-3c is a composite method based on a hybrid Hartree−Fock/DFT Hamiltonian combined with a double-ζ quality basis set, augmented with a semiclassical dispersion term and a geometrical counterpoise correction, which provides a good trade-off between cost and accuracy. Truncation criteria for the evaluation of bi-electronic integrals were set to $10^{-7}$, $10^{-7}$ for the Coulomb and to $10^{-7}$, $10^{-7}$, and $10^{-25}$ for the Exchange series, respectively. The Monkhorst-Pack/Gilat shrinking factors for the diagonalization of the Kohn-Sham matrix in the reciprocal space were set to 2.

Starting from the experimental structure of ZIF-L,[2] structural disorder including solvent molecules and misplaced hydrogen atoms on the organic linkers were removed. On the resulting structure a full relaxation of both atomic positions and lattice parameters was performed.

To gain insights into the preferential crystal growth we modelled the adsorption of perylene and fluorescein on three different surfaces of ZIF-L, i.e. (1 0 0), (0 1 0) and (1 1 0). Fluorescein was studied in its dianionic form with a $Zn^{2+}$ as counterion.

The optimized orthorhombic unit cell of bulk ZIF-L has been used as starting point to build slab models with a thickness of 12 Å for the (1 0 0) and (0 1 0) faces and 9 Å for the (1 1 0) face. In addition, the dangling bonds resulting from surface cleavage were saturated to ensure the electroneutrality of the model systems. Geometry optimization of the slab models was carried out by fixing the lattice vectors parallel to the surface to the values corresponding to the conventional orthorhombic unit cell, while only the atomic positions were fully relaxed.

## Supplementary Figures and Tables

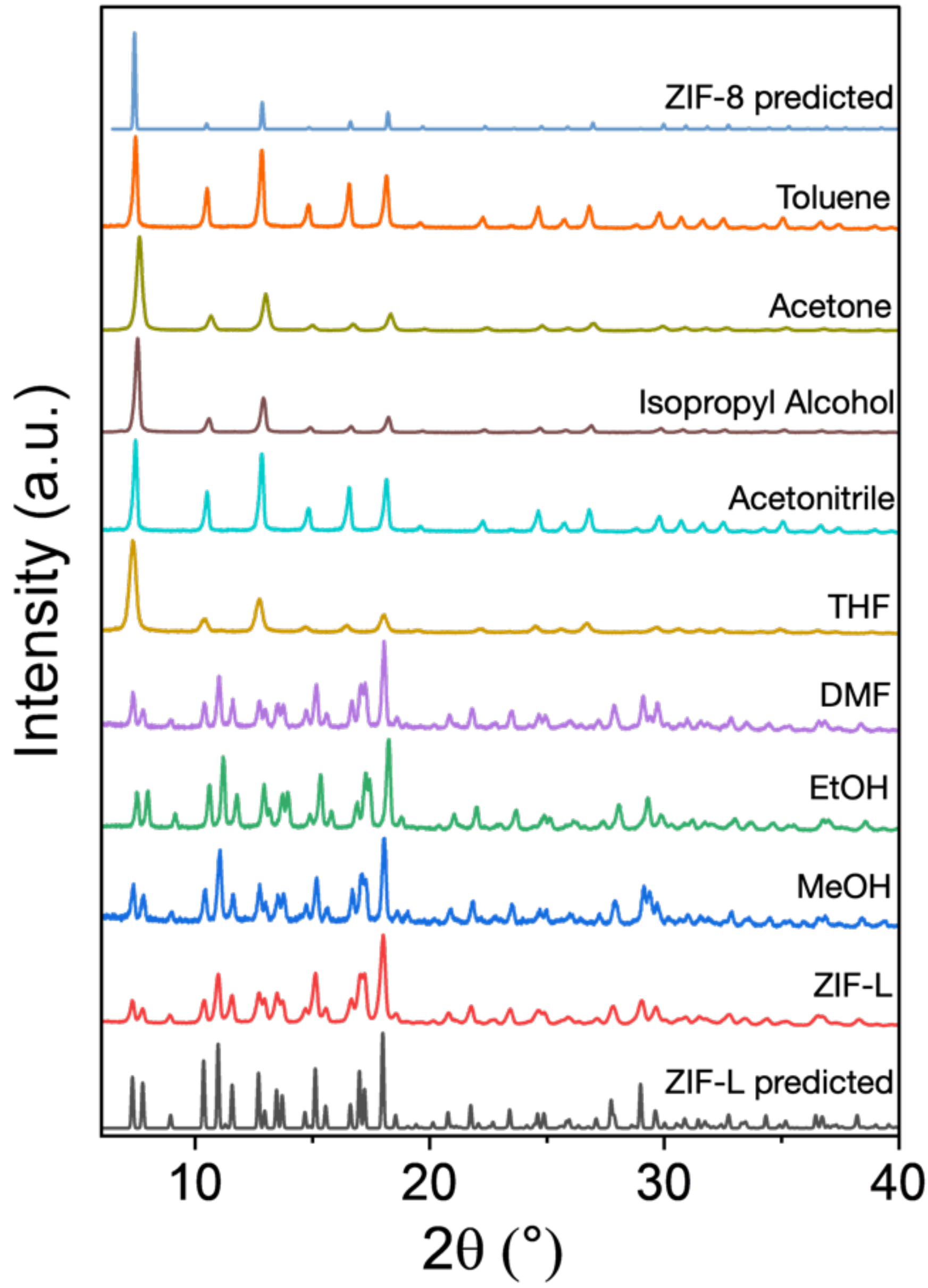

Figure S1. Powder x-ray diffraction (PXRD) of ZIF-L materials synthesised in $H_2O$ with additional solvent as indicated above (EtOH < 5mL, MeOH ≤ 10 mL) compared to simulated PXRD patterns of ZIF-L (CCDC 1509273) and ZIF-8 (CCDC 864309).[2,5]

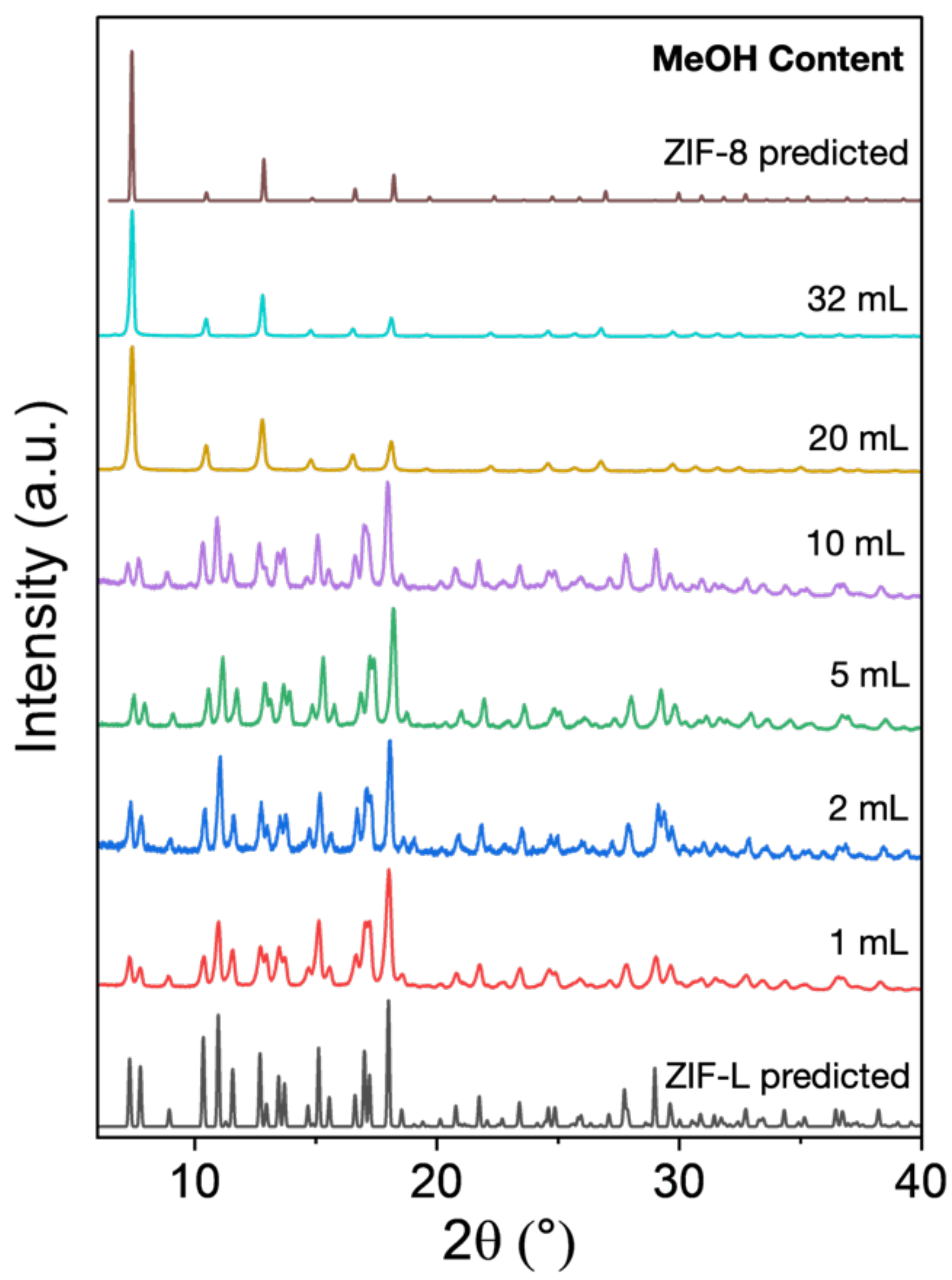

Figure S2. PXRD of ZIF-L materials synthesised in $H_2O$ with additional MeOH in quantities indicated above compared to simulated PXRD patterns of ZIF-L and ZIF-8.[2,5]

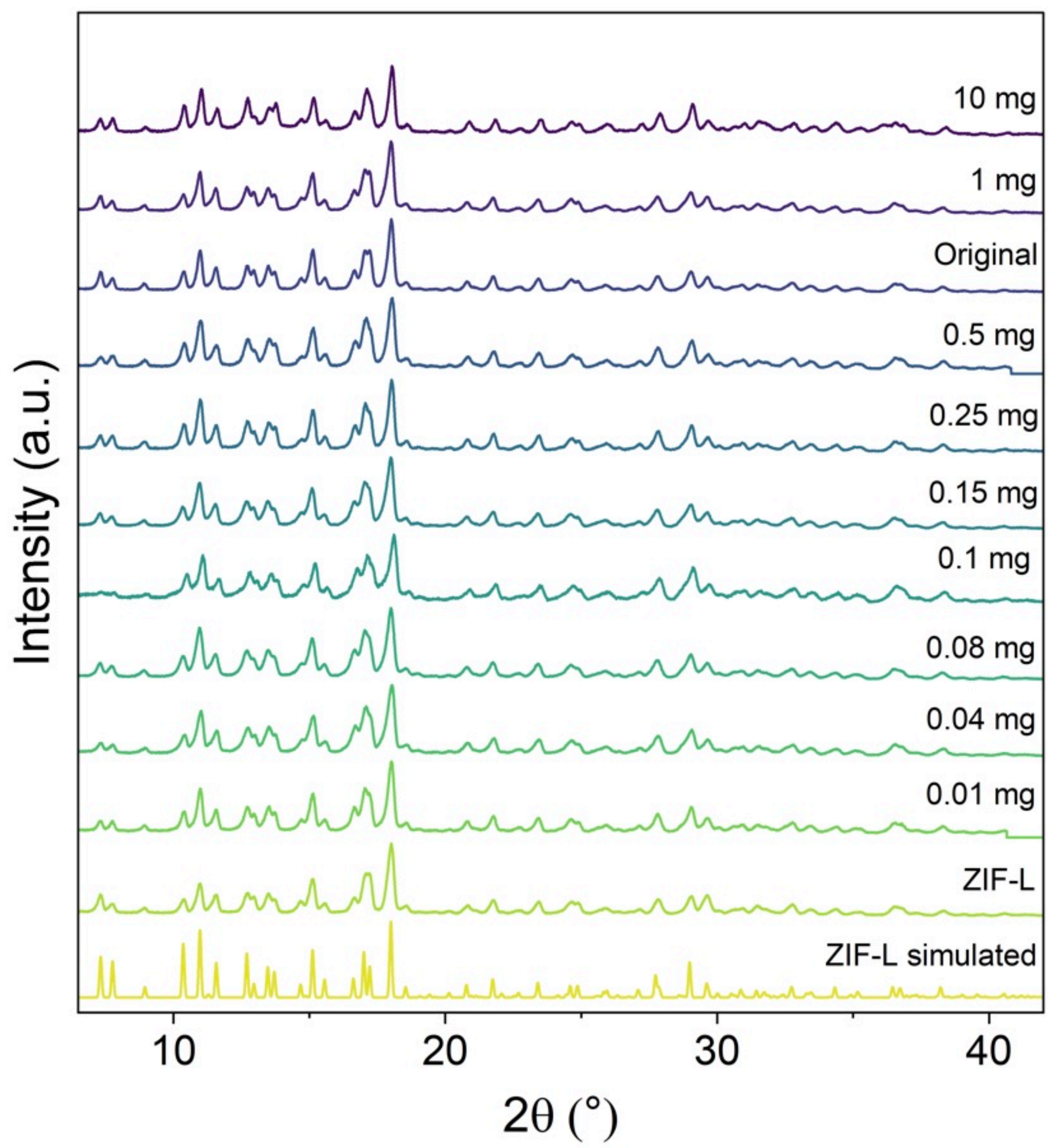

Figure S3. PXRD of F@ZIF-L materials (F = fluorescein) with varying F guest content during synthesis (80 mL $H_2O$/ 4mL MeOH) compared to simulated PXRD patterns of ZIF-L.[2,5]

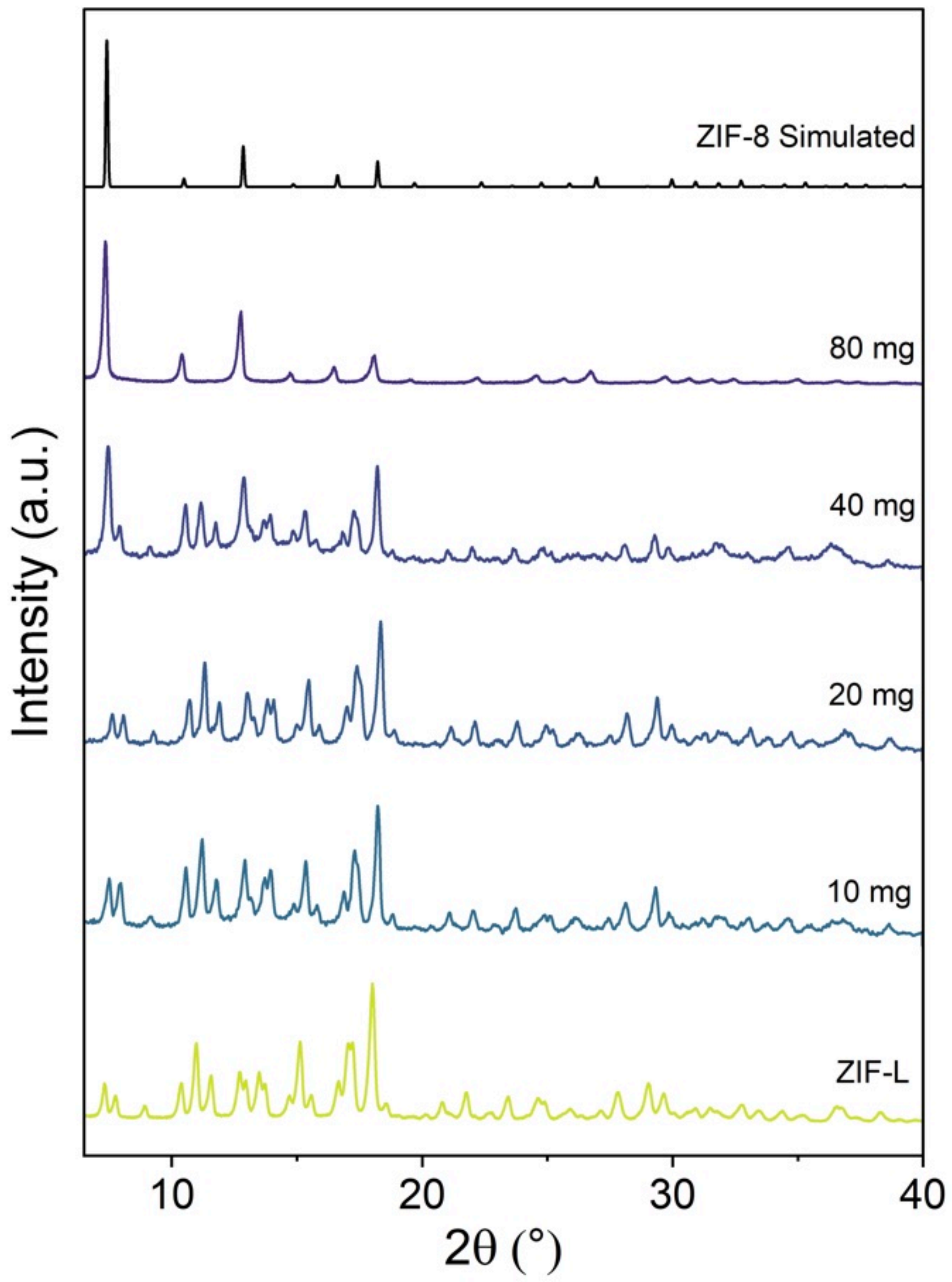

Figure S4. PXRD of F@ZIF-L materials with varying F guest content during synthesis (80 mL $H_2O$/ 4-10 mL MeOH depending on quantity of F requiring dissolution) compared to simulated PXRD pattern of ZIF-8.[2,5]

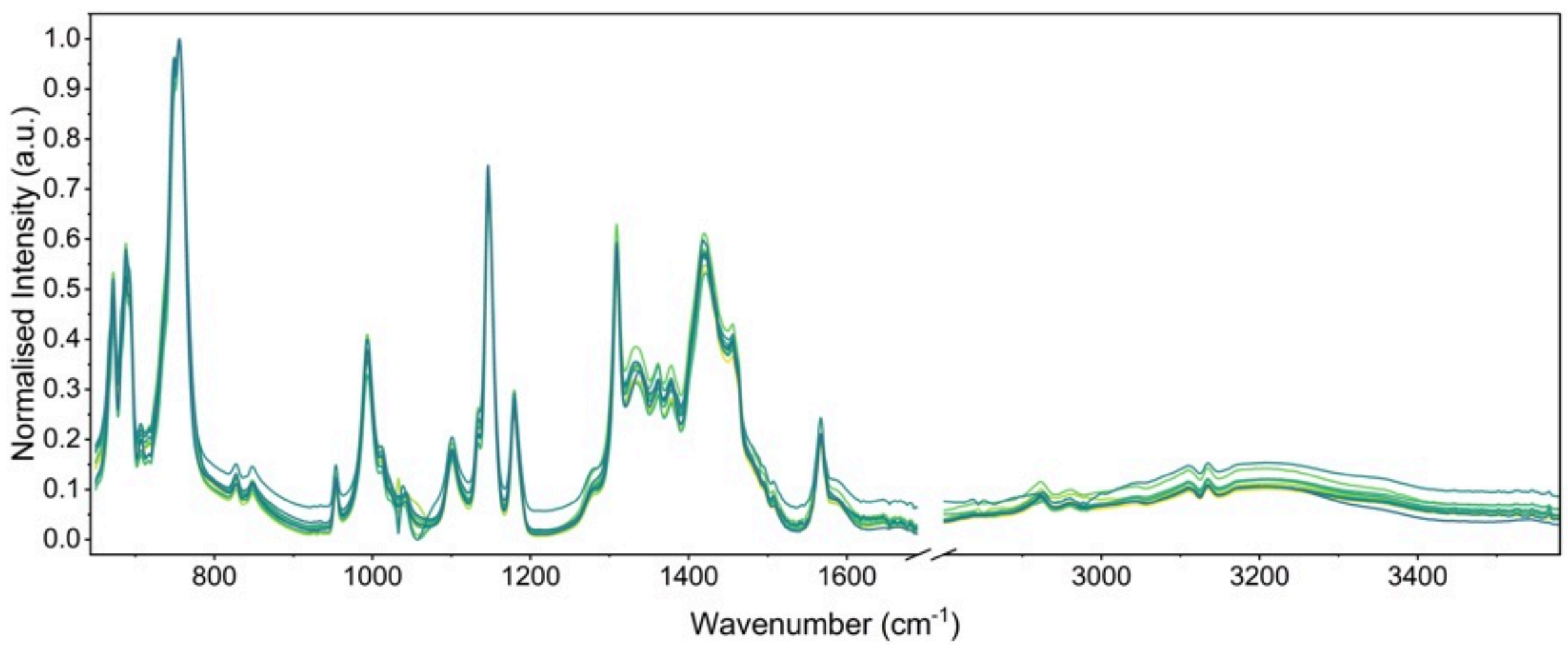

Figure S5. Attenuated total reflectance-Fourier transform infrared (ATR-FTIR) of F@ZIF-L samples with up to 5 mg of F during synthesis (light green to blue indicates least to most guest content) compared to ZIF-L (yellow).

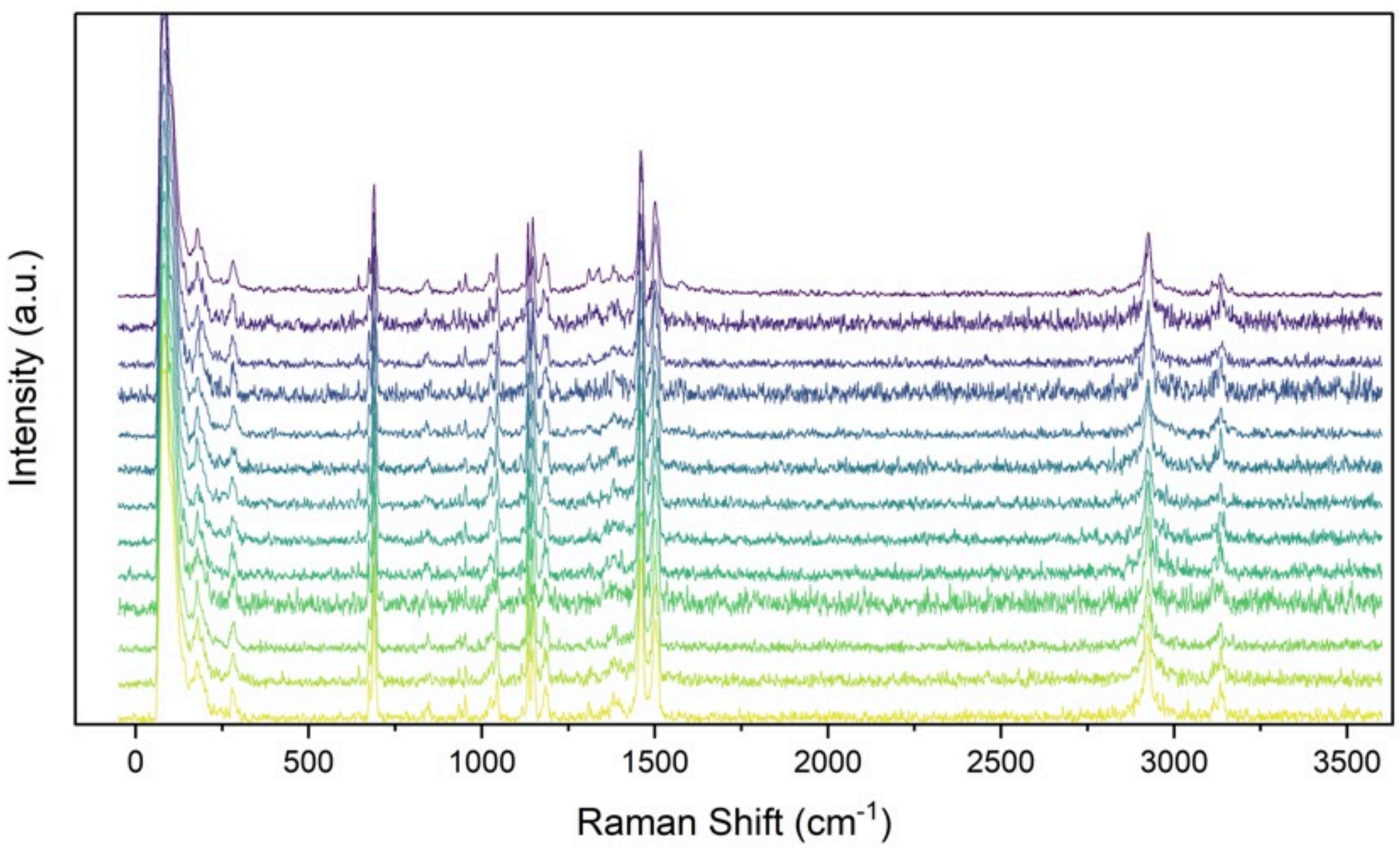

Figure S6. Raman spectra of F@ZIF-L samples with up to 5 mg of F during synthesis (light green to blue indicates least to most guest content) compared to ZIF-L (yellow).

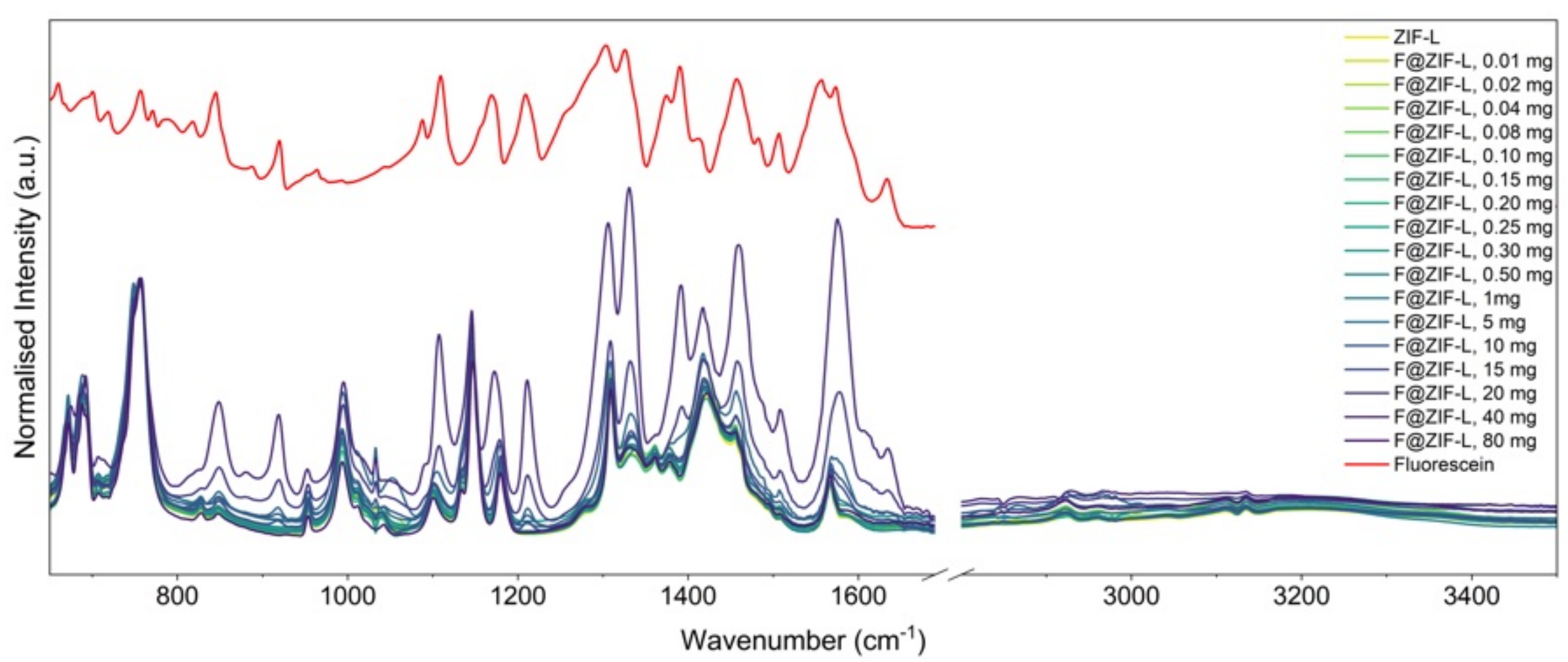

Figure S7. ATR-FTIR of F@ZIF-L samples compared to fluorescein (red spectrum).

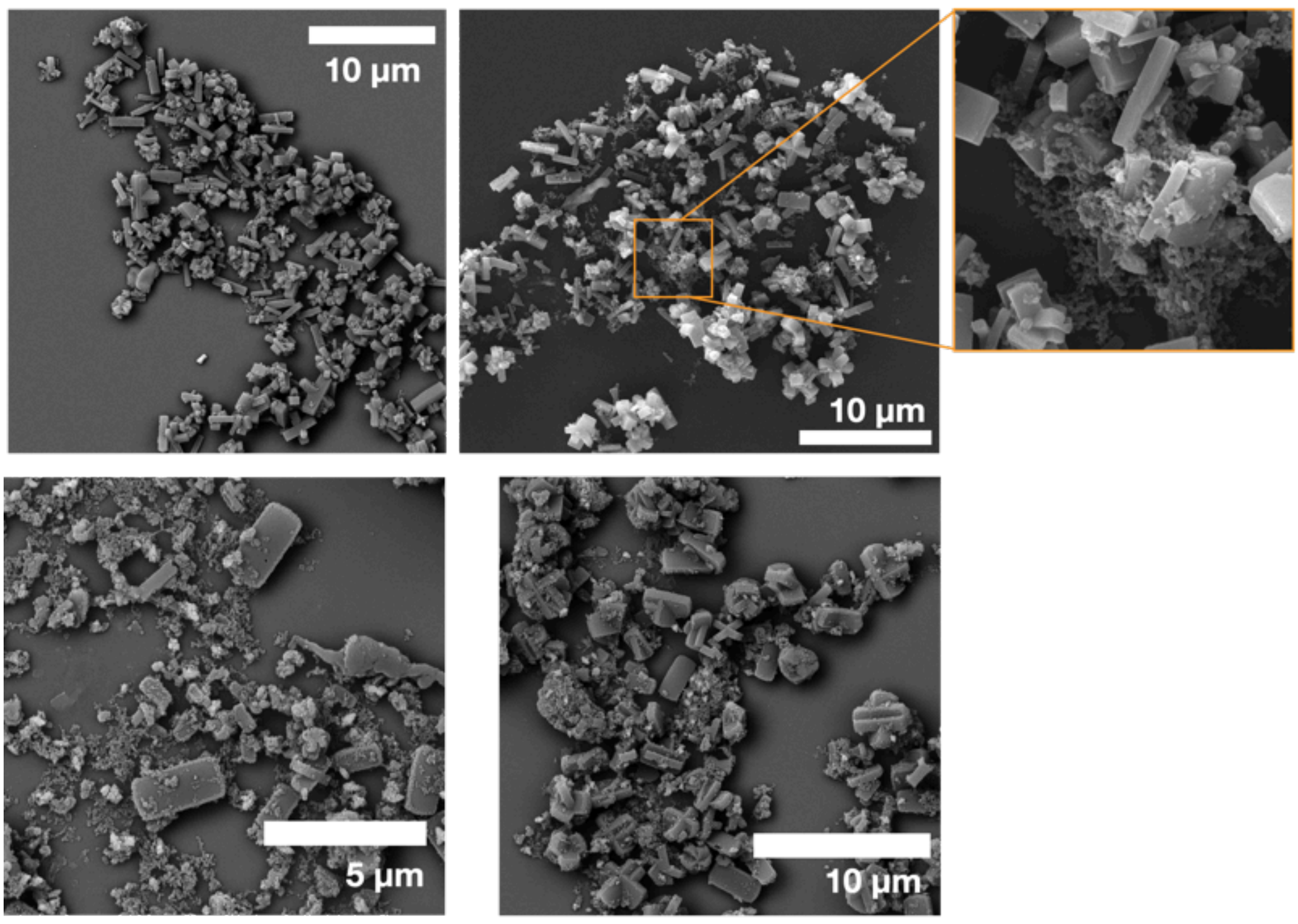

Figure S8. Field emission scanning electron microscopy (FE-SEM) of 10 mg (above), 20 mg (bottom left), and 40 mg (bottom right) F@ZIF-L samples showing a mixture of ZIF-L and F particle aggregates.

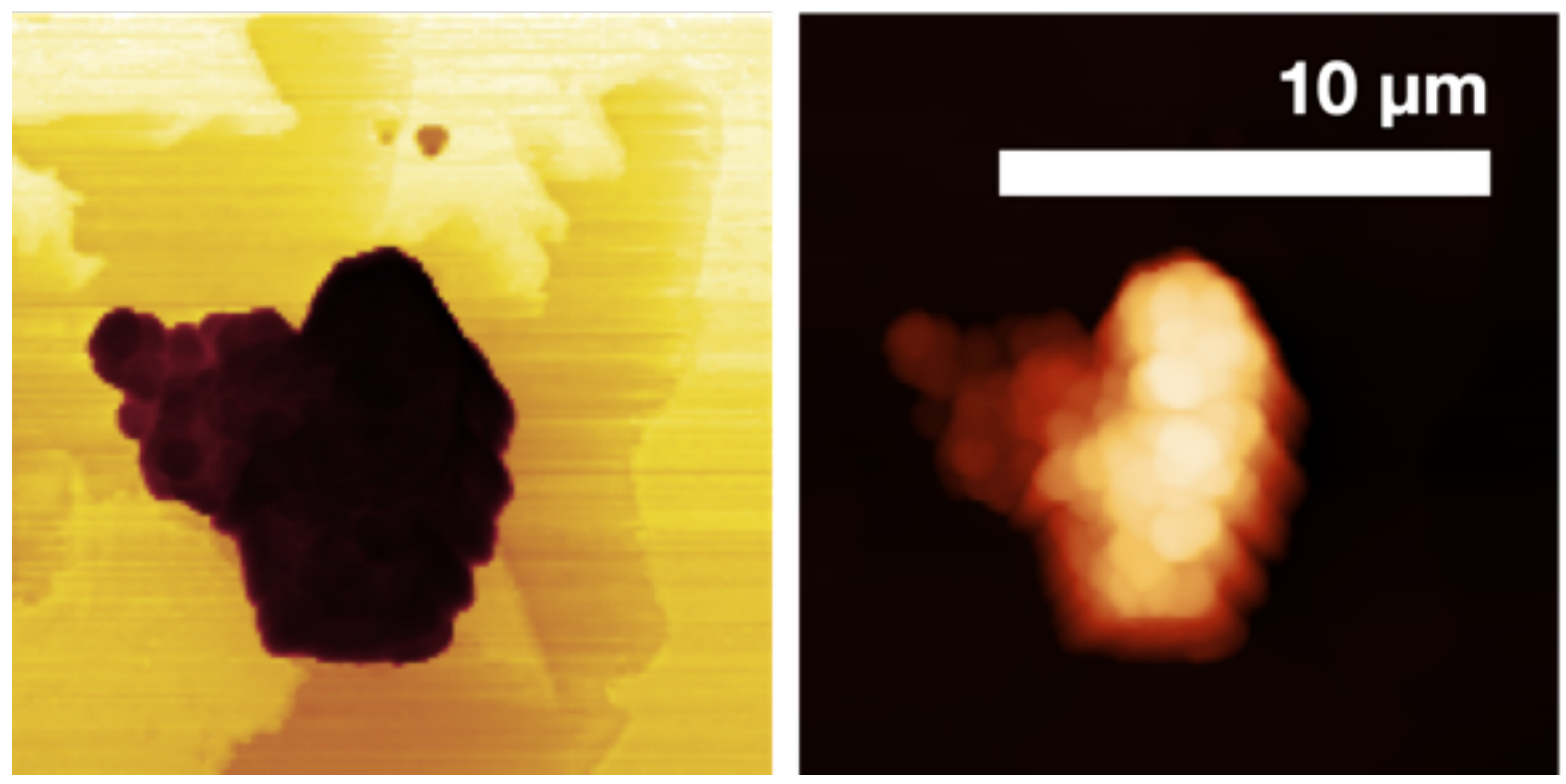

Figure S9. Atomic force microscopy (AFM) height profile (right) and optical amplitude from nanoFTIR scan (left) of a F aggregate in F@ZIF-L (10 mg) sample.

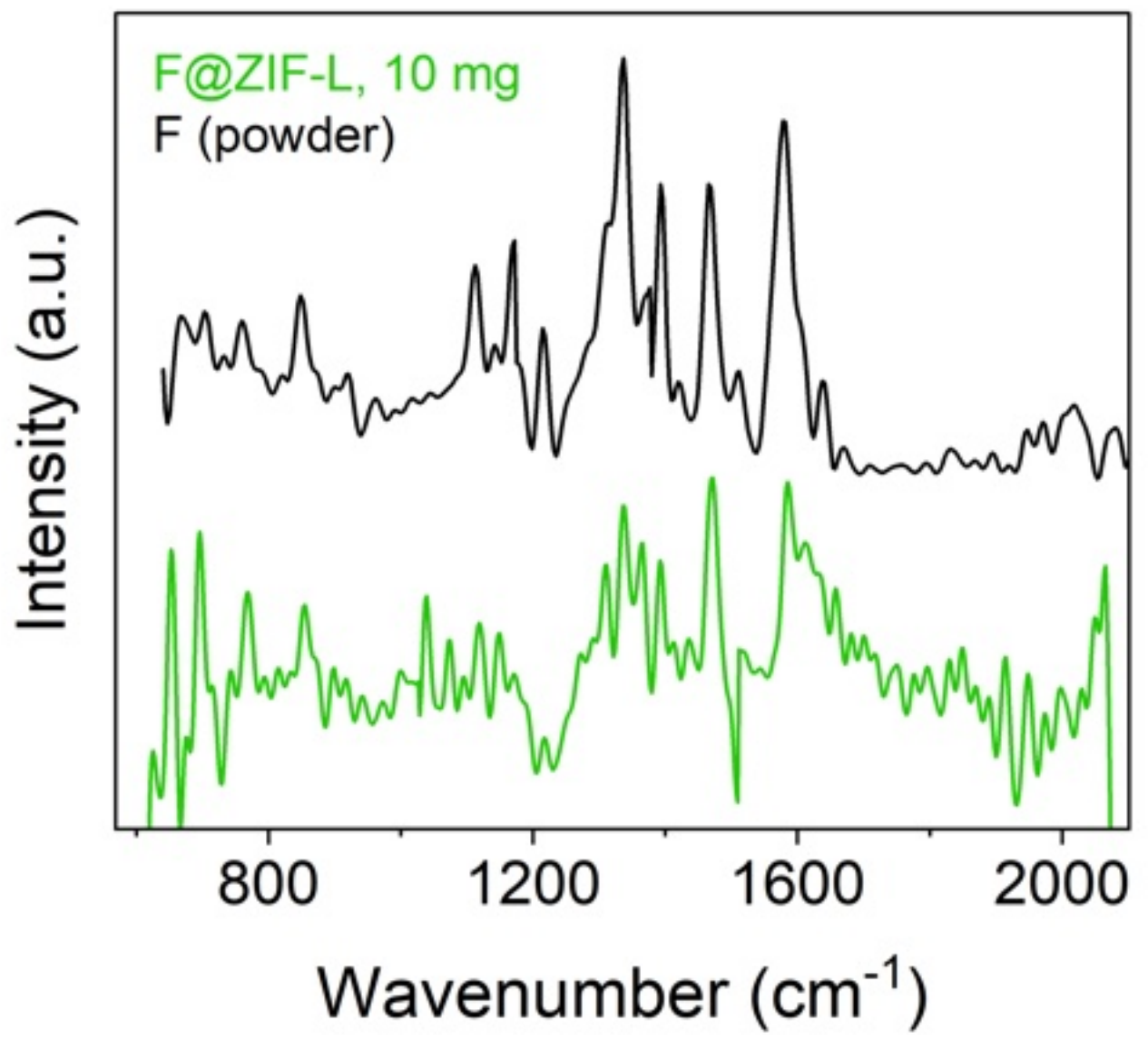

Figure S10. NanoFTIR spectra of F@ZIF-L aggregate particle of F (above), compared with nanoFTIR spectra of solid powder of F.

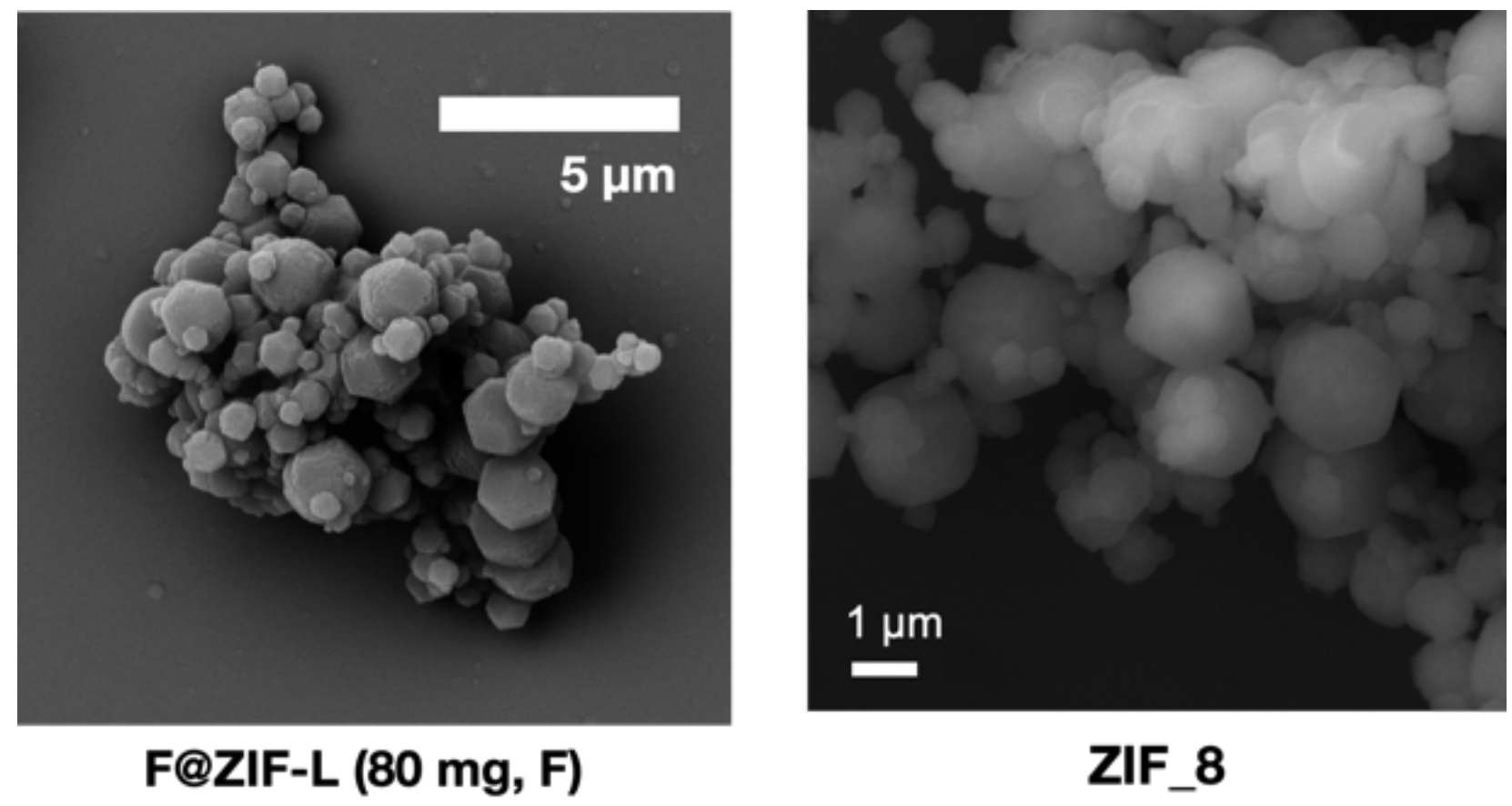

Figure S11. SEM micrographs showing morphology of F@ZIF-L (80 mg, F) compared to ZIF-8.

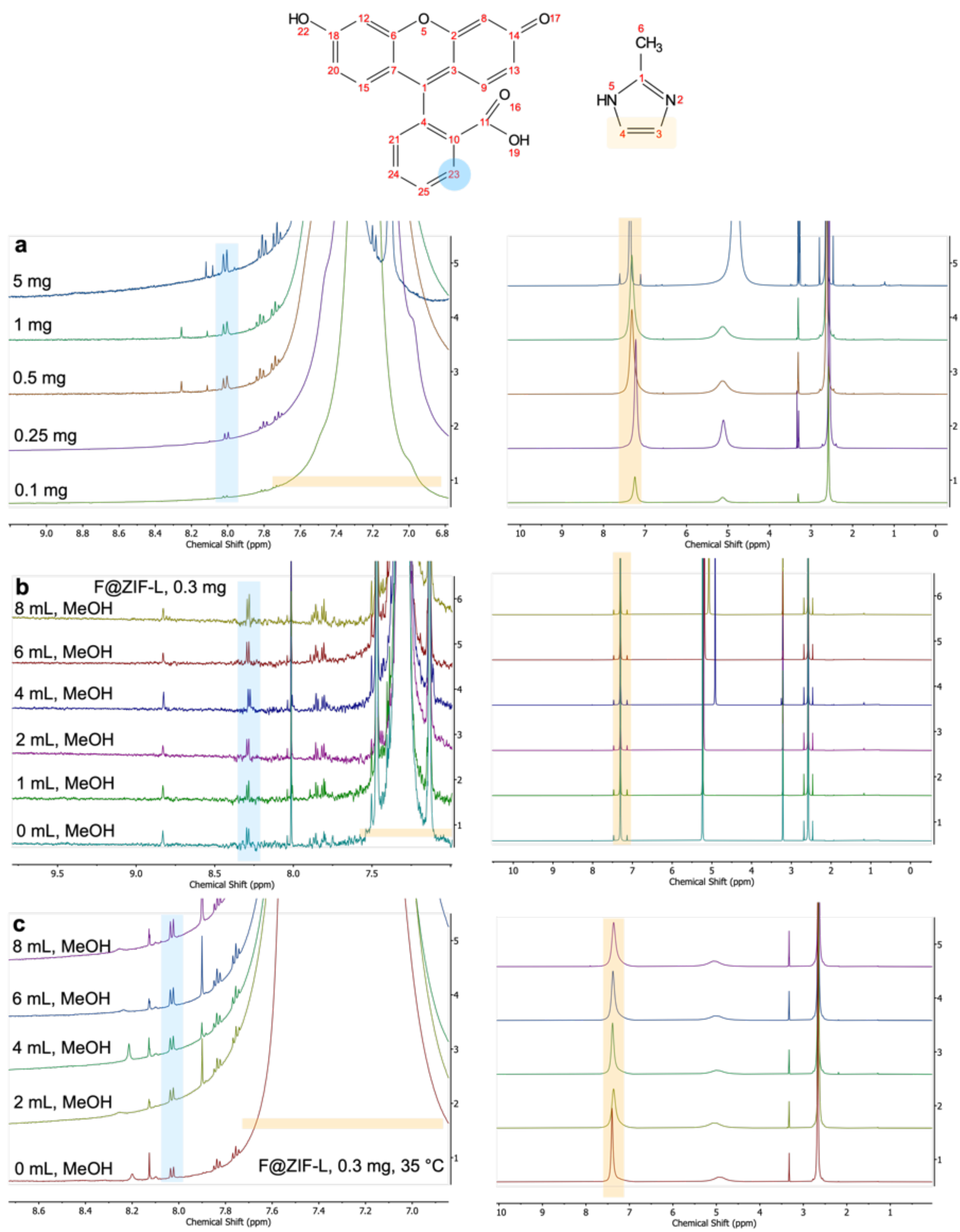

Figure S12. a) Fluorescein (left, above) and Hmim (right, above) indicating protons responsible for characteristic bands highlighted blue and orange, respectively. Below: select NMR spectra of F@ZIF-L samples with various guest loading, digested in DCl/$D_2O$ (35 wt%) with methanol-d4. b) NMR spectra of F@ZIF-L samples (0.3 mg, F) with various quantities of MeOH during synthesis. c) NMR spectra of F@ZIF-L samples (0.3 mg, F) reacted at 35 °C with various quantities of MeOH during synthesis.

Table S1. $N_2$ sorption isotherm (77.3 K) of F@ZIF-L compared to reported values.

| Material | ZIF-L BET Surface Area ($m^2/g$)* | guest@ZIF-L BET Surface Area ($m^2/g$) | % loss | Micropore Pore Volume ($cm^3/g$) | Micropore Pore Volume with guest ($cm^3/g$) | % loss |
|---|---|---|---|---|---|---|
| Phosphate@ZIFL[6] | 17.302 | 11.378 | 34% | 0.047 | 0.07 | -49% |
| Carbon dot@ZIF-L[7] | 9.9567 | 7.8031 | 22% | 0.012659 | 0.009824 | 22% |
| Carbon dot@ZIF-L[8] | 26.859 | 27.374 | -2% | 0.076 | 0.082 | -8% |
| Carbon dot@ZIFL[9] | 44.261 | 37.551 | 15% | 0.081 | 0.167 | -106% |
| F@ZIF-L | 15.4413 | 10.861 | 30% | 0.015 | 0.0053 | 65% |

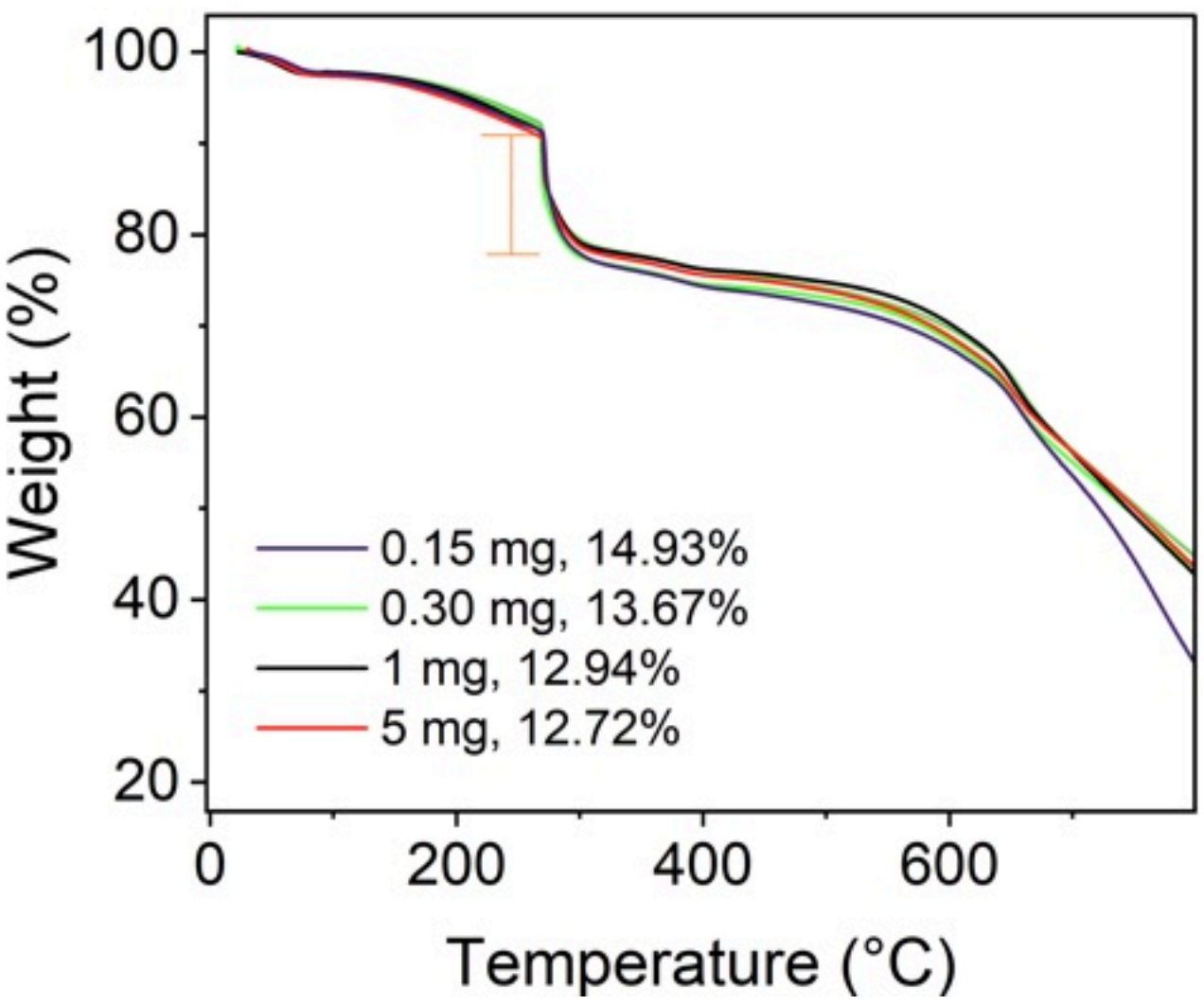

Figure S13. Thermogravimetric analysis (TGA) of F@ZIF-L samples with various guest loadings. Orange segment indicates mass loss from free Hmim ligand, with corresponding % losses indicated in legend for each sample.

MeOH Quantity

Amount of Fluorescein (mg)

$H_2O$ Only

4 mL

0.01

0.05

1 mL

2 mL

8 mL

0.1

0.3

0.5

1

5

Figure S14. Matrix of F@ZIF-L particle morphology from AFM imaging with varying F and MeOH content during synthesis, showing widening of particles with increasing MeOH content. Scale bars indicate 1 μm.

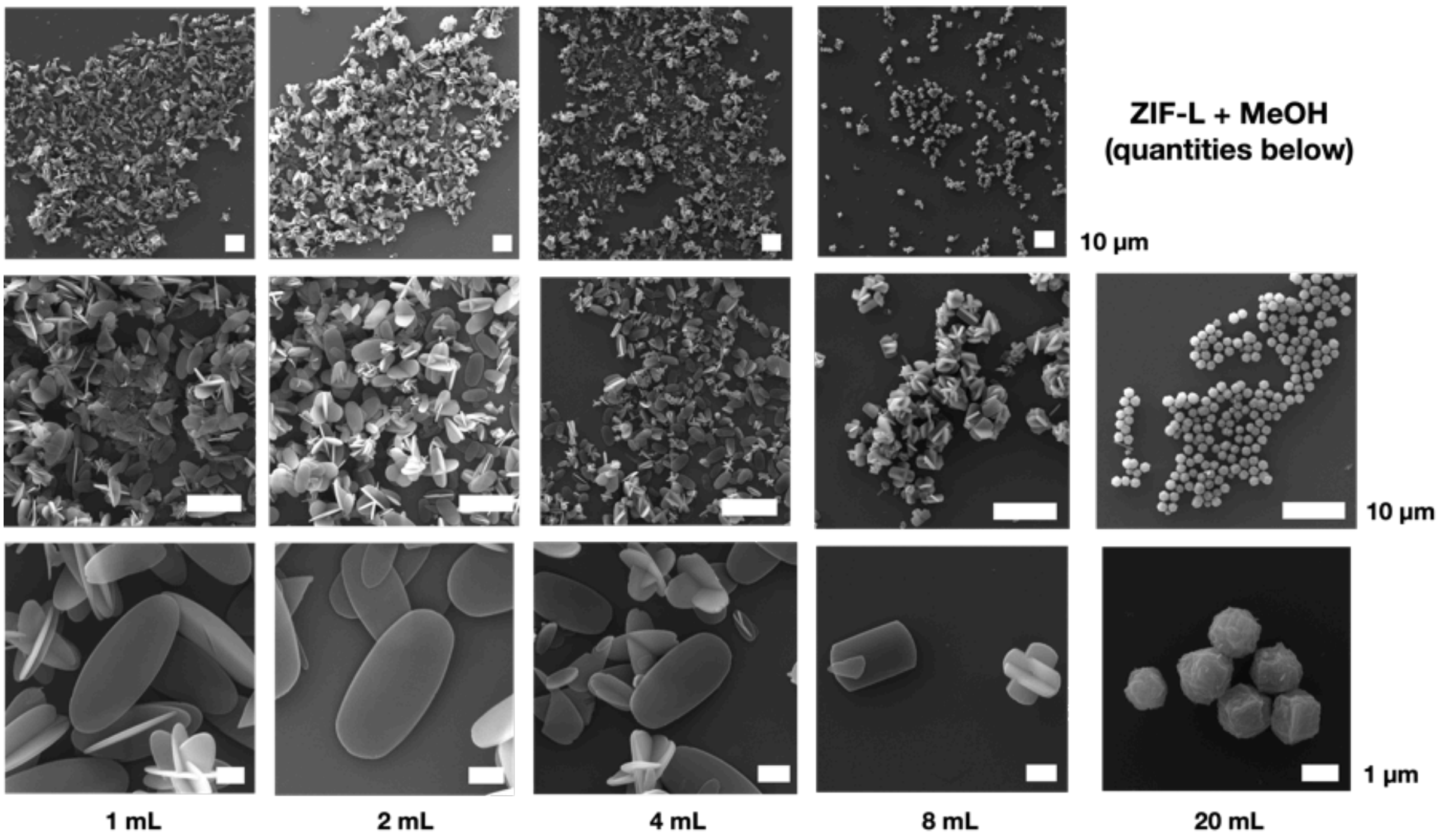

Figure S15. FE-SEM of ZIF-L synthesised with the addition of MeOH during synthesis to the quantities indicated above. Scale bar length indicated to the right of each row.

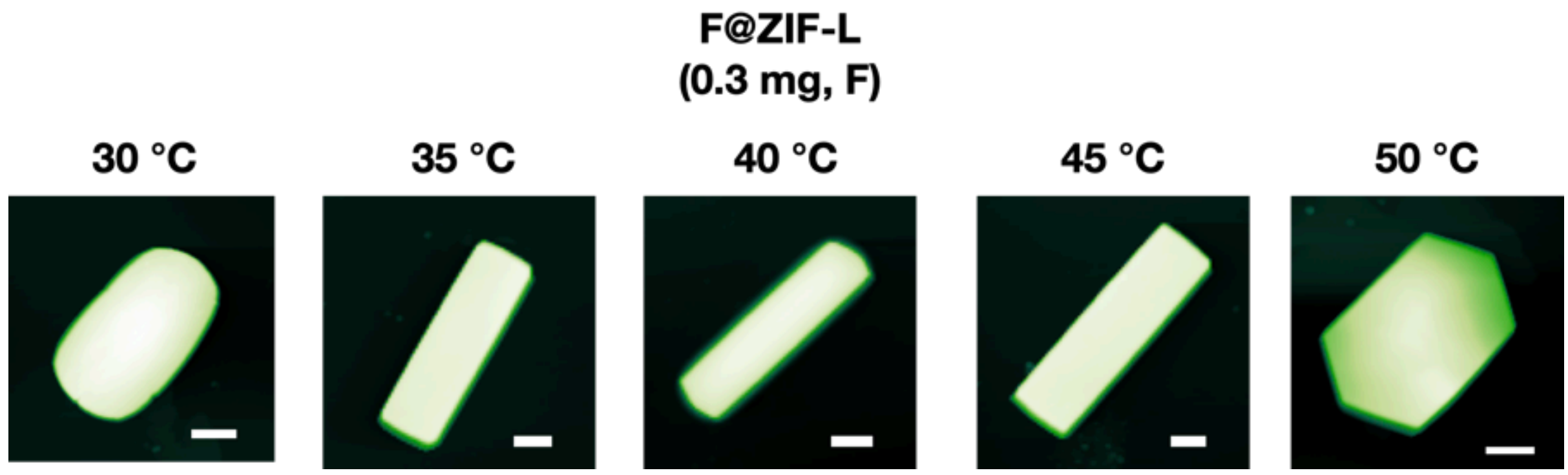

Figure S16. AFM particle morphology of F@ZIF-L (0.3 mg, F with 6 mL MeOH) at various reaction temperatures. Scale bars indicate 1 µm.

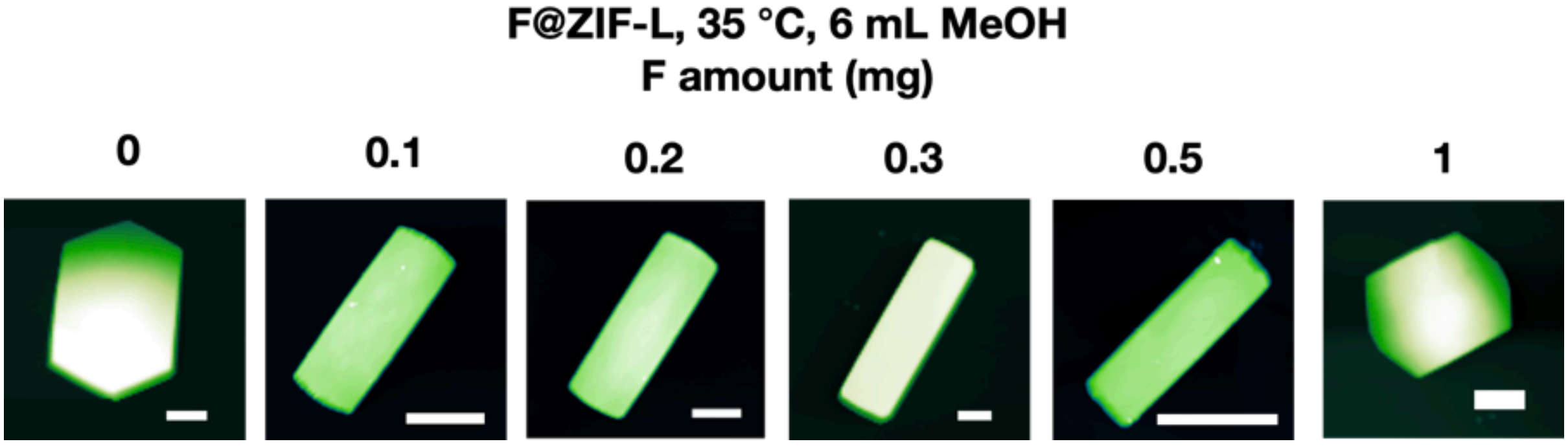

Figure S17. AFM morphology of F@ZIF-L synthesised at 35 °C, 6 mL MeOH. Scale bars indicate 1 μm.

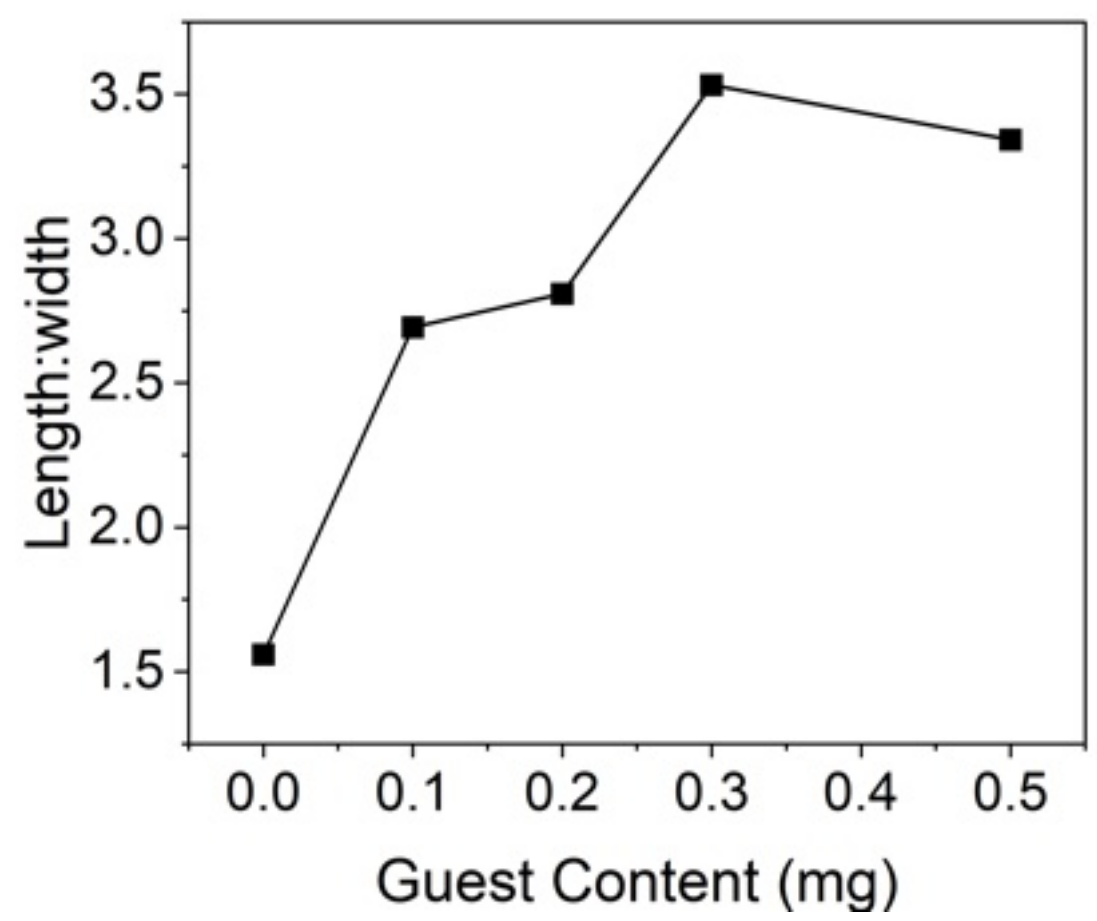

Figure S18. Dimensions of F@ZIF-L particles in Figure S17 with varying guest loading.

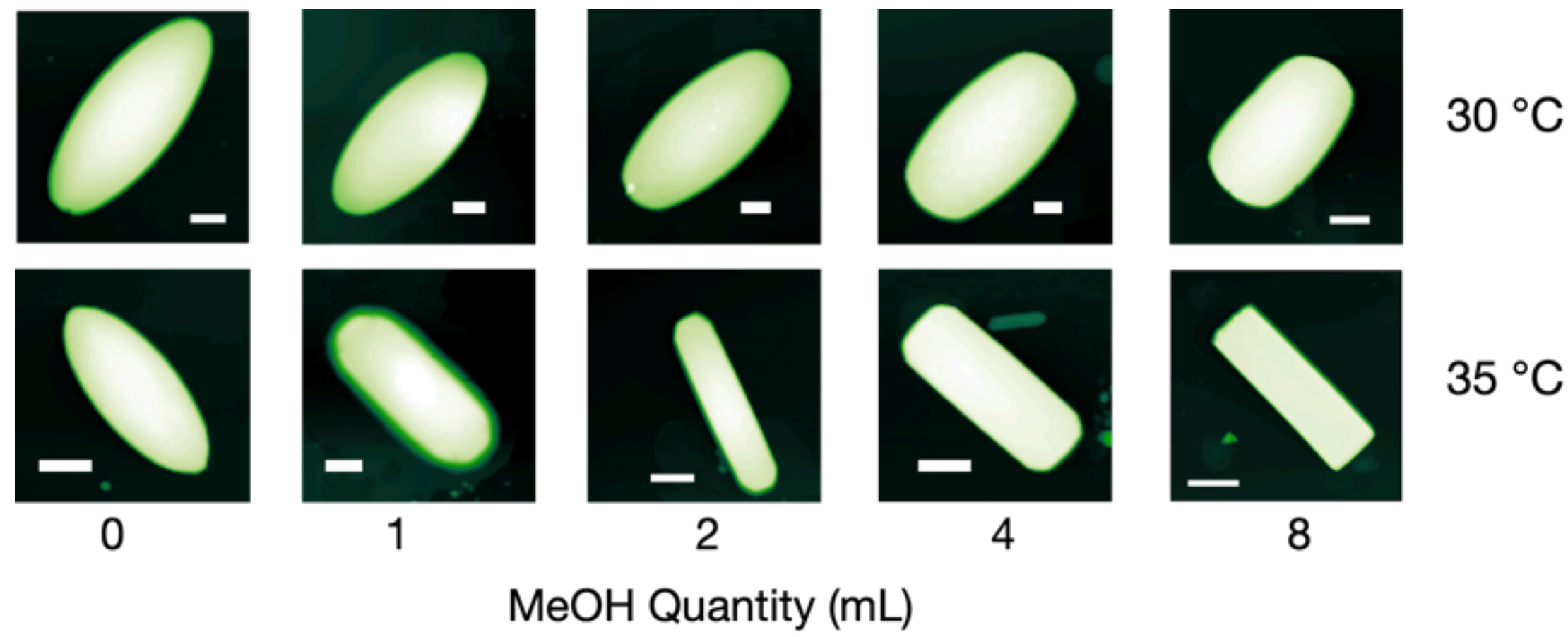

Figure S19. AFM morphology of F@ZIF-L synthesised with 0.3 mg of F at various MeOH quantities (mL, below) and temperatures (right). Scale bars indicate 1 μm.

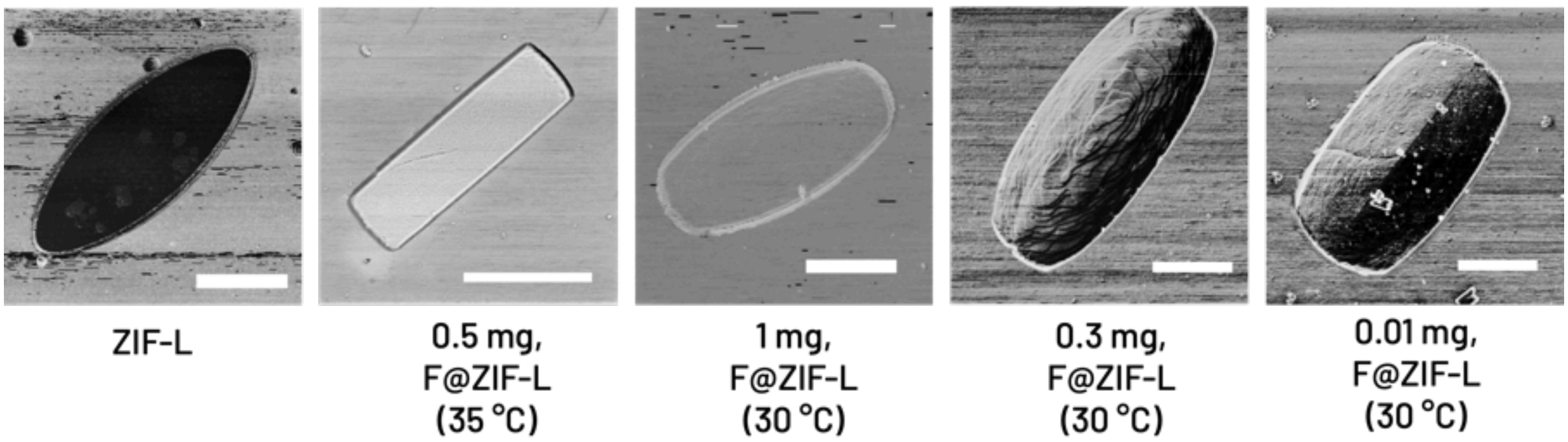

Figure S20. Mechanical phase AFM images of F@ZIF-L particles compared to ZIF-L, highlighting various surface topologies. Scale bars indicate 1 μm.

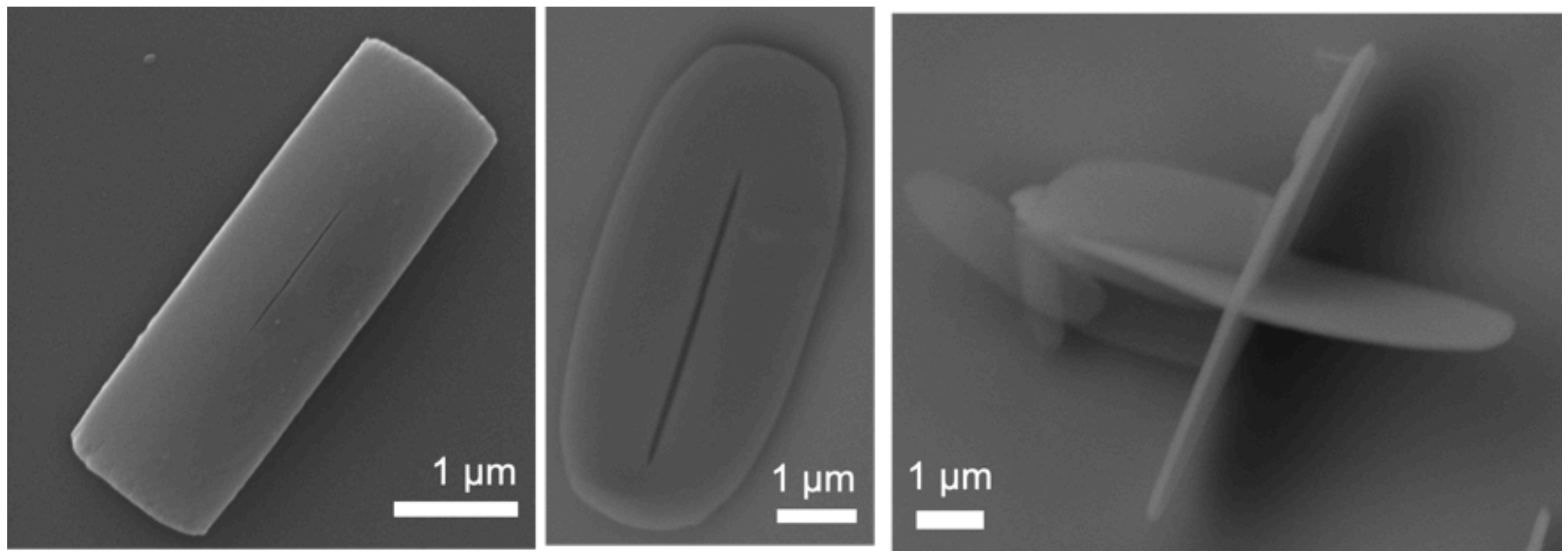

Figure S21. FE-SEM of F@ZIF-L windmill morphologies and single particles exhibiting growth sites of intersected particles.

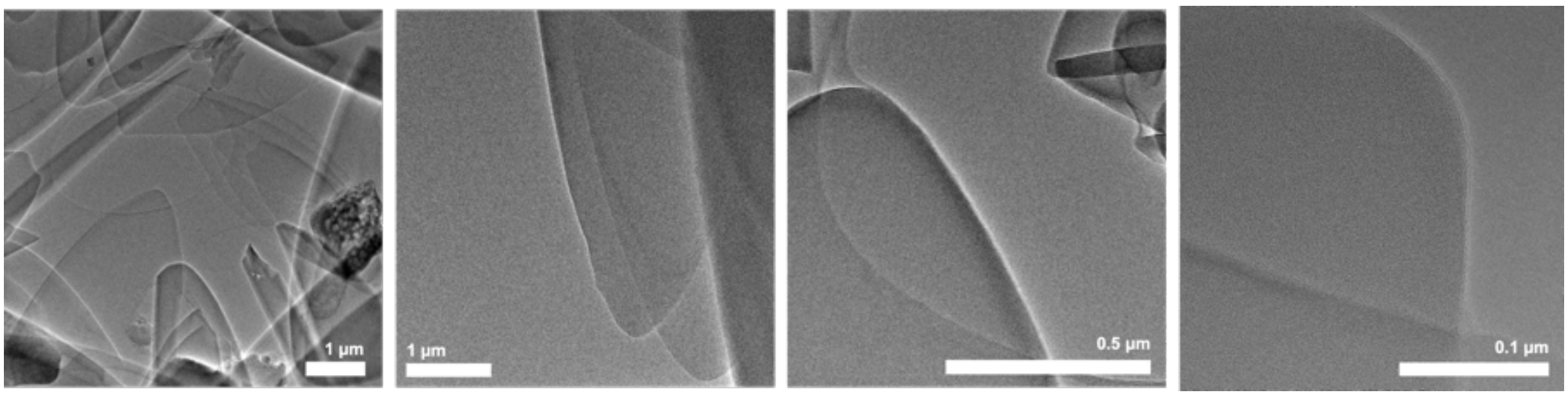

Figure S22. HAADF-STEM of ZIF-L.

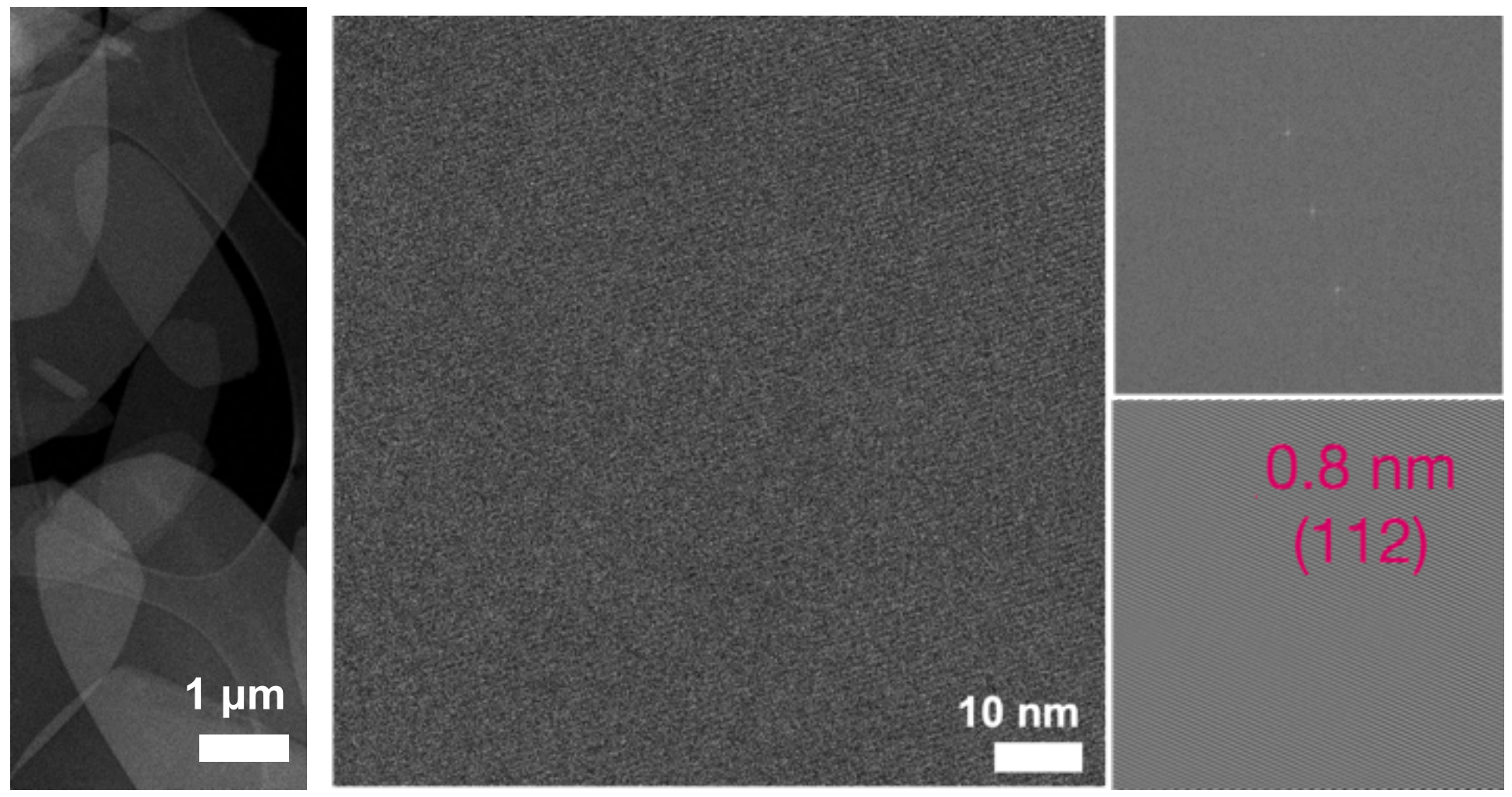

Figure S23. Additional HAADF-STEM of F@ZIF-L (left), with a BF image towards the centre of the particle's end (i.e. away from an edge) (middle), confirming uniform diffraction by Fourier transform analysis (right).

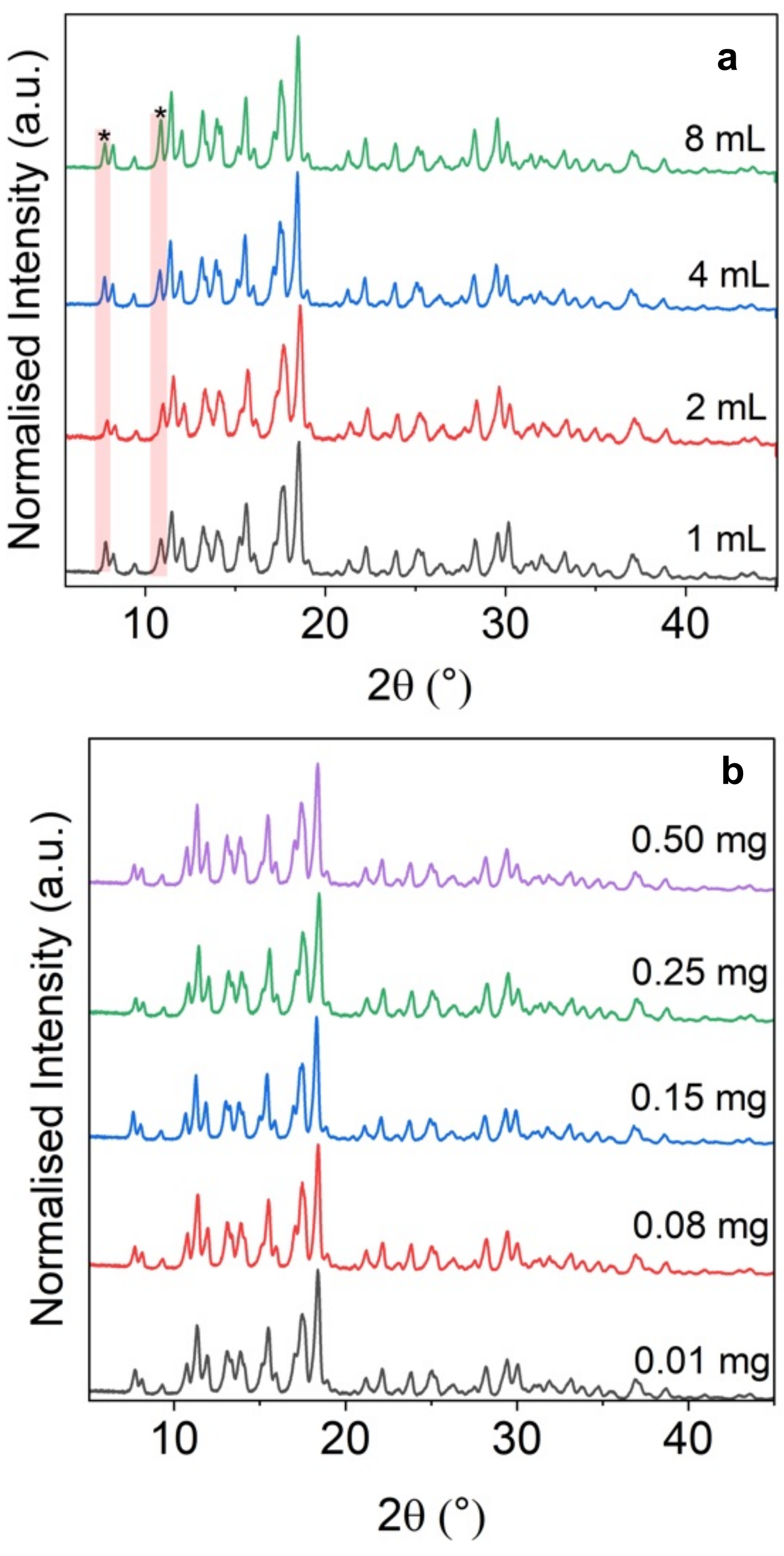

Figure S24. a) PXRD patterns of F@ZIF-L (0.3 mg, F) with increasing MeOH content during synthesis, showing decrease in (200) but increase in (020) reflections. b) PXRD patterns of F@ZIF-L with $H_2O$ only but varied F content, exhibiting no major peak shifts or intensity adjustments.

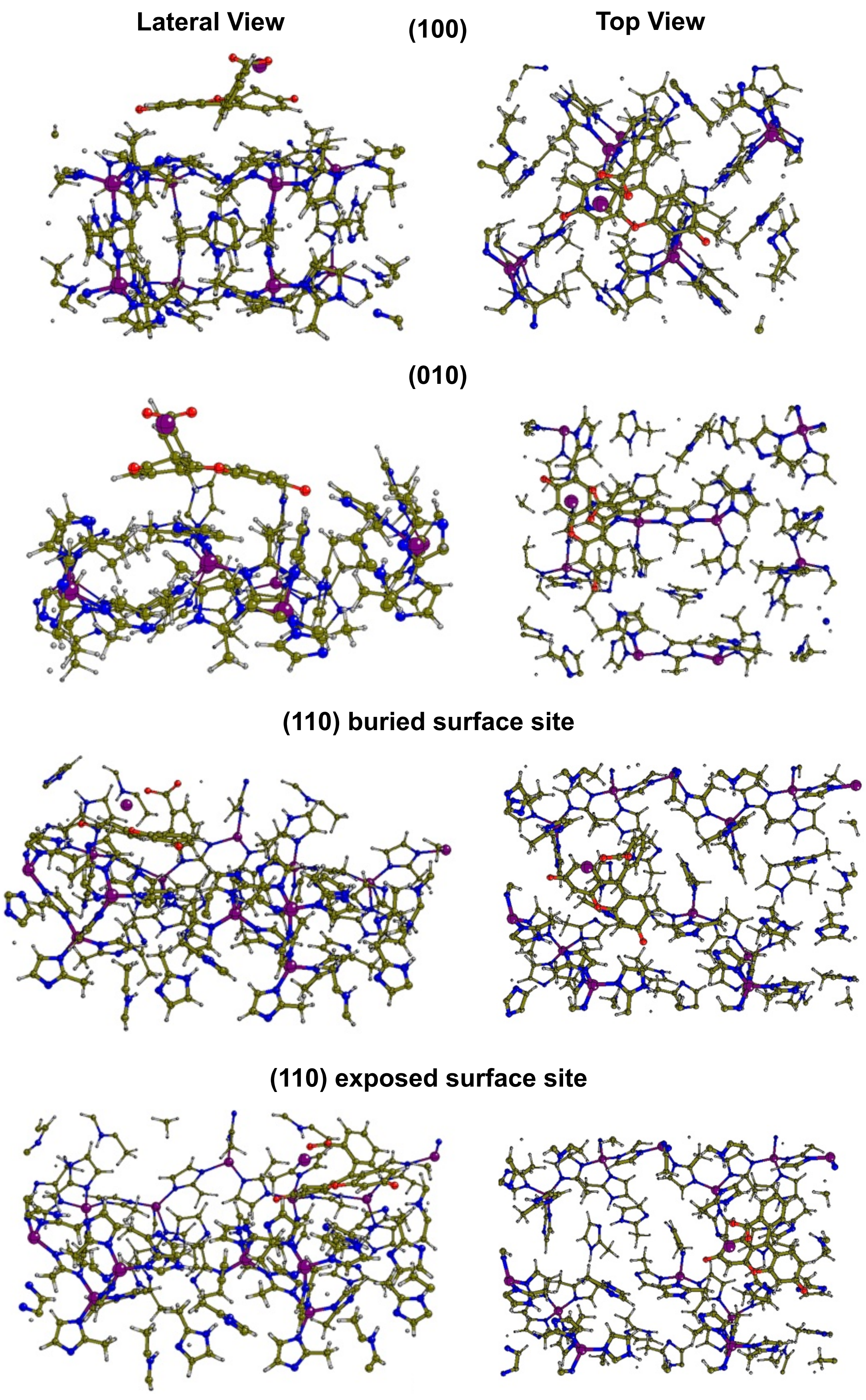

Figure S25. Modelling of fluorescein (F) guest adsorption on F@ZIF-L surfaces (Zn = purple, N = blue, C = yellow, O = red, O = white).

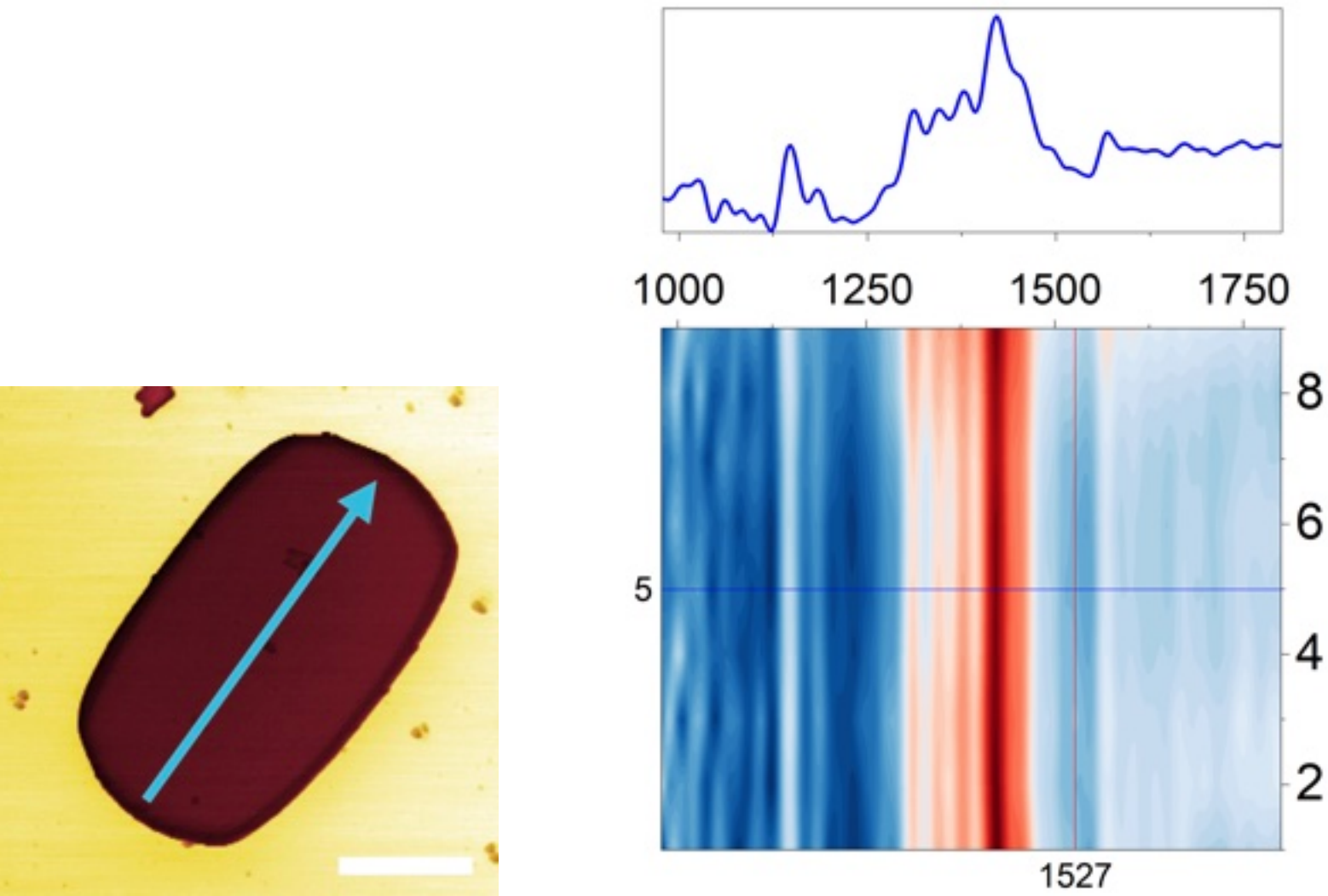

Figure S26. Line scan (9 points) of F@ZIF-L rounded particle (left, scale bar indicating 1 μm) using longer wavelength laser to identify consistency of entire ring stretching bands, along with in-plane ring bending.

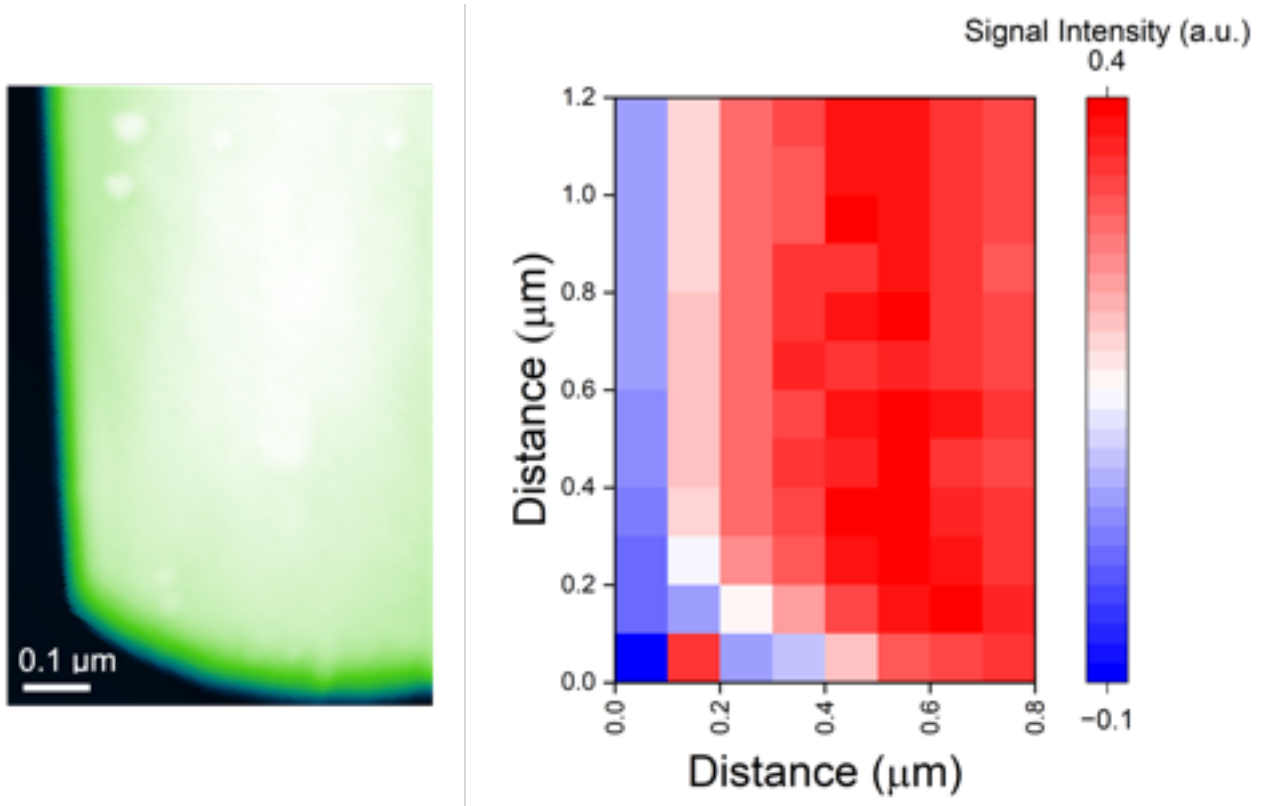

Figure S27. 12x8 nanoFTIR spectral map of F@ZIF-L (rounded rectangle morphology), showing AFM height profile (left) and intensity of 780 $cm^{-1}$ spectral band (right).

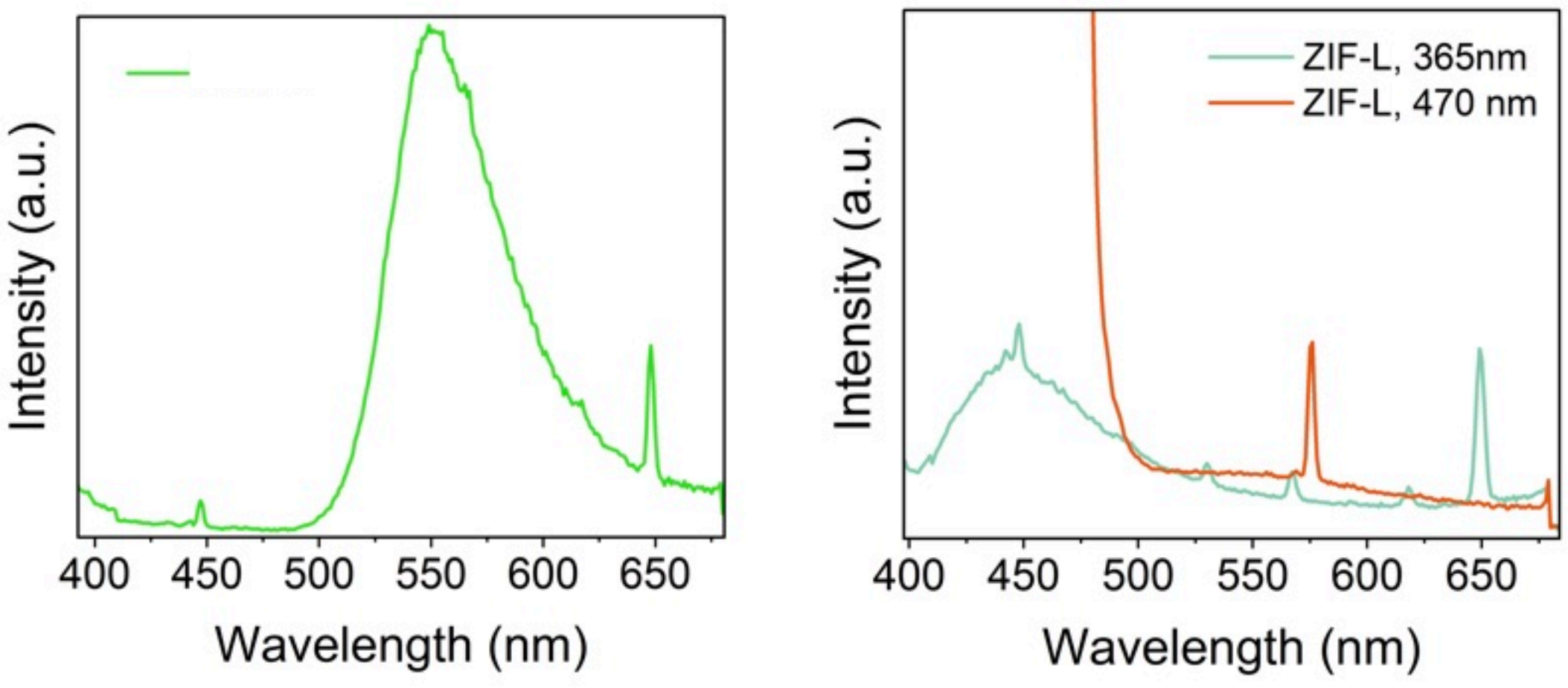

Figure S28. Left: F@ZIF-L emission spectra at 365 nm, showing minimal contribution from ZIF-L. Right: ZIF-L emission spectra when excited at 365 nm (blue) and 470 nm (orange).

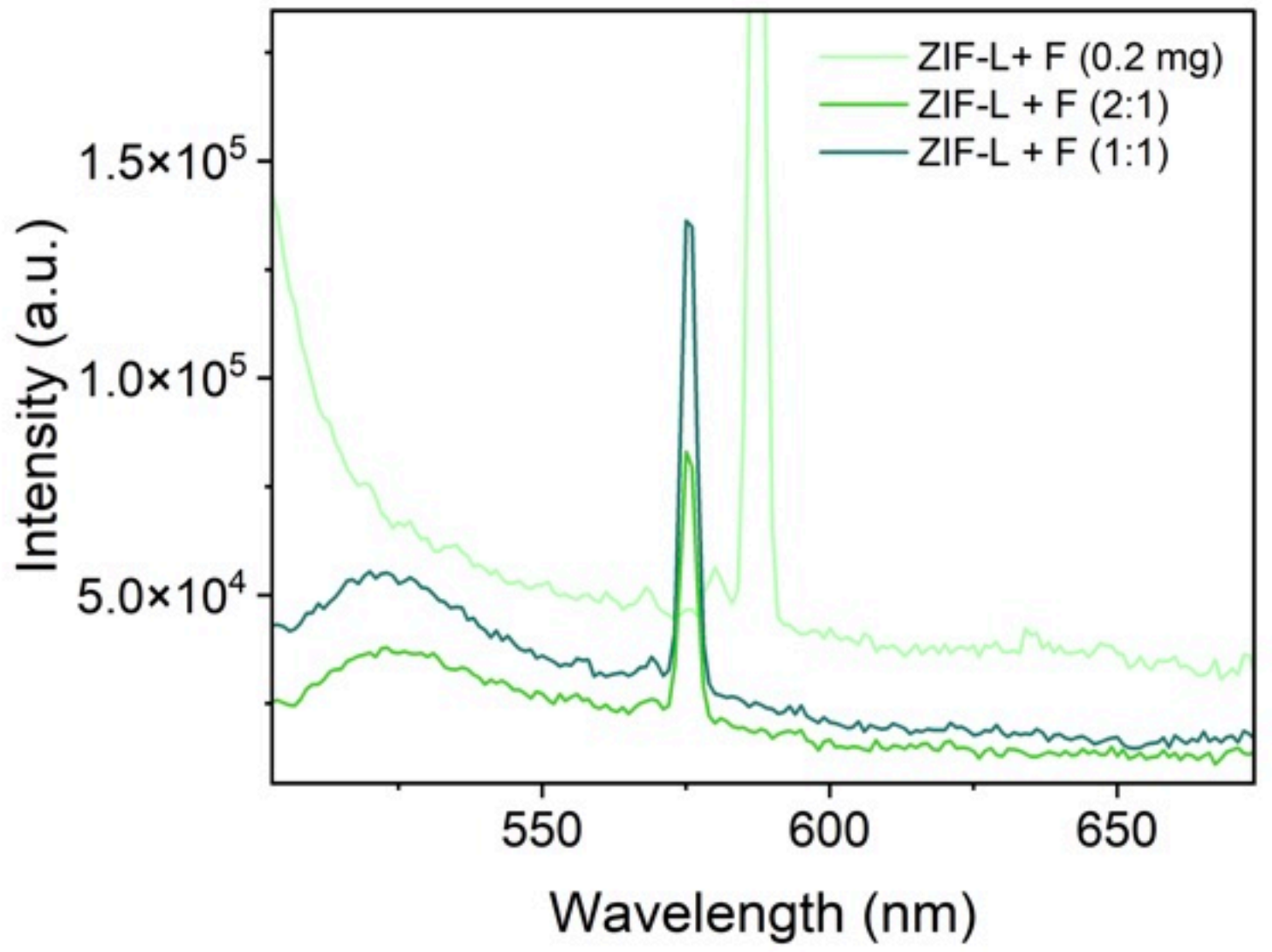

Figure S29. Emission (excited at 470 nm) of physically mixed samples of ZIF-L and F (solid powder), in a ratio equivalent to F@ZIF-L (0.2 mg, F), 2:1 and 1:1 ZIF-L: F by weight.

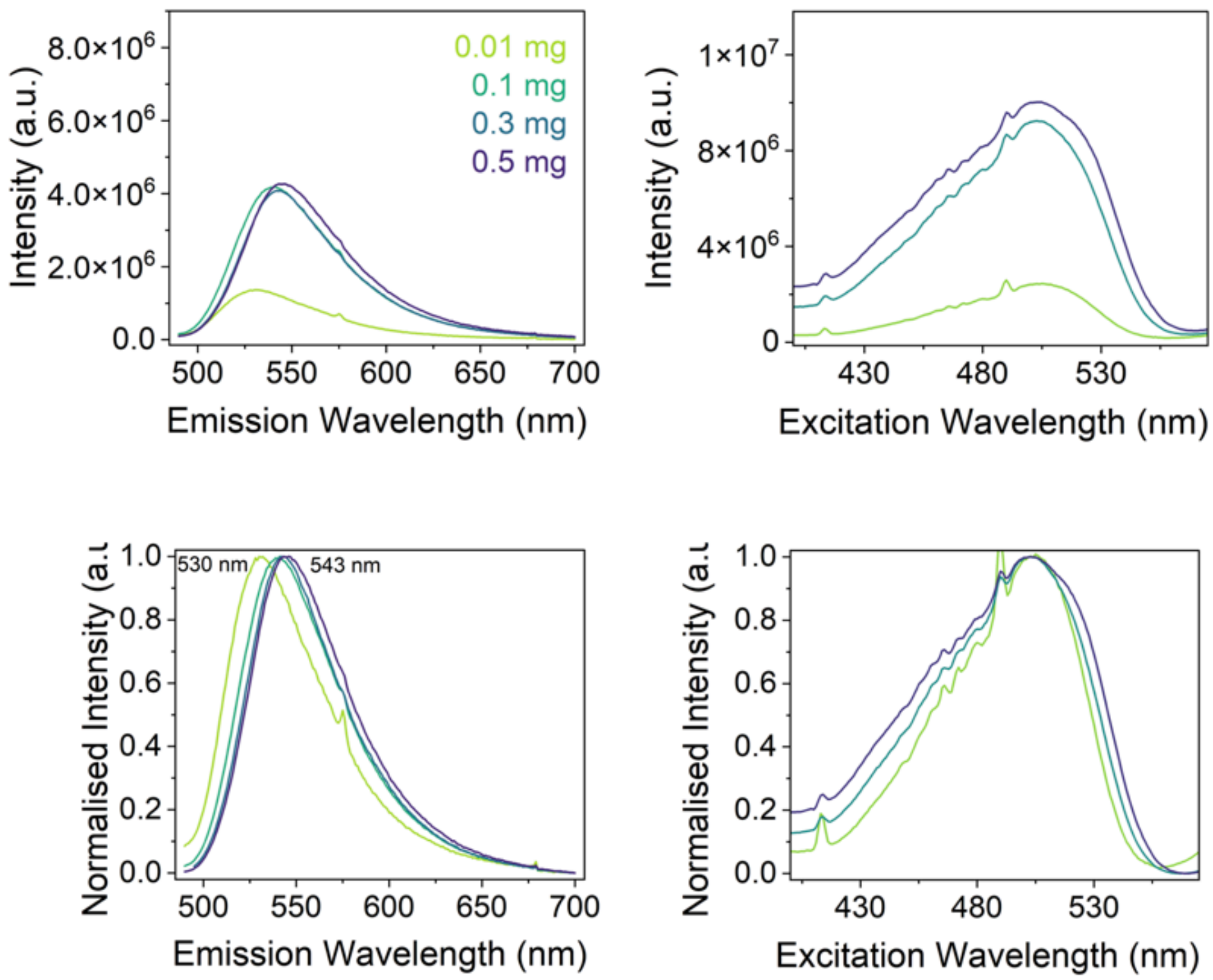

Figure S30. Emission (left, excited at 470 nm) and excitation (right, observed at 600 nm) spectra of F@ZIF-L samples with various F content, synthesised in $H_2O$ only. Data is presented with as measured intensity (above) and normalised (below).

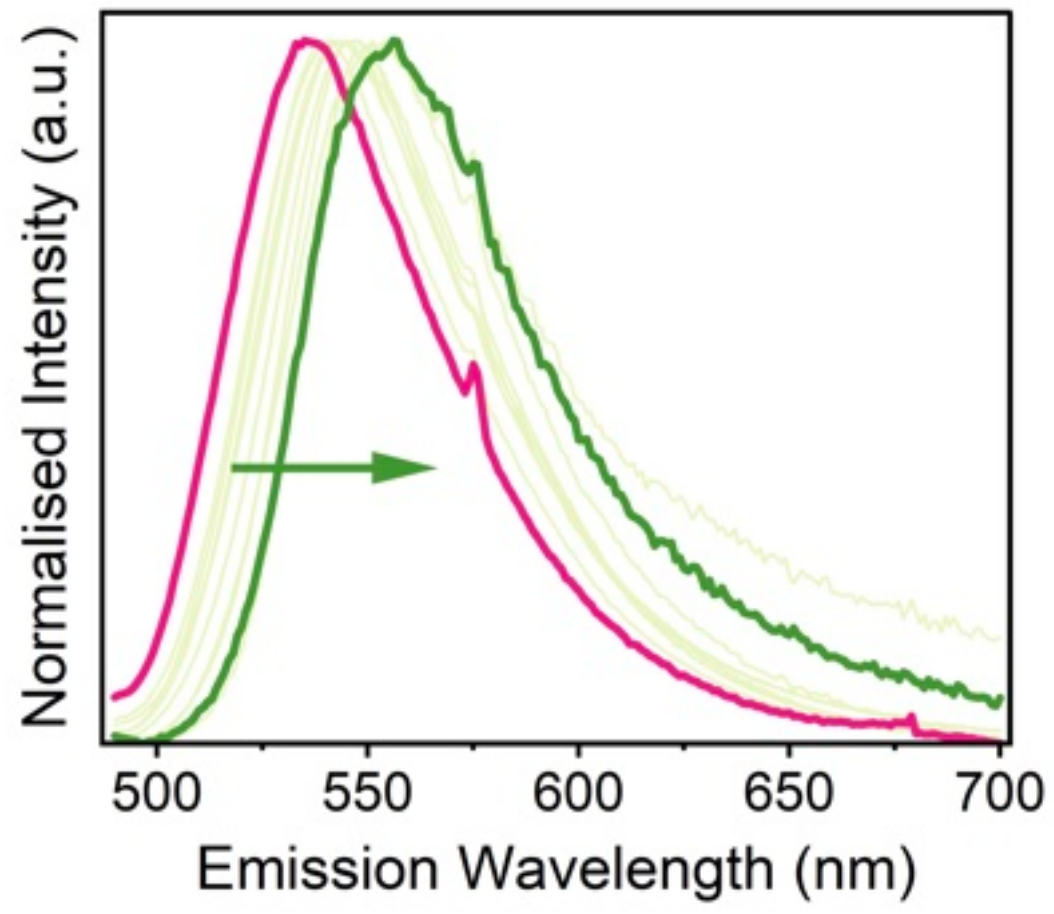

Figure S31. Normalised emission spectra of various F@ZIF-L samples excited at 470 nm (lowest loading pink, highest loading dark green), showing red shift of emission maxima.

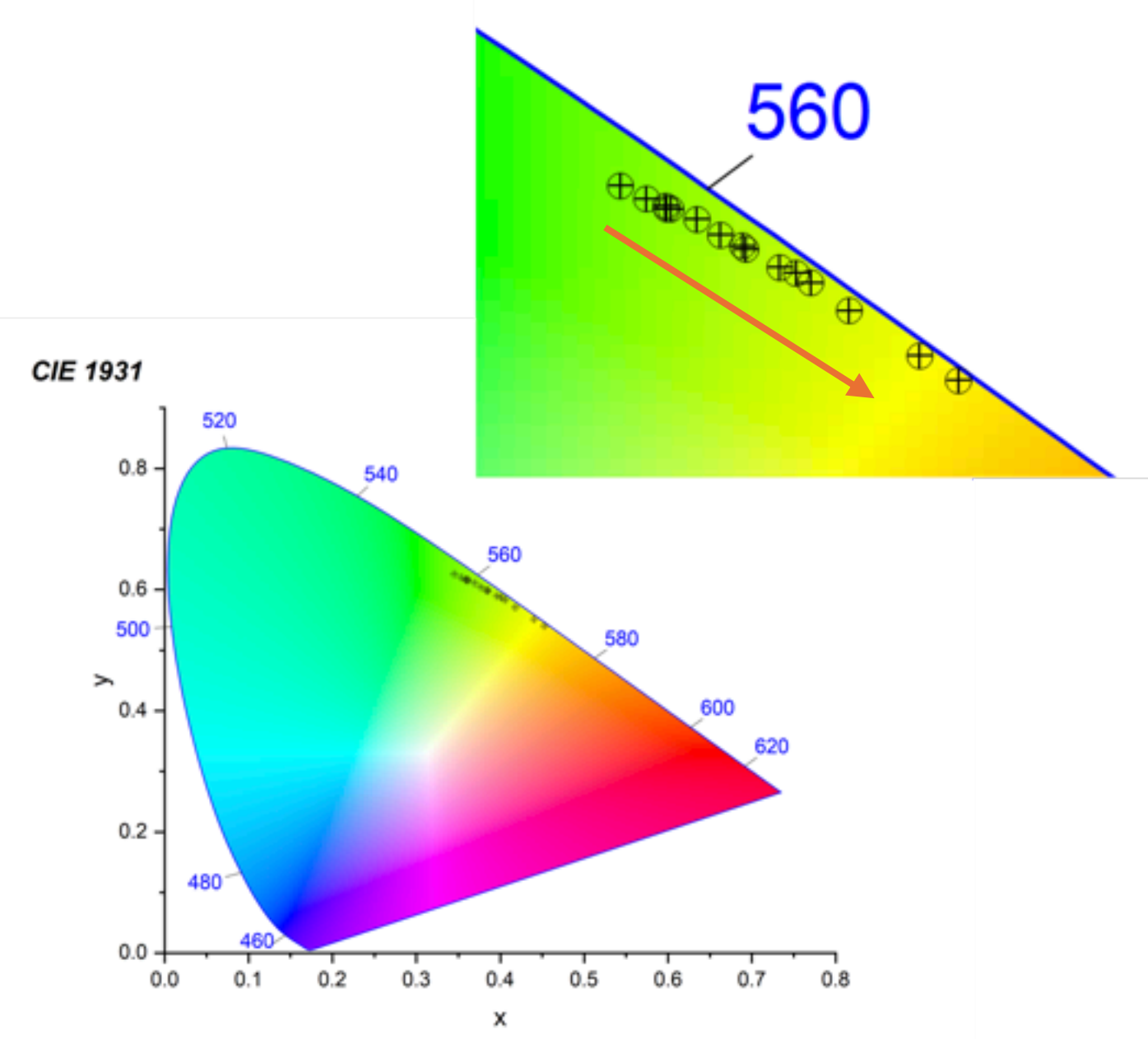

Figure S32. Emission chromaticity of various F@ZIF-L samples, sequentially shifting with increasing F loading from bright green to yellow.

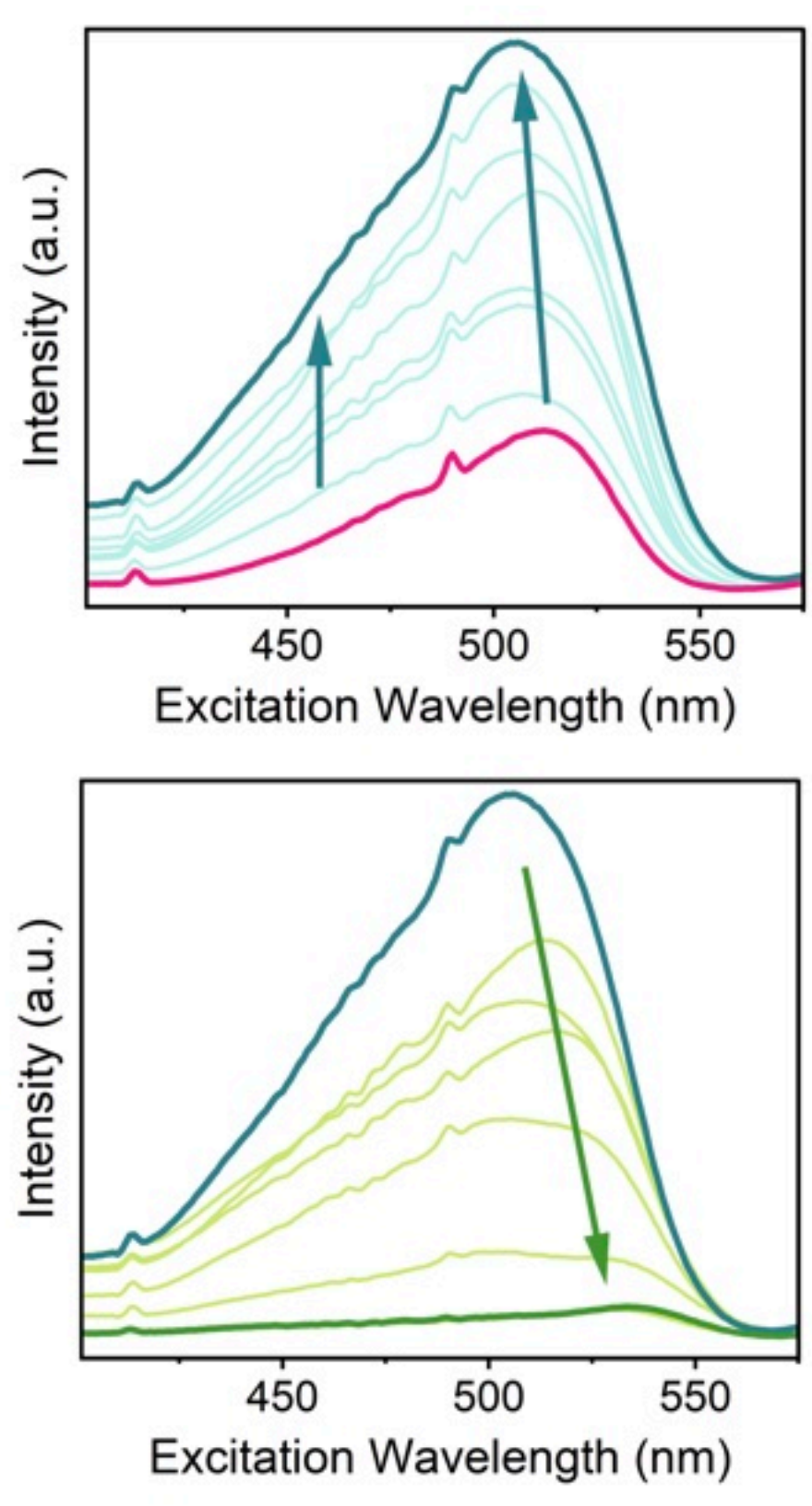

Figure S33. Excitation spectra (observed at 600 nm) of F@ZIF-L samples with varying F content. Least F loading sample in pink, most F loading sample in dark green.

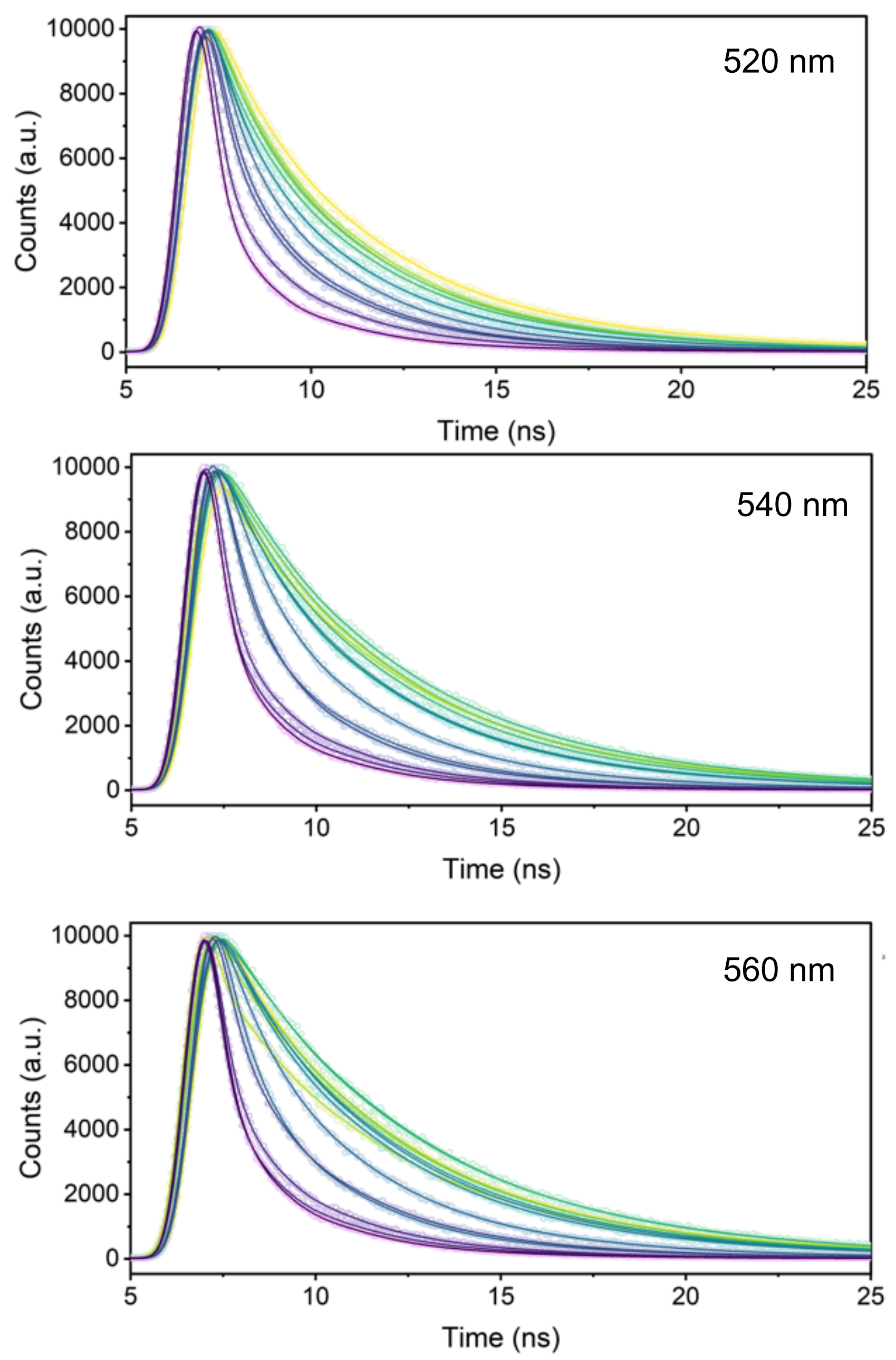

Figure S34. Lifetime decay data with exponential decay fits for F@ZIF-L samples, observed at 3 different wavelengths. Yellow-Green-Blue-Purple indicates lowest-highest F loaded F@ZIF-L samples.

Table S2 – Lifetime data for F@ZIF-L samples with varying guest loading.

| F (mg) | λobs [nm] | τ1[ns] | a1 | c1[%] | τ2[ns] | a2 | c2[%] | τ3[ns] | a3 | c3[%] | χ2 |
|---|---|---|---|---|---|---|---|---|---|---|---|
| 0.01 | 520 | 0.3901 | 0.0152 | 2.8 | 3.1023 | 0.0342 | 50.03 | 5.821 | 0.0172 | 47.17 | 1.051 |
| 0.01 | 540 | 0.07714 | 0.00551 | 1.87 | 3.525 | 0.02616 | 40.6 | 5.992 | 0.0218 | 57.52 | 1.062 |
| 0.01 | 560 | 0.04814 | 0.00914 | 1.83 | 3.499 | 0.02765 | 40.2 | 6.061 | 0.02302 | 57.98 | 1.043 |
| 0.02 | 520 | 0.0708 | 0.1432 | 4.94 | 2.6794 | 0.03 | 39.16 | 5.2478 | 0.0218 | 55.89 | 1.0293 |
| 0.02 | 540 | 0.0473 | 0.2103 | 4.16 | 3.14 | 0.0192 | 25.22 | 5.5949 | 0.0302 | 70.62 | 1.1374 |
| 0.02 | 560 | 0.0566 | 0.3099 | 7.74 | 3.3 | 0.0189 | 27.53 | 5.8306 | 0.0252 | 64.73 | 1.0314 |
| 0.04 | 520 | 0.4661 | 0.015 | 3.49 | 2.7911 | 0.0303 | 42.32 | 4.931 | 0.022 | 54.19 | 1.0432 |
| 0.04 | 540 | | | | 2.452 | 0.0153 | 15.62 | 5.1114 | 0.0396 | 84.38 | 1.0891 |
| 0.04 | 560 | 0.1627 | 0.0457 | 3.06 | 3.3015 | 0.018 | 24.49 | 5.4735 | 0.0322 | 72.46 | 1.1143 |
| 0.06 | 520 | 0.2595 | 0.0277 | 3.72 | 2.6681 | 0.0322 | 44.45 | 5.0007 | 0.0201 | 51.83 | 1.0744 |
| 0.06 | 540 | 0.0982 | 0.0437 | 1.81 | 3.3222 | 0.0233 | 32.58 | 5.3592 | 0.0291 | 65.62 | 1.0235 |
| 0.06 | 560 | 0.029 | 0.2629 | 3.13 | 3.2776 | 0.0199 | 26.78 | 5.4743 | 0.0311 | 70.08 | 1.0247 |
| 0.08 | 520 | 0.5181 | 0.0183 | 4.79 | 2.6859 | 0.031 | 42 | 5.1602 | 0.0204 | 53.2 | 0.996 |
| 0.08 | 540 | 0.7164 | 0.0025 | 0.72 | 3.1455 | 0.0179 | 22.13 | 5.4911 | 0.0357 | 77.15 | 1.0887 |
| 0.08 | 560 | | | | 2.9708 | 0.0126 | 13.89 | 5.557 | 0.0418 | 86.11 | 1.1823 |
| 0.1 | 520 | 0.2777 | 0.0332 | 4.92 | 2.3168 | 0.0279 | 34.49 | 4.8009 | 0.0237 | 60.59 | 1.0517 |
| 0.1 | 540 | 0.4178 | 0.0122 | 2.24 | 3.259 | 0.0276 | 39.54 | 5.5097 | 0.0241 | 58.22 | 1.1004 |
| 0.1 | 560 | 0.4655 | 0.0075 | 1.49 | 3.0112 | 0.0225 | 28.71 | 5.4677 | | 69.8 | 1.1012 |
| 0.15 | 520 | 0.4123 | 0.0303 | 7.33 | 2.2628 | 0.0313 | 41.48 | 4.6347 | 0.0188 | 51.19 | 1.1152 |
| 0.15 | 540 | 0.8985 | 0.011 | 4.77 | 3.1451 | 0.0297 | 45.25 | 5.2438 | 0.0197 | 49.99 | 1.0404 |
| 0.15 | 560 | 0.8451 | 0.0071 | 2.72 | 3.164 | 0.0268 | 38.13 | 5.308 | 0.0248 | 59.15 | 1.0704 |
| 0.2 | 520 | 0.2947 | 0.0351 | 6.12 | 1.982 | 0.029 | 34.02 | 4.4204 | 0.0229 | 59.86 | 1.0738 |
| 0.2 | 540 | 0.1944 | 0.0223 | 2.09 | 2.3762 | 0.0221 | 25.23 | 4.8297 | 0.0313 | 72.68 | 1.1609 |
| 0.2 | 560 | 0.8385 | 0.0058 | 2.13 | 3.1631 | 0.0242 | 33.55 | 5.3148 | 0.0276 | 64.32 | 1.022 |
| 0.25 | 520 | 0.3946 | 0.0415 | 10.77 | 2.121 | 0.0338 | 47.14 | 4.4176 | 0.0145 | 42.08 | 1.1156 |
| 0.25 | 540 | 0.786 | 0.0194 | 8.9 | 2.5851 | 0.0348 | 52.52 | 4.7307 | 0.014 | 38.58 | 1.182 |
| 0.25 | 560 | 0.4791 | 0.0143 | 3.8 | 2.3019 | 0.033 | 42.07 | 4.5466 | 0.0215 | 54.13 | 1.0472 |
| 0.35 | 525 | 0.2538 | 0.068 | 12.77 | 1.5111 | 0.0292 | 32.65 | 3.8348 | 0.0192 | 54.57 | 1.0288 |
| 0.35 | 545 | 0.3629 | 0.0471 | 11.74 | 1.7406 | 0.0271 | 32.46 | 4.0326 | 0.0201 | 55.79 | 1.1092 |
| 0.35 | 565 | 0.5263 | 0.0375 | 13.13 | 2.0296 | 0.0262 | 35.48 | 4.2213 | 0.0183 | 51.39 | 1.0692 |
| 0.5 | 520 | 0.4172 | 0.0523 | 16.45 | 1.8511 | 0.0316 | 44.05 | 4.2578 | 0.0123 | 39.5 | 1.0501 |
| 0.5 | 540 | 0.5701 | 0.0412 | 17.59 | 1.9735 | 0.0324 | 47.94 | 4.4155 | 0.0104 | 34.47 | 1.0233 |
| 0.5 | 560 | 0.5997 | 0.039 | 16.7 | 2.0933 | 0.0323 | 48.2 | 4.5099 | 0.0109 | 35.1 | 1.1015 |
| 1 | 520 | 0.2875 | 0.0715 | 16.55 | 1.6849 | 0.0291 | 39.44 | 3.8766 | 0.0141 | 44.01 | 1.2166 |
| 1 | 540 | 0.4319 | 0.0558 | 17.68 | 2.0457 | 0.0275 | 41.27 | 4.2754 | 0.0131 | 41.05 | 1.1023 |
| 1 | 560 | 0.3555 | 0.0544 | 13.74 | 1.8 | 0.0267 | 34.2 | 4.1368 | 0.0177 | 52.06 | 1.1357 |
| 5 | 520 | 0.2297 | 0.1306 | 28.37 | 1.6913 | 0.0257 | 41.13 | 4.1853 | 0.0077 | 30.5 | 1.2668 |
| 5 | 540 | 0.2291 | 0.1275 | 28.04 | 1.5125 | 0.0231 | 33.48 | 3.7035 | 0.0108 | 38.48 | 1.0831 |
| 5 | 560 | 0.2307 | 0.1186 | 25.88 | 1.418 | 0.0239 | 32.08 | 3.6898 | 0.012 | 42.05 | 1.1926 |
| 10 | 520 | 0.1467 | 0.1468 | 32.7 | 1.2654 | 0.0162 | 31.12 | 3.4066 | 0.007 | 36.18 | 1.0228 |
| 10 | 540 | 0.1671 | 0.175 | 30.2 | 1.2045 | 0.0232 | 28.9 | 3.2952 | 0.012 | 40.9 | 1.0901 |
| 10 | 560 | 0.1853 | 0.156 | 29.78 | 1.2214 | 0.0233 | 29.32 | 3.3772 | 0.0118 | 40.91 | 1.0741 |
| 20 | 520 | 0.1817 | 0.175 | 35.17 | 1.3101 | 0.0253 | 36.67 | 3.4839 | 0.0073 | 28.17 | 1.2132 |
| 20 | 540 | 0.2041 | 0.1459 | 32.45 | 1.185 | 0.0251 | 32.42 | 3.1388 | 0.0103 | 35.13 | 1.1723 |
| 20 | 560 | 0.228 | 0.1315 | 31.65 | 1.3132 | 0.026 | 36.11 | 3.3372 | 0.0091 | 32.24 | 1.1637 |

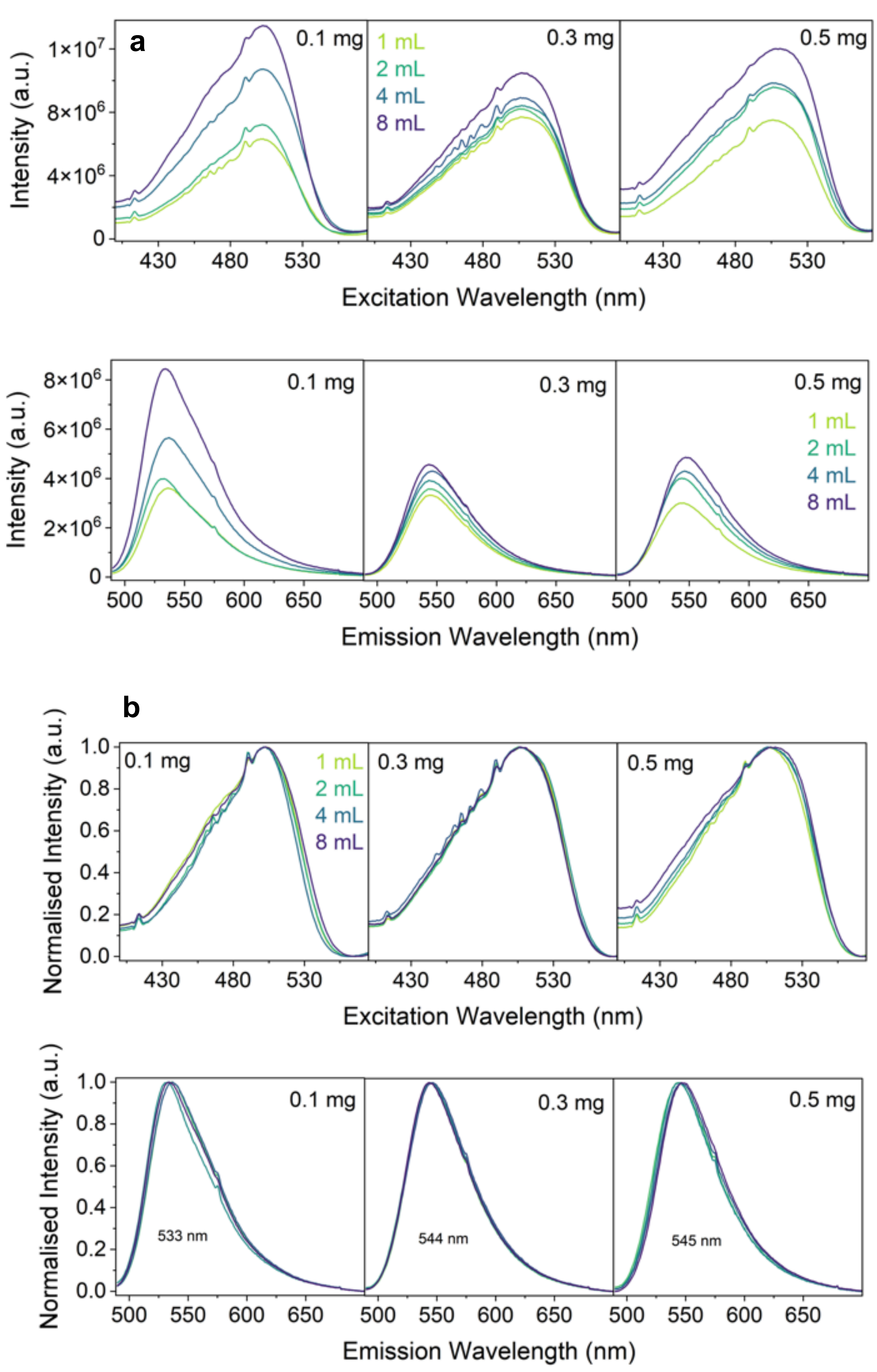

Figure S35. Emission (excited at 470 nm) and excitation (observed at 600 nm) spectra of F@ZIF-L with 0.1, 0.3 and 0.5 mg F synthesis content and various MeOH quantities (1, 2, 4, and 8 mL). Data is presented with measured intensities (a) and normalised (b).

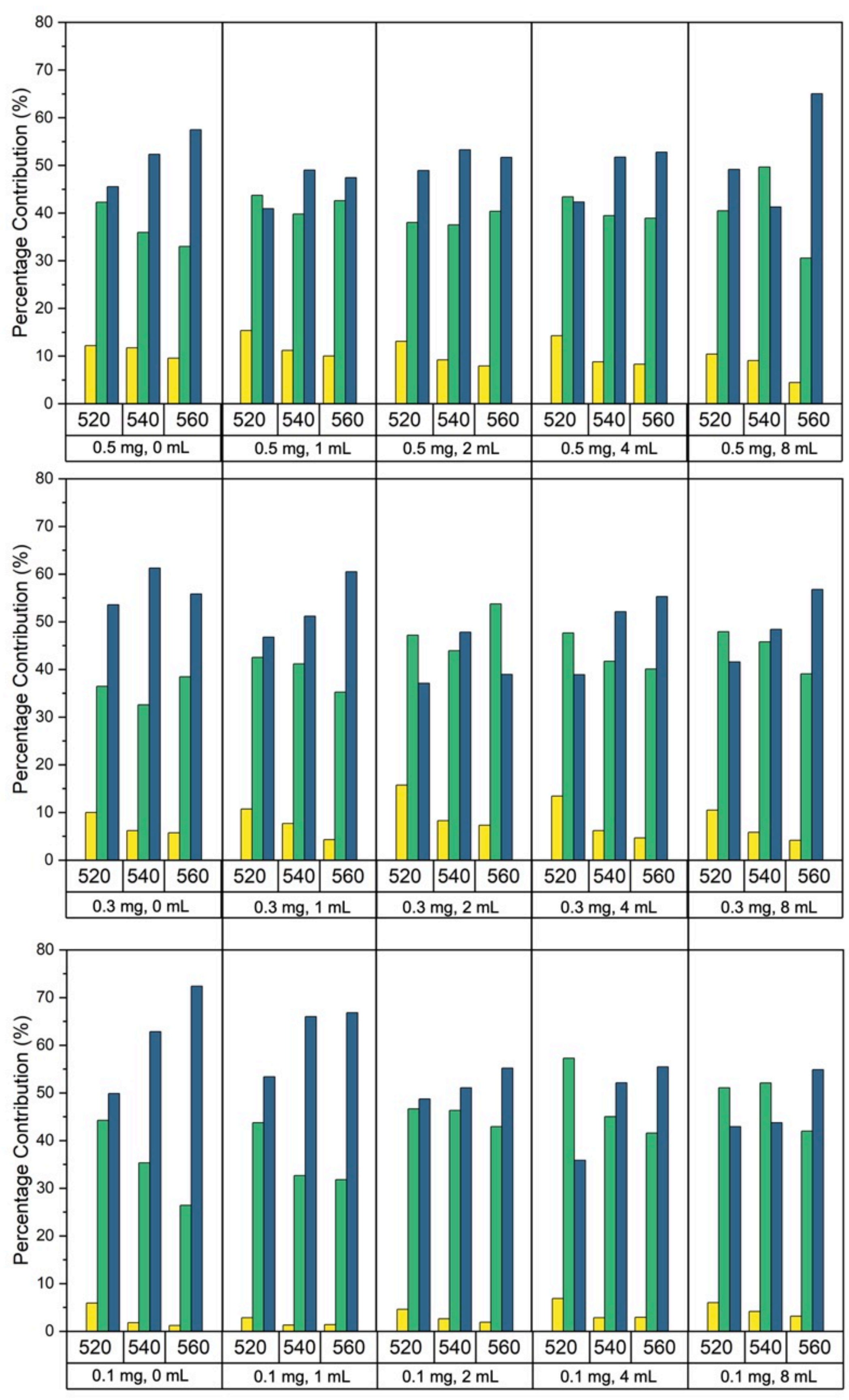

Figure S36. Lifetime decay curve fit parameters of F@ZIF-L with 0.1, 0.3 and 0.5 mg F synthesis content and various MeOH quantities (1, 2, 4, and 8 mL).

Table S3. Lifetime decay data for F@ZIF-L samples (X_YmL where X = guest content in synthesis in mg, Y = MeOH content during synthesis)

| Sample | λobs [nm] | τ1[ns] | a1 | c1[%] | τ2[ns] | a2 | c2[%] | τ3[ns] | a3 | c3[%] | χ2 |
|---|---|---|---|---|---|---|---|---|---|---|---|
| 0.1_0mL | 520 | 0.4573 | 0.0238 | 5.92 | 2.5918 | 0.0314 | 44.23 | 5.0193 | 0.0183 | 49.85 | 1.139 |
| 0.1_0mL | 540 | 0.4986 | 0.0085 | 1.82 | 3.1358 | 0.0263 | 35.32 | 5.473 | 0.0268 | 62.86 | 1.1183 |
| 0.1_0mL | 560 | 0.5133 | 0.0059 | 1.2 | 3.2149 | 0.0206 | 26.4 | 5.6183 | 0.0324 | 72.39 | 1.0899 |
| 0.1_1mL | 520 | 0.3936 | 0.0146 | 2.86 | 2.7268 | 0.0323 | 43.75 | 4.8938 | 0.0219 | 53.39 | 1.1061 |
| 0.1_1mL | 540 | 0.2668 | 0.0118 | 1.32 | 3.1996 | 0.0244 | 32.67 | 5.209 | 0.0302 | 66.01 | 1.0152 |
| 0.1_1mL | 560 | 0.389 | 0.0085 | 1.38 | 3.2818 | 0.0232 | 31.82 | 5.3673 | 0.0298 | 66.8 | 1.0349 |
| 0.1_2mL | 520 | 0.4421 | 0.0193 | 4.61 | 2.4554 | 0.0352 | 46.66 | 4.6505 | 0.0194 | 48.73 | 1.0967 |
| 0.1_2mL | 540 | 0.6847 | 0.0082 | 2.61 | 3.0932 | 0.0322 | 46.31 | 5.0924 | 0.0216 | 51.09 | 0.9869 |
| 0.1_2mL | 560 | 0.411 | 0.0103 | 1.87 | 3.076 | 0.0316 | 42.91 | 5.2598 | 0.0238 | 55.21 | 1.0634 |
| 0.1_4mL | 520 | 0.542 | 0.0221 | 6.85 | 2.6434 | 0.0379 | 57.29 | 4.9644 | 0.0126 | 35.86 | 1.0549 |
| 0.1_4mL | 540 | 0.4374 | 0.0132 | 2.85 | 2.8367 | 0.0321 | 45.02 | 4.9362 | 0.0214 | 52.13 | 1.0192 |
| 0.1_4mL | 560 | 0.4379 | 0.0141 | 2.95 | 2.8951 | 0.0299 | 41.58 | 5.0419 | 0.0229 | 55.47 | 1.055 |
| 0.1_8mL | 520 | 0.3982 | 0.027 | 6.01 | 2.5379 | 0.036 | 51.07 | 4.9702 | 0.0155 | 42.92 | 1.1153 |
| 0.1_8mL | 540 | 0.5414 | 0.0156 | 4.17 | 3.0309 | 0.0348 | 52.08 | 5.3913 | 0.0165 | 43.75 | 1.0814 |
| 0.1_8mL | 560 | 0.4432 | 0.015 | 3.16 | 2.8519 | 0.0308 | 41.97 | 5.2664 | 0.0218 | 54.87 | 1.0798 |
| | | | | | | | | | | | |
| 0.3_0mL | 520 | 0.3845 | 0.0421 | 9.97 | 2.1352 | 0.0277 | 36.48 | 4.638 | 0.0187 | 53.55 | 1.1168 |
| 0.3_0mL | 540 | 0.4655 | 0.0248 | 6.19 | 2.3258 | 0.0262 | 32.57 | 4.8565 | 0.0236 | 61.24 | 1.0987 |
| 0.3_0mL | 560 | 0.6184 | 0.0185 | 5.73 | 2.7857 | 0.0275 | 38.45 | 5.2607 | 0.0212 | 55.82 | 1.1058 |
| 0.3_1mL | 520 | 0.4291 | 0.04 | 10.72 | 2.2744 | 0.0299 | 42.51 | 4.777 | 0.0157 | 46.77 | 1.0942 |
| 0.3_1mL | 540 | 0.7347 | 0.0202 | 7.68 | 2.8052 | 0.0283 | 41.15 | 5.13 | 0.0192 | 51.17 | 1.2143 |
| 0.3_1mL | 560 | 0.5858 | 0.0152 | 4.28 | 2.7341 | 0.0268 | 35.25 | 5.1845 | 0.0242 | 60.47 | 1.0887 |
| 0.3_2mL | 520 | 0.5018 | 0.0448 | 15.74 | 2.2055 | 0.0305 | 47.18 | 4.252 | 0.0124 | 37.08 | 1.1162 |
| 0.3_2mL | 540 | 0.6179 | 0.023 | 8.29 | 2.4312 | 0.031 | 43.91 | 4.4103 | 0.0186 | 47.8 | 1.11 |
| 0.3_2mL | 560 | 0.7331 | 0.0181 | 7.34 | 2.8674 | 0.0339 | 53.73 | 4.934 | 0.0143 | 38.93 | 1.1056 |
| 0.3_4mL | 520 | 0.5619 | 0.0364 | 13.45 | 2.3147 | 0.0313 | 47.65 | 4.4955 | 0.0132 | 38.91 | 1.0753 |
| 0.3_4mL | 540 | 0.5371 | 0.0213 | 6.19 | 2.51 | 0.0307 | 41.72 | 4.7248 | 0.0204 | 52.1 | 1.0326 |
| 0.3_4mL | 560 | 0.5744 | 0.016 | 4.64 | 2.6885 | 0.0296 | 40.07 | 4.9412 | 0.0223 | 55.3 | 1.0463 |
| 0.3_8mL | 520 | 0.438 | 0.0368 | 10.5 | 2.1788 | 0.0338 | 47.92 | 4.3375 | 0.0147 | 41.58 | 1.1738 |
| 0.3_8mL | 540 | 0.5058 | 0.0213 | 5.83 | 2.645 | 0.032 | 45.78 | 4.6284 | 0.0194 | 48.4 | 1.0837 |
| 0.3_8mL | 560 | 0.3891 | 0.0213 | 4.15 | 2.6992 | 0.0289 | 39.08 | 4.7425 | 0.0239 | 56.77 | 1.1539 |
| | | | | | | | | | | | |
| 0.5_0mL | 520 | 0.3669 | 0.049 | 12.17 | 2.0519 | 0.0304 | 42.28 | 4.1911 | 0.016 | 45.55 | 1.1655 |
| 0.5_0mL | 540 | 0.5133 | 0.0363 | 11.74 | 2.1894 | 0.026 | 35.93 | 4.2847 | 0.0194 | 52.33 | 1.0684 |
| 0.5_0mL | 560 | 0.4852 | 0.0323 | 9.54 | 2.1324 | 0.0255 | 33 | 4.3487 | 0.0217 | 57.46 | 1.1327 |
| 0.5_1mL | 520 | 0.4622 | 0.0471 | 15.34 | 2.1835 | 0.0284 | 43.73 | 4.4136 | 0.0132 | 40.93 | 1.1557 |
| 0.5_1mL | 540 | 0.5284 | 0.0338 | 11.19 | 2.2695 | 0.028 | 39.79 | 4.5363 | 0.0172 | 49.02 | 1.1344 |
| 0.5_1mL | 560 | 0.6329 | 0.0267 | 9.99 | 2.5472 | 0.0283 | 42.59 | 4.766 | 0.0168 | 47.42 | 1.0682 |
| 0.5_2mL | 520 | 0.3696 | 0.0505 | 13.08 | 1.8744 | 0.0289 | 38 | 4.3072 | 0.0162 | 48.92 | 1.1599 |
| 0.5_2mL | 540 | 0.4711 | 0.0319 | 9.21 | 2.1514 | 0.0285 | 37.53 | 4.574 | 0.019 | 53.27 | 0.9731 |
| 0.5_2mL | 560 | 0.5264 | 0.0271 | 7.93 | 2.4624 | 0.0294 | 40.37 | 4.8907 | 0.019 | 51.69 | 1.1871 |
| 0.5_4mL | 520 | 0.3679 | 0.0538 | 14.25 | 1.9256 | 0.0313 | 43.42 | 4.336 | 0.0136 | 42.34 | 1.1013 |
| 0.5_4mL | 540 | 0.4102 | 0.0342 | 8.79 | 2.0581 | 0.0306 | 39.48 | 4.4852 | 0.0184 | 51.73 | 1.0909 |
| 0.5_4mL | 560 | 0.5307 | 0.0268 | 8.29 | 2.3367 | 0.0286 | 38.92 | 4.7185 | 0.0192 | 52.79 | 1.1025 |
| 0.5_8mL | 520 | 0.4039 | 0.0389 | 10.42 | 1.8885 | 0.0323 | 40.45 | 4.0083 | 0.0185 | 49.13 | 1.0365 |
| 0.5_8mL | 540 | 0.683 | 0.0236 | 9.04 | 2.7201 | 0.0326 | 49.66 | 4.6655 | 0.0158 | 41.3 | 1.1493 |
| 0.5_8mL | 560 | 0.4309 | 0.0191 | 4.43 | 2.2153 | 0.0256 | 30.54 | 4.4158 | 0.0274 | 65.03 | 1.1447 |

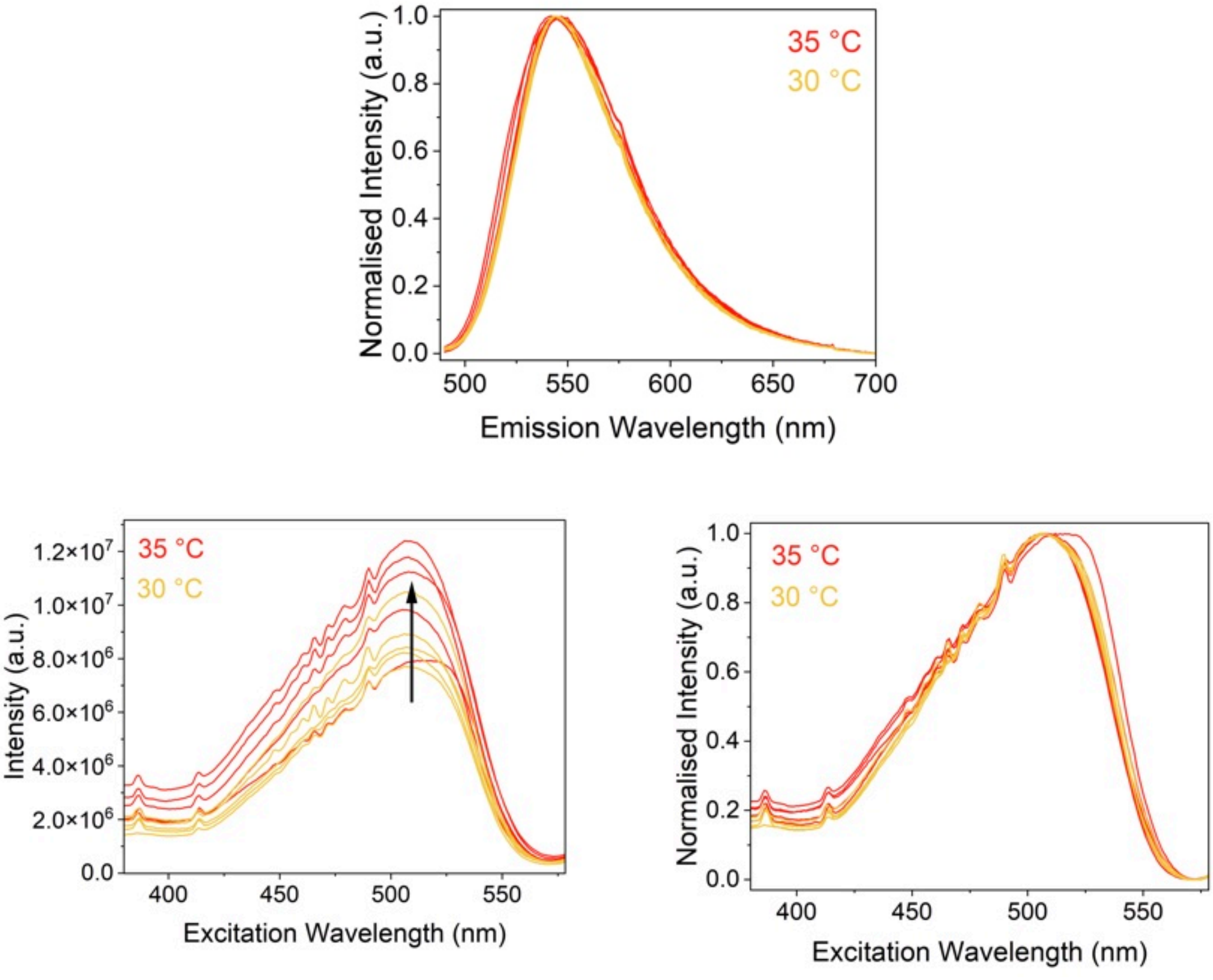

Figure S37. Above: normalised emission spectra (excited at 470 nm) of F@ZIF-L samples synthesised with different temperatures. Below: excitation spectra (observed at 600 nm) of F@ZIF-L samples synthesised with different temperatures. Arrow indicates effect of increasing MeOH content during synthesis (0, 1, 2, 4, and 8 mL of MeOH).

Table S4. Lifetime decay data for F@ZIF-L (0.3 mg, F) samples synthesised at 30 °C and 35 °C with varying MeOH content.

| Sample | λobs [nm] | τ1[ns] | a1 | c1[%] | τ2[ns] | a2 | c2[%] | τ3[ns] | a3 | c3[%] | χ2 |
|---|---|---|---|---|---|---|---|---|---|---|---|
| 0.3_35_H2O | 540 | 0.6365 | 0.0242 | 8.25 | 2.75 | 0.0293 | 43.18 | 5.3806 | 0.0169 | 48.57 | 1.1449 |
| 0.3_35_1 | 540 | 0.8607 | 0.0185 | 8.16 | 2.9854 | 0.0298 | 45.43 | 5.1421 | 0.0177 | 46.41 | 1.0988 |
| 0.3_35_2 | 540 | 0.5711 | 0.0211 | 6.22 | 2.7267 | 0.0285 | 39.99 | 4.9037 | 0.0213 | 53.79 | 1.0531 |
| 0.3_35_4 | 540 | 0.656 | 0.0129 | 3.92 | 3.0143 | 0.0278 | 38.69 | 5.2366 | 0.0237 | 57.4 | 1.0526 |
| 0.3_35_8 | 540 | 0.4579 | 0.0247 | 6.49 | 2.1815 | 0.0258 | 32.43 | 4.3032 | 0.0247 | 61.08 | 1.0663 |
| 0.3_30_H2O | 540 | 0.4655 | 0.0248 | 6.19 | 2.3258 | 0.0262 | 32.57 | 4.8565 | 0.0236 | 61.24 | 1.0987 |
| 0.3_30_1 | 540 | 0.4534 | 0.021 | 5.31 | 2.1563 | 0.0295 | 35.53 | 4.4167 | 0.024 | 59.16 | 1.1296 |
| 0.3_30_2 | 540 | 0.6179 | 0.023 | 8.29 | 2.4312 | 0.031 | 43.91 | 4.4103 | 0.0186 | 47.8 | 1.11 |
| 0.3_30_4 | 540 | 0.5371 | 0.0213 | 6.19 | 2.51 | 0.0307 | 41.72 | 4.7248 | 0.0204 | 52.1 | 1.0326 |
| 0.3_30_8 | 540 | 0.5058 | 0.0213 | 5.83 | 2.645 | 0.032 | 45.78 | 4.6284 | 0.0194 | 48.4 | 1.0837 |

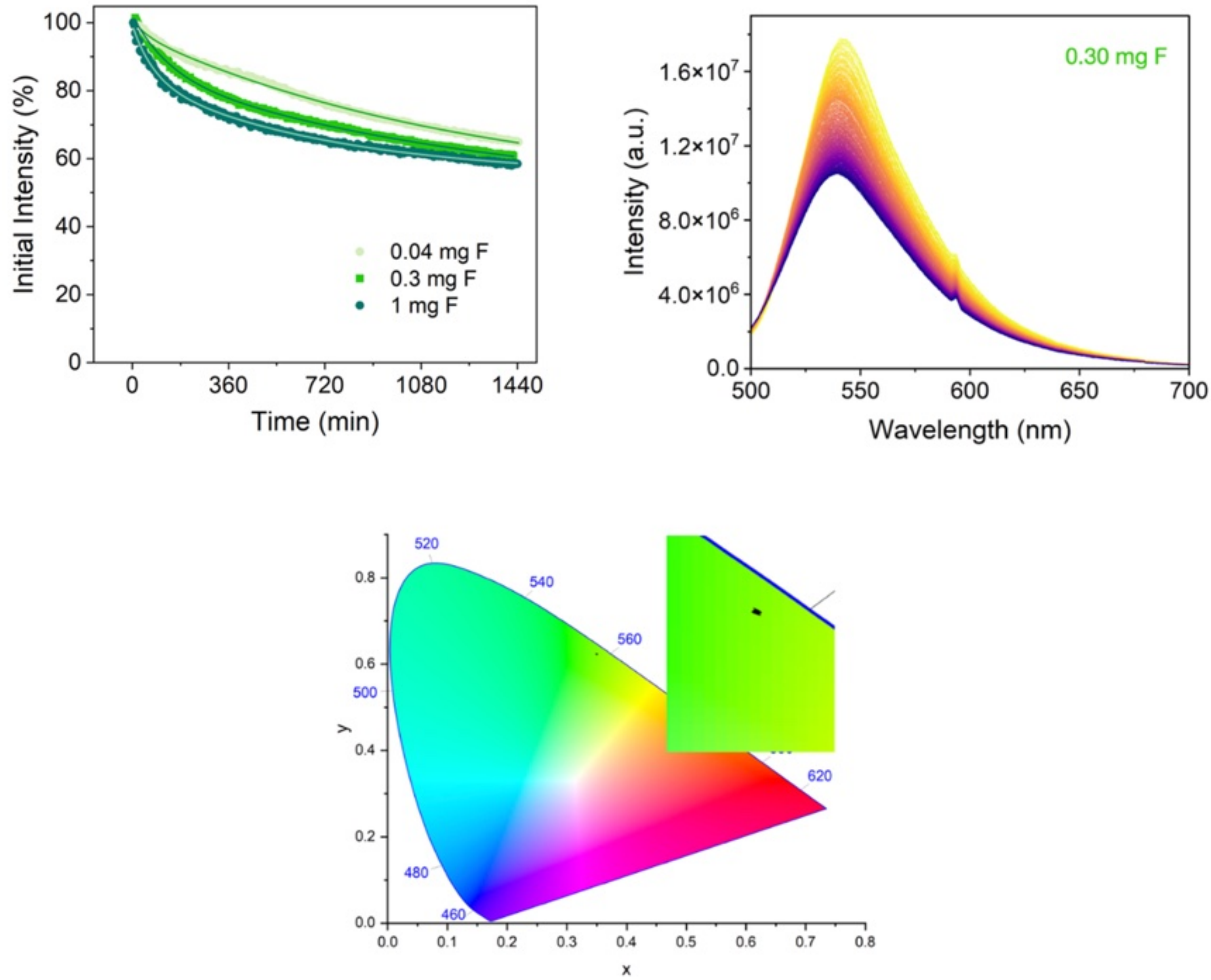

Figure S38. Above: loss of intensity over 24-hour photostability test, with each F@ZIF-L sample exposed to 150 W excitation source tuned to each sample's absorption maximum (left), example procession of emission band over 24 hours (yellow = 0 minutes, purple = 24 hours) (right). Below: Negligible change in chromaticity over 24 hours confirming retention of emission colour.

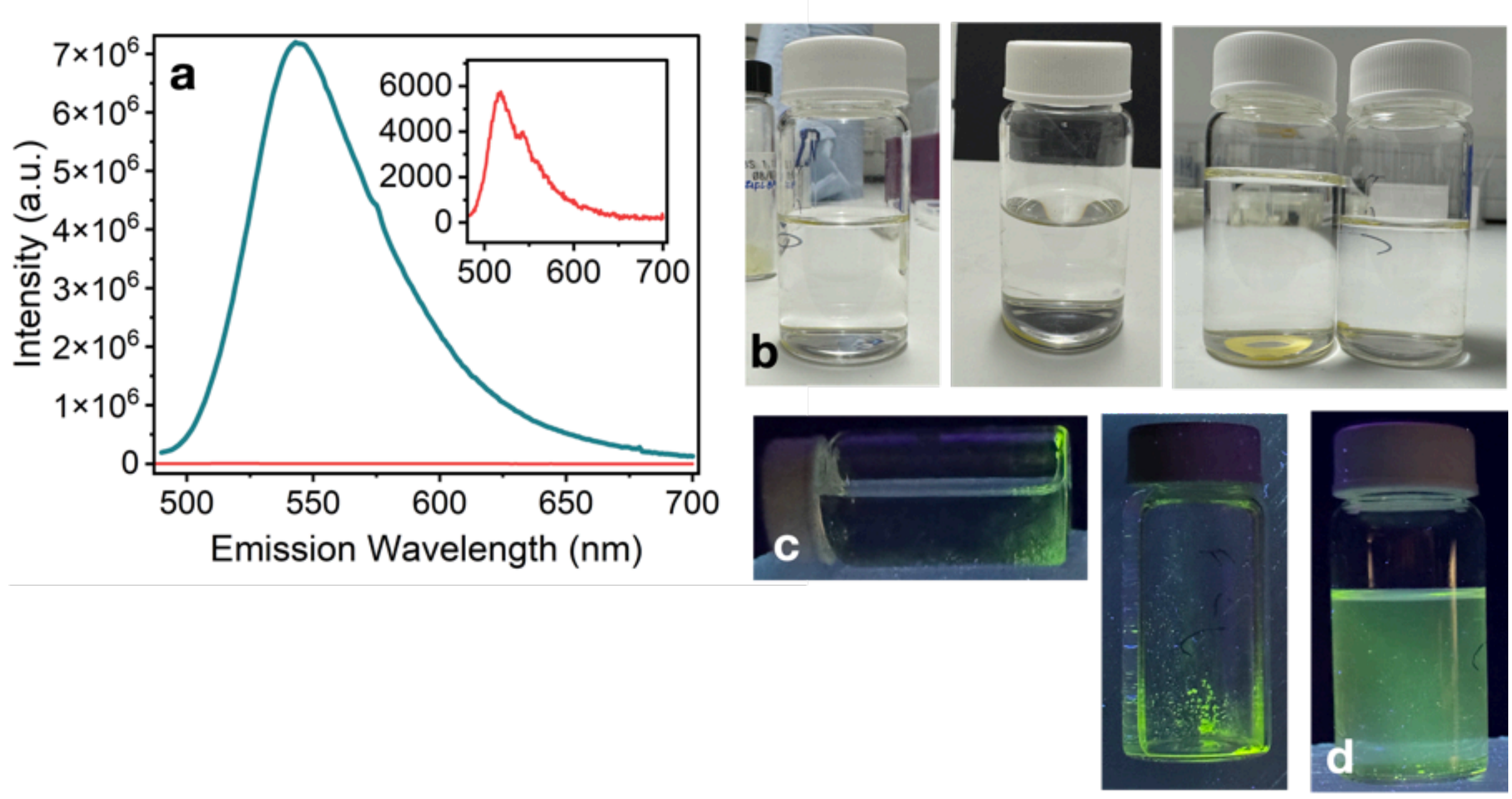

Figure S39. a) Emission of F@ZIF-L after 13 months suspended in 15 mL MeOH (blue), compared with the emission from the MeOH (red). Inset: close-up of the MeOH spectra to indicate trace quantity of F present (0.0071 wt.%). b) 20 mL sample vials after 13 month showing lack of particle dispersion or tinting of MeOH. c) Sample vials from (b) under 365 nm UV with minimal disturbance, showing emission only from F@ZIF-L particles. d) Emission from vial after sample was mixed by shaking for 30 seconds.

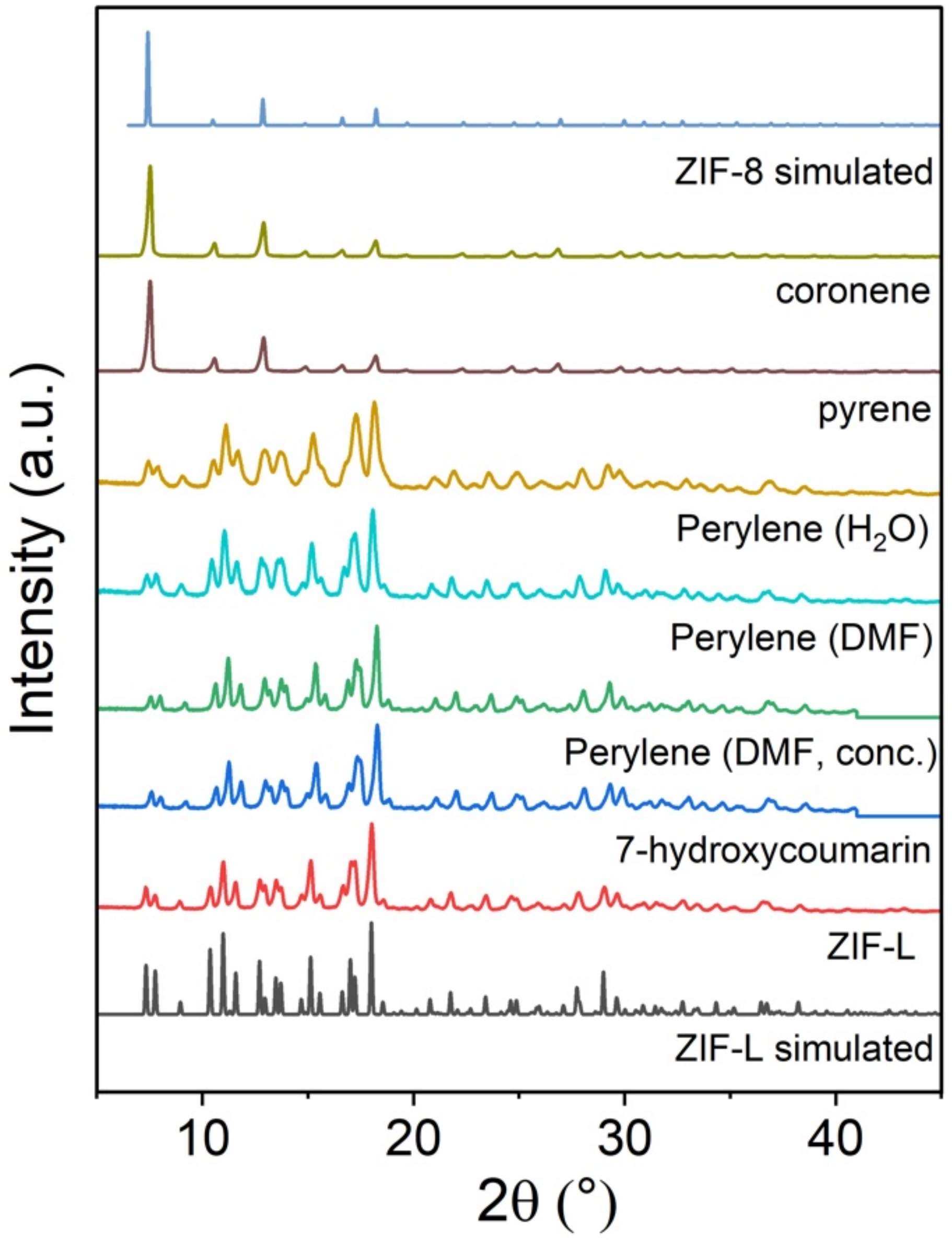

Figure S40. PXRD of various guest@ZIF-L samples with guest indicated for each corresponding spectra, compared to ZIF-L and ZIF-8.[2,5]

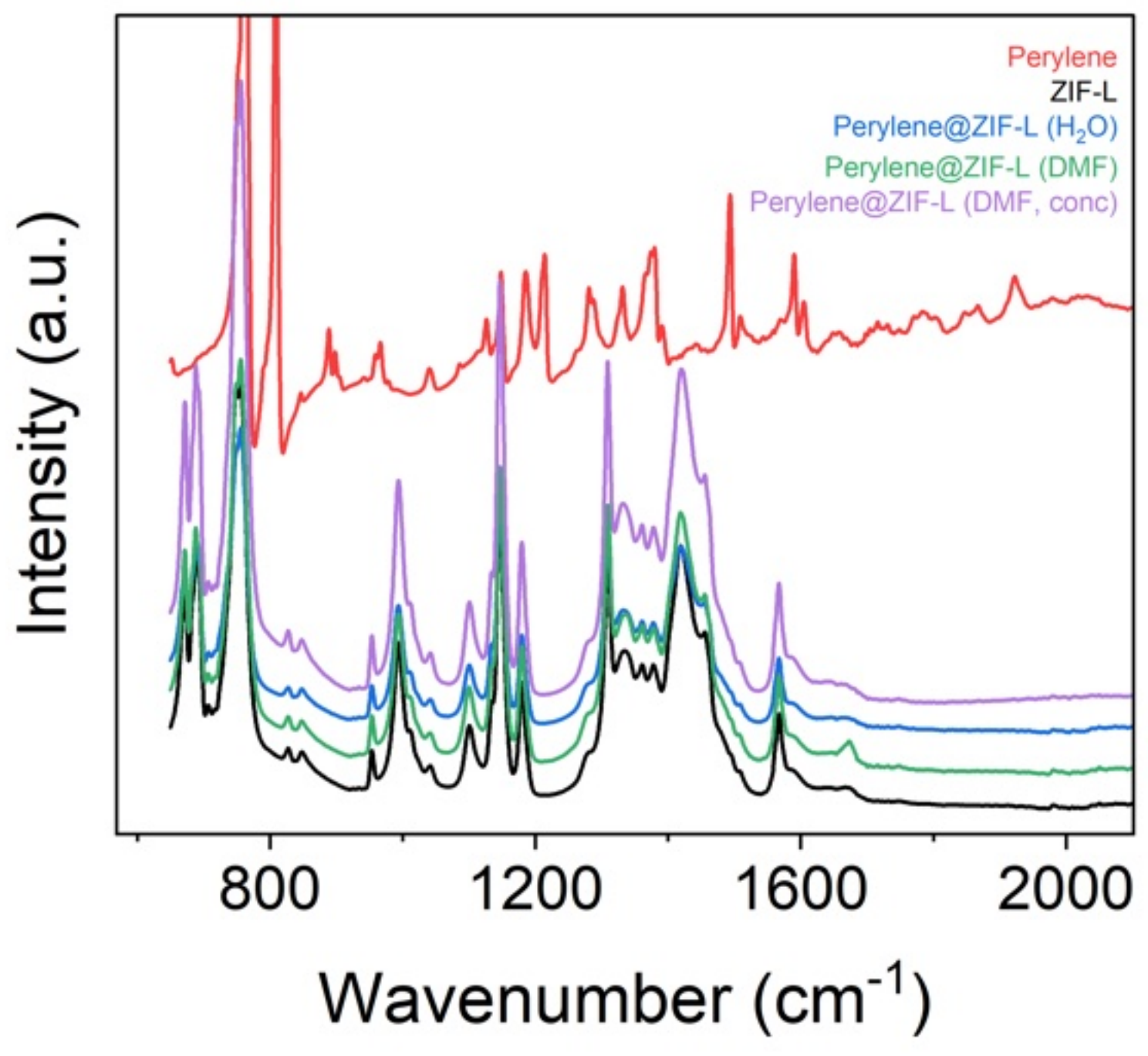

Figure S41. ATR-FTIR of perylene@ZIF-L samples.

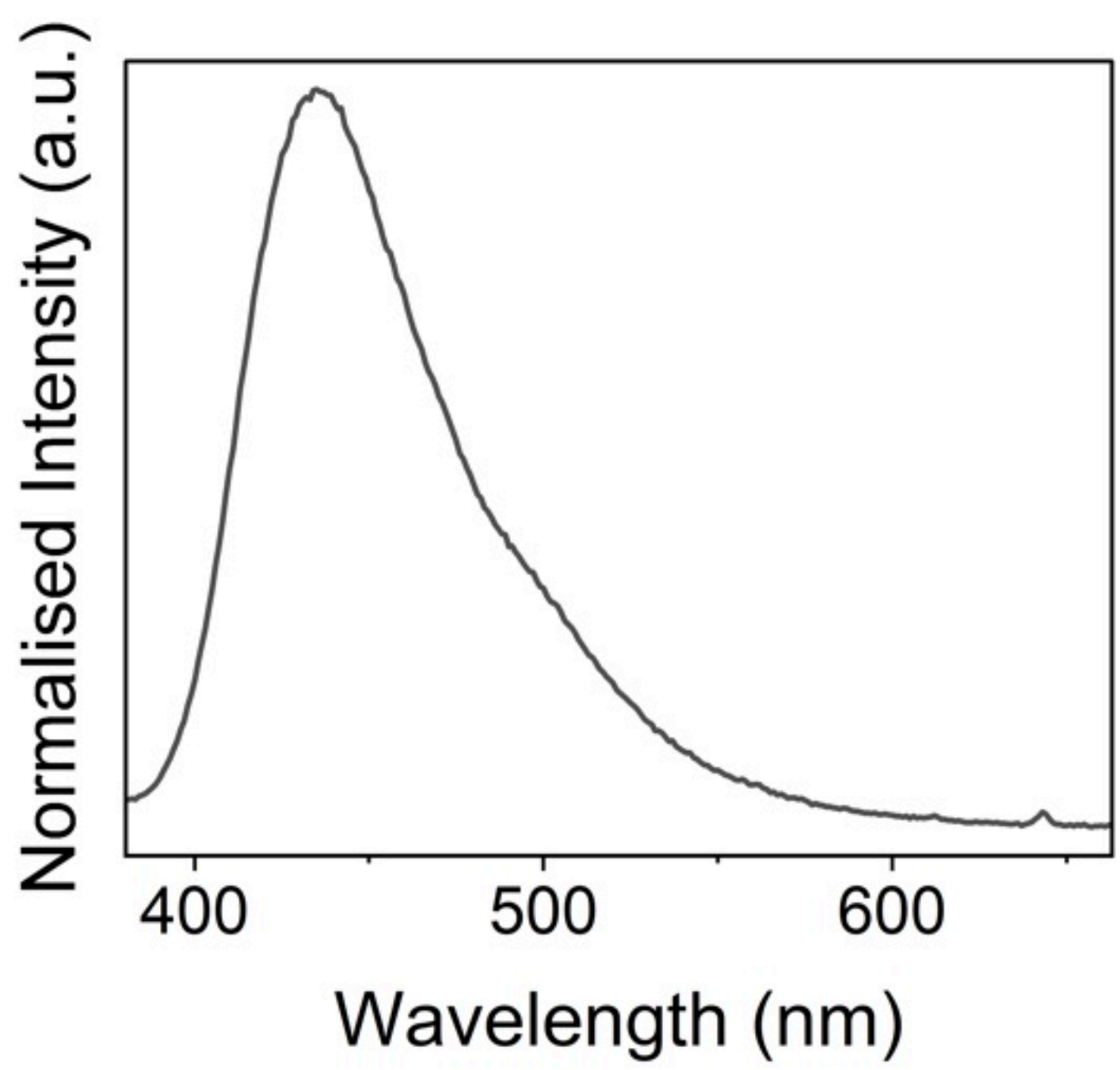

Figure S42. Emission spectra of 7-hydroxycoumarin@ZIF-L when excited at 350 nm.

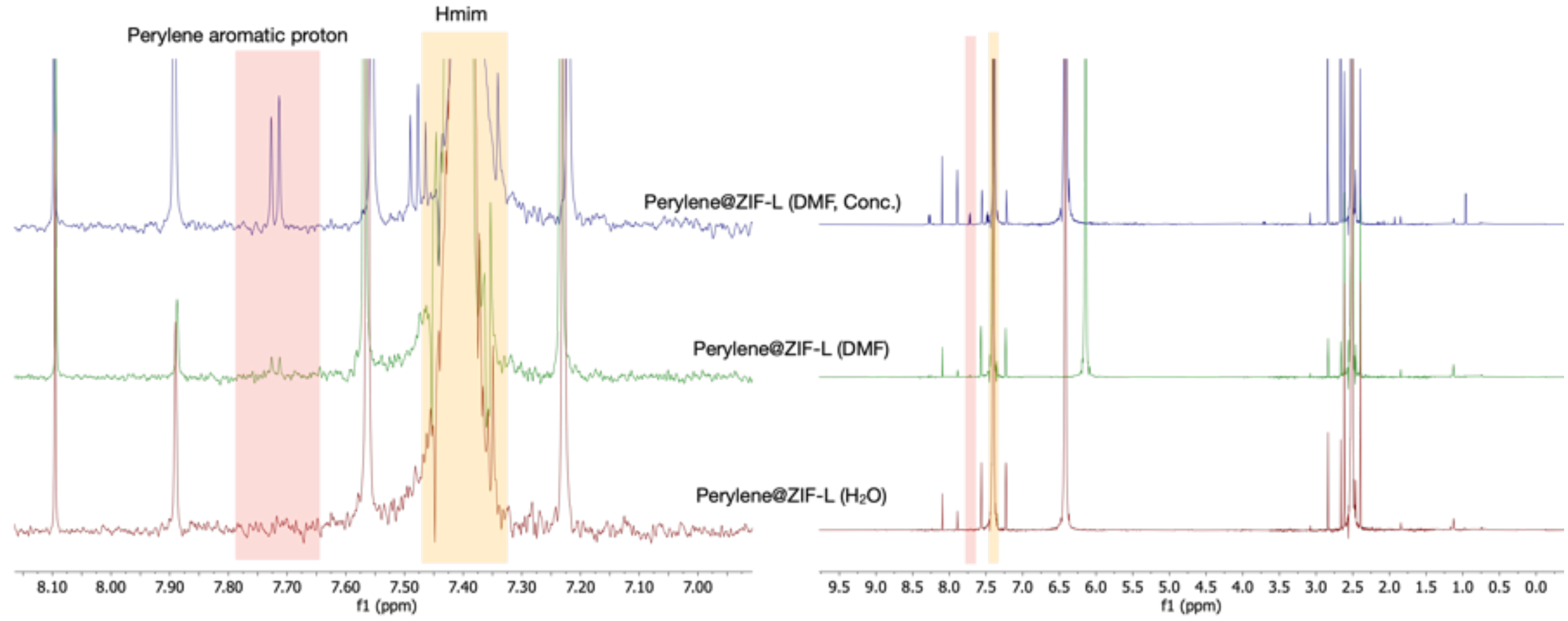

Figure S43. NMR digest of perylene@ZIF-L samples, with characteristic peaks from Hmim and perylene highlighted orange and red, respectively.

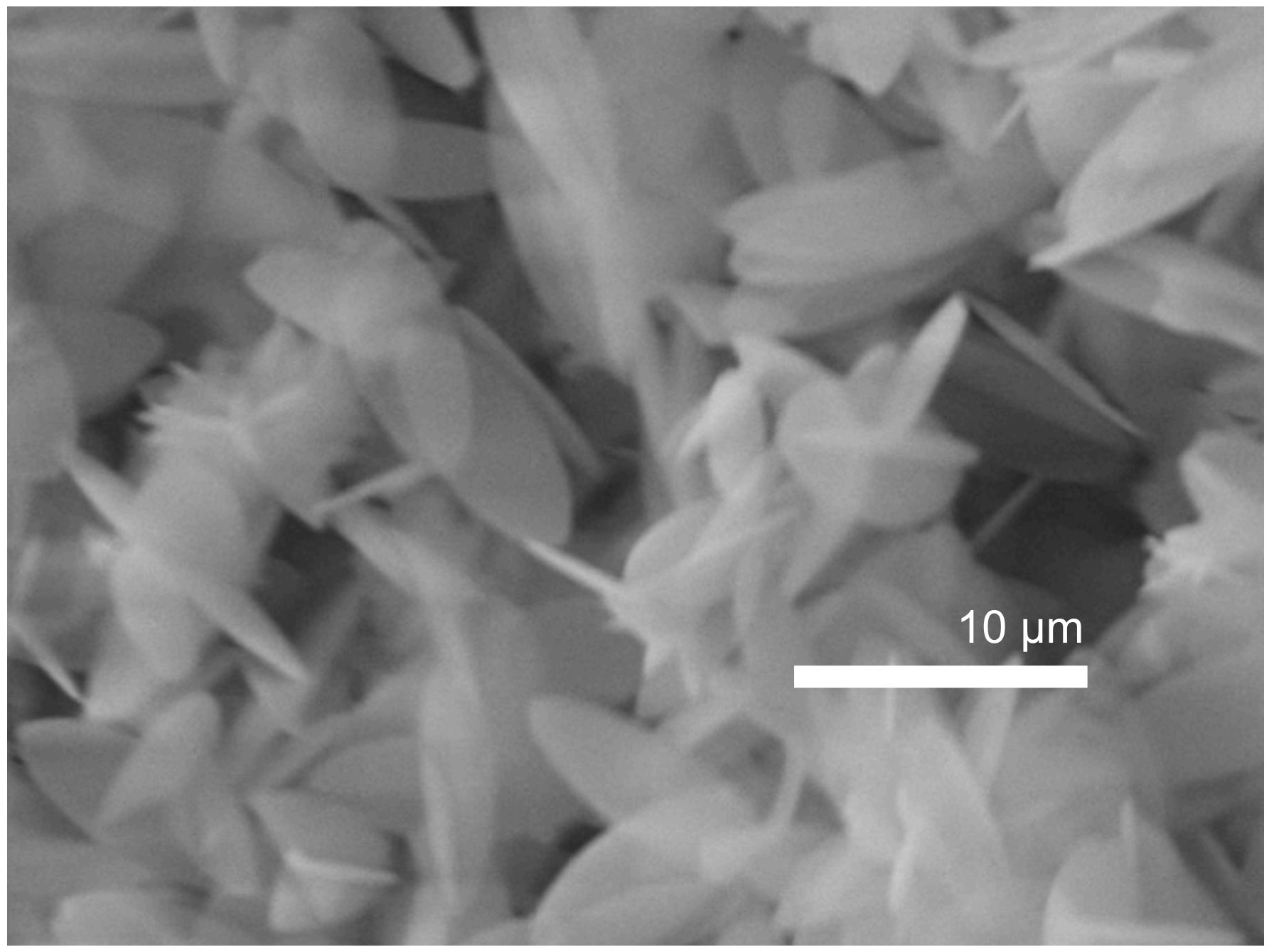

Figure S44 – SEM of perylene@ZIF-L synthesised in $H_2O$ (minimal guest solubility).

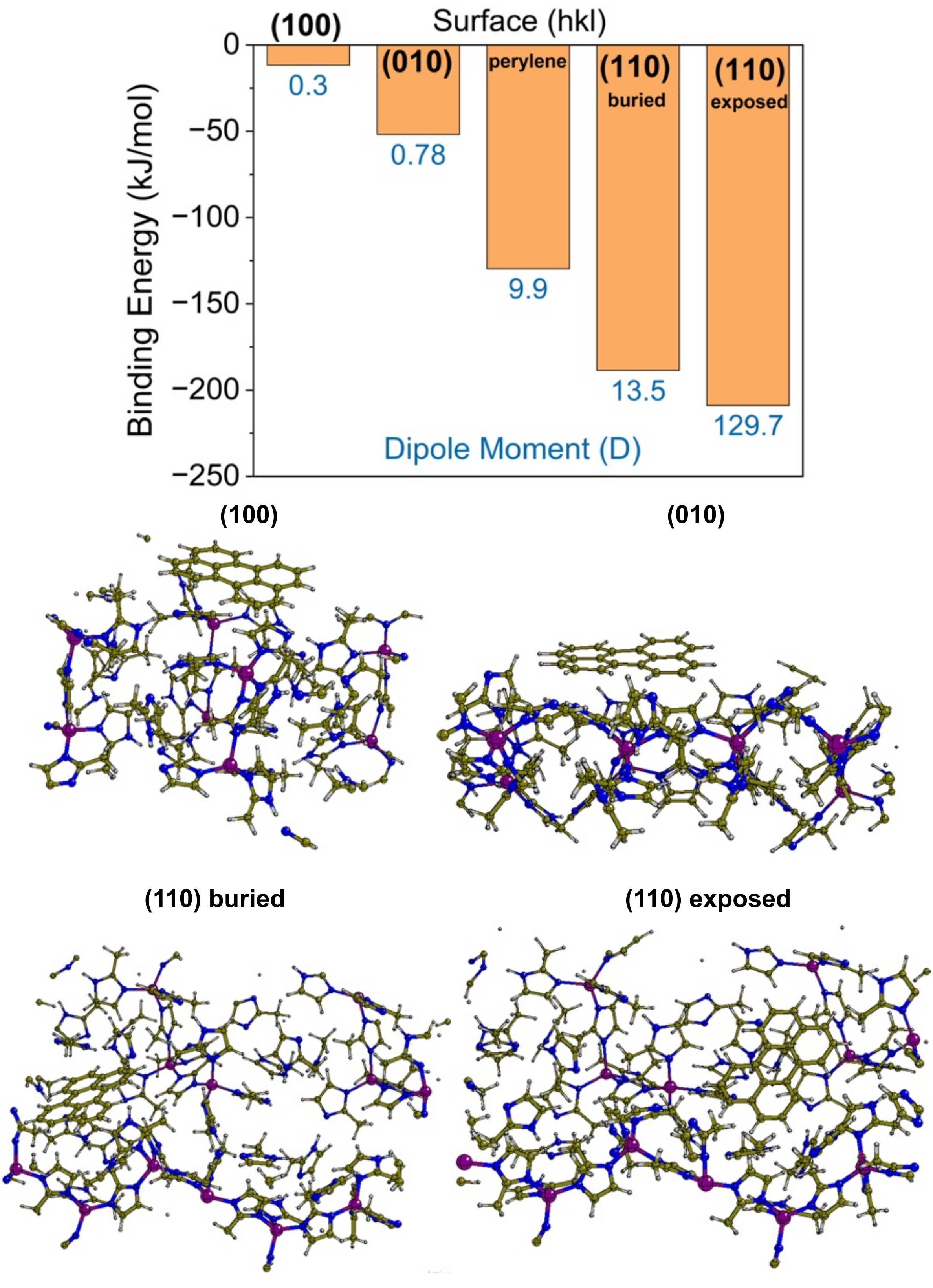

Figure S45. Binding energy and dipole moments from perylene adsorption on F@ZIF-L surface modelling (Zn = purple, N = blue, C = yellow, O = red, O = white).

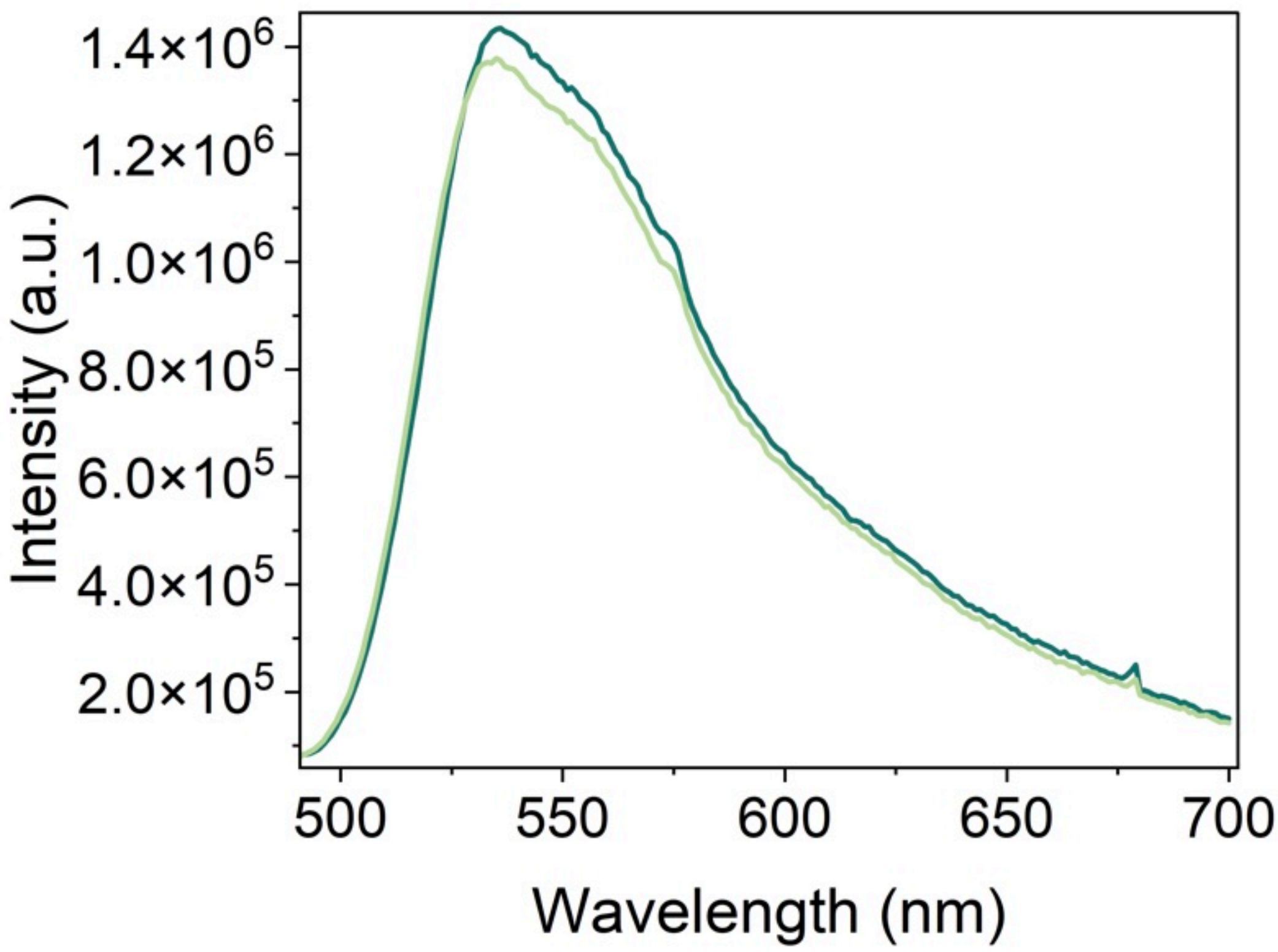

Figure S46. Emission spectra of F@ZIF-L surface on Zn film at a region towards the centre (light green) and edge (dark green). Excitation = 470 nm.